\documentclass[useAMS,usenatbib]{mn2e}

\usepackage{afterpage}
\usepackage{graphicx}
\usepackage{amsmath}
\usepackage{dcolumn}
\usepackage{longtable}

\usepackage{pdflscape}

\newcommand{\nstars}{64}

\newcommand{\mcc}[1]{\multicolumn{1}{c}{#1}}

\newcommand{\etas}{\eta_\ast}
\newcommand{\ra}{R_\mathrm{A}}
\newcommand{\rk}{R_\mathrm{K}}
\newcommand{\rs}{R_\ast}
\newcommand{\ms}{M_\ast}
\newcommand{\rp}{R_\mathrm{p}}
\newcommand{\vrot}{V_\mathrm{rot}}
\newcommand{\vcrit}{V_\mathrm{orb}}
\newcommand{\vesc}{V_\mathrm{esc}}
\newcommand{\vsini}{v \sin i}
\newcommand{\vinf}{V_\infty}

\newcommand{\beq}{B_\mathrm{eq}}

\newcommand{\bp}{B_\mathrm{p}}
\newcommand{\mbl}{\langle B_{z}\rangle}

\newcommand{\mdot}{\dot{M}_{B=0}}
\newcommand{\qbar}{\bar{Q}}
\newcommand{\gae}{\Gamma_e}
\newcommand{\alf}{\alpha_\mathrm{eff}}

\newcommand{\teff}{T_\mathrm{eff}}

\newcommand{\ls}{L_\star}
\newcommand{\lx}{L_\mathrm{X}}
\newcommand{\lum}{\log(L_\star/L_\odot)}
\newcommand{\lxls}{\log(\lx/\ls)}
\newcommand{\logg}{\log g}

\newcommand{\tauj}{\tau_\mathrm{J}}
\newcommand{\tsmax}{t_{\mathrm{s,max}}}
\newcommand{\ts}{t_\mathrm{s}}

\newcommand{\sm}{\,s$^{-1}$}

\newcommand{\rsol}{R_\odot}
\newcommand{\msol}{M_\odot}

\newcommand{\oriunc}{$\theta^1$\,Ori\,C}
\newcommand{\sigorie}{$\sigma$\,Ori\,E}

\newcommand{\beqn}{\begin{equation}}
\newcommand{\eeqn}{\end{equation}}
\newcommand{\beqa}{\begin{eqnarray}}
\newcommand{\eeqa}{\end{eqnarray}}

\usepackage{color}
\definecolor{darkblue}{rgb}{0.0, 0.0, 0.60}

\usepackage{hyperref}
\hypersetup{bookmarks=true, colorlinks=true, linkcolor=darkblue, citecolor=darkblue, filecolor=darkblue, urlcolor=darkblue}

\title[Classification of Massive Star Magnetospheres]{A Magnetic Confinement vs.\ Rotation Classification of Massive-Star Magnetospheres
}

\author[Petit et al.]{%
V. Petit$^{1}$\thanks{E-mail: VPetit@wcupa.edu}, 
S.~P. Owocki$^{2}$,
G.~A. Wade$^{3}$,
D.~H. Cohen$^{4}$,
J.~O. Sundqvist$^{2}$,
\newauthor
M. Gagn\'e$^{1}$,
J. Ma\'iz Apell\'aniz$^{5}$,
M.~E. Oksala$^{6}$,
D.~A. Bohlender$^{7}$,
Th. Rivinius$^{8}$,
\newauthor
H.~F. Henrichs$^{9}$,
E. Alecian$^{10}$,
R.~H.~D Townsend$^{11}$,
A. ud-Doula$^{12}$
\newauthor
and the MiMeS Collaboration
\\
$^{1}$Dept. of Geology \& Astronomy, West Chester University, West Chester, PA 19383\\
$^{2}$Dept. of Physics \& Astronomy, University of Delaware, Bartol Research Institute, Newark, Delaware 19716, USA\\
$^{3}$Dept. of Physics, Royal Military College of Canada, PO Box 17000, Stn Forces, Kingston, Ontario K7K 7B4, Canada\\
$^{4}$Dept. of Physics \& Astronomy, Swarthmore College, Swarthmore, PA 19081, USA\\
$^{5}$Instituto de Astrof\'isica de Andaluc\'ia-CSIC, Glorieta de la Astronom\'ia s/n, E-18008 Granada, Spain\\
$^{6}$Astronomick\'y \'ustav, Akademie v\v{e}d \v{C}esk\'e republiky, Fri\v{c}ova 298, 251\,65 Ond\v{r}ejov, Czech Republic\\
$^{7}$National Research Council of Canada, Herzberg Institute of Astrophysics, 5071 West Saanich Road, Victoria, BC V9E 2E7, Canada\\
$^{8}$ESO - European Organisation for Astronomical Research in the Southern Hemisphere, Casilla 19001, Santiago 19, Chile\\
$^{9}$Astronomical Institute Anton Pannekoek, University of Amsterdam, Science Park 904, 1098 XH Amsterdam, Netherlands\\
$^{10}$LESIA, Observatoire de Paris, CNRS UMR 8109, UPMC, Universit\'e Paris Diderot, 5 place Jules Janssen, 92190 Meudon, France\\
$^{11}$Department of Astronomy, University of Wisconsin-Madison, 475 N Charter Street, Madison, WI 53706, USA\\
$^{12}$Penn State Worthington Scranton, Dunmore, PA 18512, USA
}
\begin{document}

%
%
%


\def\jnl@style{\it}
\def\aaref@jnl#1{{\jnl@style#1}}

\def\aaref@jnl#1{{\jnl@style#1}}

\def\aj{\aaref@jnl{AJ}}                   
\def\araa{\aaref@jnl{ARA\&A}}             
\def\apj{\aaref@jnl{ApJ}}                 
\def\apjl{\aaref@jnl{ApJ}}                
\def\apjs{\aaref@jnl{ApJS}}               
\def\ao{\aaref@jnl{Appl.~Opt.}}           
\def\apss{\aaref@jnl{Ap\&SS}}             
\def\aap{\aaref@jnl{A\&A}}                
\def\aapr{\aaref@jnl{A\&A~Rev.}}          
\def\aaps{\aaref@jnl{A\&AS}}              
\def\azh{\aaref@jnl{AZh}}                 
\def\baas{\aaref@jnl{BAAS}}               
\def\jrasc{\aaref@jnl{JRASC}}             
\def\memras{\aaref@jnl{MmRAS}}            
\def\mnras{\aaref@jnl{MNRAS}}             
\def\pra{\aaref@jnl{Phys.~Rev.~A}}        
\def\prb{\aaref@jnl{Phys.~Rev.~B}}        
\def\prc{\aaref@jnl{Phys.~Rev.~C}}        
\def\prd{\aaref@jnl{Phys.~Rev.~D}}        
\def\pre{\aaref@jnl{Phys.~Rev.~E}}        
\def\prl{\aaref@jnl{Phys.~Rev.~Lett.}}    
\def\pasp{\aaref@jnl{PASP}}               
\def\pasj{\aaref@jnl{PASJ}}               
\def\qjras{\aaref@jnl{QJRAS}}             
\def\skytel{\aaref@jnl{S\&T}}             
\def\solphys{\aaref@jnl{Sol.~Phys.}}      
\def\sovast{\aaref@jnl{Soviet~Ast.}}      
\def\ssr{\aaref@jnl{Space~Sci.~Rev.}}     
\def\zap{\aaref@jnl{ZAp}}                 
\def\nat{\aaref@jnl{Nature}}              
\def\iaucirc{\aaref@jnl{IAU~Circ.}}       
\def\aplett{\aaref@jnl{Astrophys.~Lett.}} 
\def\apspr{\aaref@jnl{Astrophys.~Space~Phys.~Res.}}
\def\bain{\aaref@jnl{Bull.~Astron.~Inst.~Netherlands}} 
\def\fcp{\aaref@jnl{Fund.~Cosmic~Phys.}}  
\def\gca{\aaref@jnl{Geochim.~Cosmochim.~Acta}}   
\def\grl{\aaref@jnl{Geophys.~Res.~Lett.}} 
\def\jcp{\aaref@jnl{J.~Chem.~Phys.}}      
\def\jgr{\aaref@jnl{J.~Geophys.~Res.}}    
\def\jqsrt{\aaref@jnl{J.~Quant.~Spec.~Radiat.~Transf.}}
\def\memsai{\aaref@jnl{Mem.~Soc.~Astron.~Italiana}}
\def\nphysa{\aaref@jnl{Nucl.~Phys.~A}}   
\def\physrep{\aaref@jnl{Phys.~Rep.}}   
\def\physscr{\aaref@jnl{Phys.~Scr}}   
\def\planss{\aaref@jnl{Planet.~Space~Sci.}}   
\def\procspie{\aaref@jnl{Proc.~SPIE}}   

\let\astap=\aap
\let\apjlett=\apjl
\let\apjsupp=\apjs
\let\applopt=\ao

\date{Accepted 2012 November 1.  Received 2012 November 1; in original form 2012 October 2 }

\pagerange{\pageref{firstpage}--\pageref{lastpage}} \pubyear{2012}

\maketitle

\label{firstpage}

\begin{abstract}

Building on results from the Magnetism in Massive Stars (MiMeS) project, this paper shows how a two-parameter classification of massive-star magnetospheres in terms of the magnetic wind confinement (which sets the Alfv\'{e}n radius $\ra$) and stellar rotation (which sets the Kepler co-rotation radius $\rk$) provides a useful organisation of both observational signatures and theoretical predictions. 
We compile the first comprehensive study of inferred and observed values for relevant stellar and magnetic parameters of \nstars\ confirmed magnetic OB stars with $\teff \ga 16$\,kK.
Using these parameters, we locate the stars in the magnetic confinement-rotation diagram, a log-log plot of $\rk$ vs.\ $\ra$. 
This diagram can be subdivided into regimes of {\em centrifugal magnetospheres} (CM), with $\ra > \rk$, vs. {\em dynamical magnetospheres} (DM), with $\rk > \ra$.
We show how key observational diagnostics, like the presence and characteristics of H$\alpha$ emission, depend on a star's position within the diagram, as well as other parameters, especially the expected wind mass-loss rates.
In particular, we identify two distinct populations of magnetic stars with H$\alpha$ emission: namely, slowly rotating O-type stars with narrow emission consistent with a DM, and more rapidly rotating B-type stars with broader emission associated with a CM.
For O-type stars, the high mass-loss rates are sufficient to accumulate enough material for line emission even within the relatively short free-fall timescale associated with a DM: this high mass-loss rate also leads to a rapid magnetic spindown of the stellar rotation. For the B-type stars, the longer confinement of a CM is required to accumulate sufficient emitting material from their relatively weak winds, which also lead to much longer spindown timescales.
Finally, we discuss how other observational diagnostics, e.g. variability of UV wind lines or X-ray emission, relate to the inferred magnetic properties of these stars, and summarise prospects for future developments in our understanding of massive-star magnetospheres.

\end{abstract}


\section{Introduction}

Building on pioneering detections of strong (kG) fields in the chemically peculiar Ap and Bp stars \citep[e.g.][]{1947ApJ...105..105B,1980ApJS...42..421B}, new generations of spectropolarimeters have directly revealed large-scale, organised (often predominantly dipolar) magnetic fields ranging in dipolar strength\footnote{In the following all field strengths will be given as dipolar, unless explicitly noted otherwise.} from order of 0.1 to 10 kG in several dozen OB stars \citep[e.g.][]{2002MNRAS.333...55D,2006MNRAS.365L...6D,2006MNRAS.369L..61H,2008MNRAS.387L..23P,2009MNRAS.400L..94G,2010MNRAS.407.1423M}.
In recent years,  an observational consortium known as MiMeS (for Magnetism in Massive Stars) has been carrying out surveys to detect new magnetic OB stars, while also monitoring known magnetic OB stars with high resolution spectroscopy and polarimetry \citep{2010arXiv1009.3563W}.
Concurrently, theoretical models \citep{2005ApJ...630L..81T,2007MNRAS.382..139T} and magnetohydrodynamical (MHD) simulations 
\citep{2002ApJ...576..413U,2008MNRAS.385...97U,2009MNRAS.392.1022U} have explored the dynamical interaction of such fields with stellar rotation and mass loss, showing for example how suitably strong fields can channel the radiatively driven stellar wind outflow into a circumstellar \textit{magnetosphere}.
This paper aims now to provide an initial classification of the observed magnetospheric properties for a broad sample of magnetic massive stars.

The idea of a magnetosphere has been exploited to explain particular properties of some massive stars, for example the photometric light curve and H$\alpha$ variations of the He-strong star \sigorie\ \citep{1978ApJ...224L...5L}, the UV resonance line variations of magnetic Bp stars \citep{1990ApJ...365..665S}, the X-ray properties of the O-type star \oriunc\ \citep{2005ApJ...628..986G}, and the radio emission of Ap-Bp stars that correlates with the field strength \citep{1992ApJ...393..341L}.

For a few specific stars, previous work has already shown some promising agreement between theoretical predictions and key observational characteristics. 
For example, the luminosity, hardness, and rotational modulation of X-rays observed in the O-type star \oriunc\ all match well the X-rays computed in MHD simulations of magnetically confined wind shocks, which result from the collision of the wind from opposite footpoints of closed magnetic loops in its $\sim1$\,kG dipole field \citep{2005ApJ...628..986G}.
In the B2p star \sigorie, the combination of its very strong ($\sim 10$\,kG) field and moderately fast (1.2-day period) rotation leads to formation of a centrifugally supported magnetosphere with observed, rotationally modulated Balmer line emission reasonably well explained within the Rigidly Rotating Magnetosphere model \citep[RRM;][]{2005MNRAS.357..251T,2005ApJ...630L..81T}. 
Most recently, \citet{2012MNRAS.423L..21S} showed that, even in the very slowly rotating (537-day period) O-type star HD\,191612, the magnetic confinement and transient, dynamical suspension of its strong wind mass loss leads to sufficient density to likewise match the observed rotationally modulated Balmer line emission.

Building on these results, along with those from the MiMeS observational survey, this paper compiles an exhaustive list of confirmed magnetic, hot OB stars, along with their physical, rotational and magnetic properties (\S \ref{sec:stars}). 
As a basis for organising this compilation according to modelling predictions, we follow (\S \ref{sec:theo})  the two-parameter theoretical study of \citet{2008MNRAS.385...97U}, which characterised MHD simulation results according to the strength of wind magnetic confinement ($\etas$) and  fraction of stellar rotation to orbital speed at the stellar equatorial radius ($W$). 
These dimensionless parameters uniquely define associated characteristic radii, namely the Alfv\'en radius $\ra$ and Kepler co-rotation radius $\rk$.

We show in particular (\S \ref{sec:disc}) that an associated log-log plot of known magnetic stars in the $\ra$-vs.-$\rk$ (or equivalently  $\etas$-vs.-$W$) plane, the \textit{magnetic confinement-rotation diagram}, provides a particularly useful initial classification for interpreting the H$\alpha$ properties of their associated magnetospheres. Furthermore, we also explore the UV and X-ray characteristics as potential additional proxies of magnetospheres (\S \ref{sec:other}). 
We briefly review our main findings and conclusions in \S \ref{sec:conclu}.

\section{Exhaustive list of magnetic O-type and early B-type stars}
\label{sec:stars}

A central goal of this paper is to compile a comprehensive list of OB stars for which magnetic fields have been convincingly detected via the Zeeman effect, so that their magnetospheres can be classified. 

The work here is done within the context of the MiMeS project \citep{2010arXiv1009.3563W}, which aims to expand the population of known magnetic stars, confirm the detection of poorly studied magnetic OB stars, and provide a modern determination of their magnetic field characteristics. These goals are being achieved through Large Program observing allocations at the Canada-France-Hawaii Telescope (CFHT), the T\'elescope Bernard-Lyot (TBL) and the ESO 3.6m Telescope to collect high resolution, high signal-to-noise ratio spectropolarimetric observations of massive stars \citep[see][respectively]{2010arXiv1009.3563W,2012MNRAS.419..959O,2011A&A...536L...6A}.

Table \ref{tab:stars} lists our derived sample of \nstars\ magnetic stars, ordered by spectral type and temperature. Column (1) gives the numerical identification (ID) we use in the figures. Column (2) gives the HD number or a SIMBAD-friendly\footnote{\href{http://simbad.u-strasbg.fr}{http://simbad.u-strasbg.fr}} designation. A dagger indicates that a note for that particular star is available in Appendix \ref{sec:notes}. Columns (3) and (4) give a commonly used designation and the spectral type, respectively. Column (5) indicates if the star is a known single- or double-line spectroscopic binary (SB1/2), slowly pulsating B-type star (SPB), $\beta$\,Cep-type pulsator ($\beta$\,Cep), or a Herbig Be star (HeBe). 
Table \ref{tab:ref} compiles, for each star, the list of references where information can be found or how it is derived from MiMeS observations or other archival data.

\clearpage
\onecolumn
\begin{landscape}
	{\small \begin{longtable}{l l l c c D{*}{\,\pm\,}{4,4} D{*}{\,\pm\,}{4,4} D{*}{\,\pm\,}{6,6} c c c c D{*}{\,}{2,4} } 
\caption{\label{tab:stars}List of magnetic massive OB stars and their physical, rotational and magnetic properties.}\\
\hline
\mcc{ID} & \mcc{Star} & \mcc{Alt. name} & \mcc{Spec. type} & \mcc{Remark} & \mcc{$\teff$} & \mcc{$\log g$} & \mcc{$\lum$} & \mcc{$\rs$} & \mcc{$\ms$} & \mcc{$P$} & \mcc{$\vsini$} & \mcc{$B_p$} \\
\mcc{ } & \mcc{ } & \mcc{ } & \mcc{ } & \mcc{ } & \mcc{kK} & \mcc{cgs} & \mcc{ } & \mcc{$\rsol$} & \mcc{$\msol$} & \mcc{d} & \mcc{km\,\sm} & \mcc{kG} \\
\mcc{(1)} & \mcc{(2)} & \mcc{(3)} & \mcc{(4)} & \mcc{(5)} & \mcc{(6)} & \mcc{(7)} & \mcc{(8)} & \mcc{(9)} & \mcc{(10)} & \mcc{(11)} & \mcc{(12)} & \mcc{(13)} \\
\hline
\endfirsthead
\caption{Continued}\\
\hline
\mcc{ID} & \mcc{Star} & \mcc{Alt. name} & \mcc{Spec. type} & \mcc{Remark} & \mcc{$\teff$} & \mcc{$\log g$} & \mcc{$\lum$} & \mcc{$\rs$} & \mcc{$\ms$} & \mcc{$P$} & \mcc{$\vsini$} & \mcc{$B_p$} \\
\mcc{ } & \mcc{ } & \mcc{ } & \mcc{ } & \mcc{ } & \mcc{kK} & \mcc{cgs} & \mcc{ } & \mcc{$\rsol$} & \mcc{$\msol$} & \mcc{d} & \mcc{km\,\sm} & \mcc{kG} \\
\mcc{(1)} & \mcc{(2)} & \mcc{(3)} & \mcc{(4)} & \mcc{(5)} & \mcc{(6)} & \mcc{(7)} & \mcc{(8)} & \mcc{(9)} & \mcc{(10)} & \mcc{(11)} & \mcc{(12)} & \mcc{(13)} \\
\hline
\endhead
\hline
\multicolumn{13}{r}{Following on the next page} \\
\endfoot
\hline
\multicolumn{13}{l}{$^\dagger$ Notes in Appendix.} \\
\multicolumn{13}{l}{(5) Single- or double-lined spectroscopic binary (SB1-2), slowly pulsating B-type star (SPB), $\beta$\,Cep-type pulsator ($\beta$\,Cep), Herbig Be star (HeBe).} \\
\multicolumn{13}{l}{$^s$ Parameters determined from modern spectral modelling.}\\
\multicolumn{13}{l}{$^p$ Luminosity derived from our photometric calculations with bolometric correction.}\\
\multicolumn{13}{l}{$^c$ Luminosity derived from SED fitting with \textsc{Chorizos}.}\\
\multicolumn{13}{l}{$^m$ Higher multipole components.} \\
\endlastfoot
1 & HD\,148937 &   & O6\,f?p &  &41*2\,^{s} & 4.0*0.1 & 5.8*0.1 & 15 & 60 & 7.0323 & $<$~45 & *1.0 \\ 

2 & CPD\,-28\,2561 &   & O6.5\,f?p &  &35*2\,^{s} & 4.0*0.2 & 5.5*0.2 & 14 & 43 & 70 &  & >*1.7 \\ 

3 & HD\,37022\,$^\dagger$ & $\theta^1$\,Ori\,C & O7\,Vp & SB1 &39*1\,^{s} & 4.1*0.1 & 5.3*0.1 & 9.9 & 45 & 15.424 & 24 & *1.1 \\ 

4 & HD\,191612\,$^\dagger$ &   & O6\,f?p-O8\,fp & SB2 &35*1\,^{s} & 3.5*0.1 & 5.4*0.2 & 14 & 30 & 537.2 & $<$~60 & *2.5 \\ 

5 & NGC\,1624-2 &   & O6.5\,f?cp-O8\,f?cp &  &35*2\,^{s} & 4.0*0.2 & 5.1*0.2 & 9.7 & 34 & 158.0 & $<$~3 & >*20\,^m \\ 

6 & HD\,47129\,$^\dagger$ & Plaskett's star & O7.5\,III & SB2 &33*2\,^{s} & 4.1*0.1 & 5.09*0.04 & 10 & 56 &   & 305 & >*2.8 \\ 

7 & HD\,108 &   & O8\,f?p &  &35*2\,^{s} & 3.5*0.2 & 5.7*0.1 & 19 & 43 & 18000 & $<$~50 & >*0.50 \\ 

8 & ALS\,15218\,$^\dagger$ & Tr16-22 & O8.5\,V &  &34*2 & 4.0*0.2 & 5.0*0.1 & 9.0 & 28 &   & 25 & >*1.5 \\ 

9 & HD\,57682 &   & O9\,V &  &34*1\,^{s} & 4.0*0.2 & 4.8*0.2 & 7.0 & 17 & 63.571 & 15 & *1.7 \\ 

10 & HD\,37742\,$^\dagger$ & $\zeta$\,Ori\,Aa	 & O9.5\,Ib & SB2 &29*1\,^{s} & 3.2*0.1 & 5.6*0.1 & 25 & 40 & 7.0 & 110 & *0.060 \\ 

11 & HD\,149438 & $\tau$\,Sco & B0.2\,V &  &32*1\,^{s} & 4.0*0.1 & 4.5*0.1 & 5.6 & 11 & 41.033 & 5 & *0.20\,^m \\ 

12 & HD\,37061\,$^\dagger$ & NU\,Ori & B0.5\,V & SB2 &31.0*0.5\,^{s} & 4.2*0.1 & 4.4*0.1 & 5.7 & 19 &   & 225 & *0.65 \\ 

13 & HD\,63425 &   & B0.5\,V &  &29*1\,^{s} & 4.0*0.1 & 4.5*0.4 & 6.8 & 17 &   & $<$~10 & *0.46 \\ 

14 & HD\,66665 &   & B0.5\,V &  &28*1\,^{s} & 3.9*0.1 & 4.2*0.5 & 5.5 & 9.0 & 21 & $<$~10 & *0.67 \\ 

15 & HD\,46328 & $\xi^1$\,CMa	 & B1\,III & $\beta$\,Cep &27*2\,^{s} & 3.5*0.2 & 4.6*0.1 & 8.6 & 9.0 & 4.26 & $<$~15 & >*1.5 \\ 

16 & ALS\,8988 & NGC\,2244\,OI\,201 & B1 & HeBe &27*1\,^{s} & 4.18*0.06 & 4.05*0.08 & 4.7 & 12 &   & 23 & >*1.5 \\ 

17 & HD\,47777 & NGC\,2264\,83 & B1\,III & HeBe &27*2\,^{s} & 4.0*0.2 & 4.1*0.1 & 5.0 & 9.0 &   & 65 & >*2.1 \\ 

18 & HD\,205021\,$^\dagger$ & $\beta$\,Cep & B1\,IV & SB2,$\beta$\,Cep &26*1\,^{s} & 3.7*0.1 & 4.22*0.08 & 6.5 & 12 & 12.00092 & 27 & *0.36 \\ 

19 & ALS\,15211\,$^\dagger$ & Tr16-13 & B1\,V &  &26*2 & 4.0*0.2 & 4.0*0.1 & 4.9 & 9.0 &   &  & >*1.4 \\ 

20 & HD\,122451\,$^\dagger$ & $\beta$\,Cen & B1 & SB2,$\beta$\,Cep &25*2 & 3.5*0.4 & 4.4*0.2\,^{p} & 8.7 & 8.8 &   & 75 & >*0.25 \\ 

21 & HD\,127381 & $\sigma$\,Lup & B1/B2\,V &  &23*1\,^{s} & 4.0*0.1 & 3.76*0.06 & 4.8 & 9.0 & 3.0197 & 68 & *0.50 \\ 

22 & ALS\,3694 & NGC\,6193\,17 & B1 &  &20*3 & 4.0*0.4 & 3.7*0.3\,^{p} & 5.6 & 11 &   & 83 & >*6.0 \\ 

23 & HD\,163472 & V\,2052\,Oph & B1/B2\,V & $\beta$\,Cep &25*1\,^{s} & 4.2*0.1 & 3.8*0.1 & 4.1 & 10 & 3.638833 & 68 & *0.40 \\ 

24 & HD\,96446\,$^\dagger$ & V\,430\,Car	 & B1\,IVp/B2\,Vp &  &21.6*0.8\,^{s} & 4.0*0.1 & 3.6*0.2 & 4.5 & 8.0 & 5.73 & 3 & *6.5 \\ 

25 & HD\,66765 &   & B1/B2\,V &  &20*2 & 3.9*0.2 & 3.6*0.2\,^{c} & 5.3 & 7.5 & 1.61 & 100 & >*2.1 \\ 

26 & HD\,64740 & HR\,3089	 & B1.5\,Vp &  &24*1\,^{s} & 4.0*0.1 & 4.1*0.3 & 6.3 & 11 & 1.33026 & 160 & *16 \\ 

27 & ALS\,15956 & Col\,228\,30 & B1.5\,V &  &23*3 & 3.6*0.2 & 4.3*0.2\,^{c} & 9.1 & 11 &   &  & >*1.5 \\ 

28 & ALS\,9522 & NGC\,6611\,W601	 & B1.5\,Ve & HeBe &22*2\,^{s} & 3.8*0.3 & 4.0*0.1 & 6.4 & 10 &   & 190 & >*4.0 \\ 

29 & HD\,36982 & LP\,Ori & B1.5\,Vp &  &22*2 & 4.0*0.2 & 3.1*0.2\,^{p} & 2.5 & 2.2 &   & 80 & *0.91 \\ 

30 & HD\,37017\,$^\dagger$ & V\,1046\,Ori & B1.5-2.5\,IV-Vp & SB2 &21*2 & 4.1*0.2 & 3.4*0.2 & 3.9 & 7.2 & 0.90119 & 90 & >*6.0 \\ 

31 & HD\,37479 & $\sigma$\,Ori\,E & B2\,Vp &  &23*2 & 4.0*0.2 & 3.6*0.2\,^{p} & 3.9 & 5.0 & 1.1908 & 170 & *9.6\,^m \\ 

32 & HD\,149277\,$^\dagger$ &   & B2\,IV/V & SB2 &22*3 & 4.0*0.4 & 4.0*0.3\,^{p} & 7.0 & 17 &   & 15 & >*4.7 \\ 

33 & HD\,184927 & V\,1671\,Cyg & B2\,Vp &  &22*1\,^{s} & 3.9*0.1 & 3.6*0.2 & 4.3 & 5.5 & 9.530 & 14 & *3.9\,^m \\ 

34 & HD\,37776\,$^\dagger$ & V\,901\,Ori & B2\,Vp &  &22*1 & 4.0*0.1 & 3.5*0.1 & 3.8 & 5.5 & 1.538756 & 95 & *15\,^m \\ 

35 & HD\,136504\,$^\dagger$ & $\epsilon$\,Lup & B2\,IV-V & SB2,$\beta$\,Cep &22*2 & 4.0*0.2 & 3.8*0.2\,^{c} & 5.3 & 8.6 &   & 42 & >*0.60 \\ 

36 & HD\,156424 &   & B2\,V &  &22*3 & 4.0*0.3 & 3.7*0.4\,^{p} & 4.8 & 8.5 &   & 15 & >*0.65 \\ 

37 & HD\,156324 &   & B2\,V &  &22*3 & 4.0*0.3 & 3.7*0.4\,^{p} & 5.1 & 9.4 &   & 60 & >*1.8 \\ 

38 & HD\,121743 & $\phi$\,Cen & B2\,IV & $\beta$\,Cep &22*3 & 4.0*0.3 & 3.7*0.2\,^{p} & 4.7 & 8.0 &   & 80 & >*0.53 \\ 

39 & HD\,3360 & $\zeta$\,Cas & B2\,IV & SPB &20.4*0.9\,^{s} & 3.8*0.1 & 3.7*0.2 & 5.9 & 8.3 & 5.37045 & 17 & >*0.34 \\ 

40 & HD\,186205\,$^\dagger$ &   & B2\,Vp &  &20*3 & 4.0*0.2 & 3.5*0.2\,^{c} & 4.9 & 7.4 &   & 5 & >*1.7 \\ 

41 & HD\,67621 &   & B2\,IV &  &19*3 & 4.0*0.3 & 3.3*0.2\,^{p} & 4.1 & 6.2 & 3.59 & 20 & >*0.90 \\ 

42 & HD\,200775\,$^\dagger$ & V\,3780\,Cep & B2\,Ve & SB2, HeBe &18*2 & 3.4*0.2 & 4.0*0.3 & 10 & 10 & 4.328 & 26 & *1.0 \\ 

43 & HD\,35912 & HR\,1820 & B2\,V &  &18*1 & 4.0*0.1 & 3.3*0.3\,^{p} & 4.4 & 7.2 & 0.89786 & $<$~12 & >*6.0 \\ 

44 & HD\,66522 &   & B2\,III &  &18*2 & 4.0*0.4 & 3.3*0.2\,^{p} & 4.6 & 7.6 &   & $<$~10 & *0.90 \\ 

45 & HD\,182180 & HR\,7355 & B2\,Vn &  &17*1\,^{s} & 4.2*0.2 & 3.0*0.1 & 3.7 & 6.0 & 0.5214404 & 310 & *11 \\ 

46 & HD\,55522 & HR\,2718 & B2\,IV/V &  &17.4*0.4\,^{s} & 4.2*0.1 & 3.0*0.1 & 3.3 & 5.5 & 2.729 & 70 & >*2.6 \\ 

47 & HD\,142184 & HR\,5907 & B2\,V &  &17*1\,^{s} & 4.3*0.1 & 2.8*0.1 & 3.1 & 5.5 & 0.50828 & 290 & *10 \\ 

48 & HD\,58260\,$^\dagger$ &   & B3\,Vp &  &20*2 & 3.5*0.2 & 4.1*0.2\,^{c} & 9.5 & 9.5 &   & $<$~12 & >*7.0 \\ 

49 & HD\,36485\,$^\dagger$ & $\delta$\,Ori\,C & B3\,Vp & SB2 &20*2 & 4.0*0.1 & 3.5*0.1\,^{c} & 4.5 & 7.1 & 1.47775 & 32 & *10 \\ 

50 & HD\,208057\,$^\dagger$ & 16\,Peg & B3\,V & SPB &19*3 & 3.9*0.2 & 3.6*0.2\,^{c} & 5.5 & 7.1 & 1.441 & 104 & >*0.50 \\ 

51 & HD\,306795 & NGC\,3766\,MG170 & B3\,V &  &18*2 & 3.9*0.2 & 3.2*0.3\,^{p} & 4.1 & 4.3 &   & 65 & >*5.0 \\ 

52 & HD\,25558\,$^\dagger$ & 40\,Tau & B3\,V & SPB &17*2 & 4.0*0.2 & 3.1*0.2\,^{p} & 3.9 & 5.5 &   & 28 & >*0.15 \\ 

53 & HD\,35298\,$^\dagger$ &   & B3\,Vw &  &16*2 & 3.8*0.2 & 3.2*0.2\,^{c} & 5.5 & 5.6 & 1.85336 & 260 & >*9.0 \\ 

54 & HD\,130807 & $o$\,Lup & B5 &  &18*2 & 4.1*0.1 & 3.1*0.1\,^{c} & 3.5 & 5.7 &   & 25 & >*2.0 \\ 

55 & HD\,142990\,$^\dagger$ & V\,913\,Sco & B5\,V &  &17*2 & 4.2*0.2 & 2.9*0.2\,^{p} & 3.1 & 5.7 & 0.97907 & 125 & >*7.5 \\ 

56 & HD\,37058\,$^\dagger$ & V\,359\,Ori & B3\,VpC &  &17*2 & 3.8*0.2 & 3.5*0.2\,^{c} & 5.6 & 6.6 & 14.61 & 25 & >*3.0 \\ 

57 & HD\,35502\,$^\dagger$ &   & B5\,V & SB2 &16*2 & 3.8*0.2 & 3.3*0.2\,^{c} & 5.7 & 5.7 & 0.85 & 80 & >*6.8 \\ 

58 & HD\,176582 & HR\,7185 & B5\,IV &  &16*1\,^{s} & 4.0*0.1 & 2.9*0.1 & 3.6 & 4.7 & 1.581984 & 105 & *7.0 \\ 

59 & HD\,189775 & HR\,7651  & B5\,V &  &16*2 & 3.8*0.2 & 3.2*0.2\,^{c} & 5.3 & 5.5 & 2.6048 & 85 & >*4.5 \\ 

60 & HD\,61556\,$^\dagger$ & HR\,2949 & B5\,V &  &15*2 & 4.0*0.3 & 2.6*0.1\,^{p} & 2.8 & 2.9 & 1.9093 & 70 & *4.0\,^m \\ 

61 & HD\,175362\,$^\dagger$ & Wolff's star & B5\,V &  &15*3 & 3.7*0.2 & 3.2*0.1\,^{c} & 5.8 & 5.3 & 3.6738 & 35 & >*21\,^m \\ 

62 & HD\,105382\,$^\dagger$ & HR\,4618 & B6\,III &  &17*2 & 4.0*0.2 & 3.0*0.2\,^{p} & 3.6 & 4.8 & 1.285 & 90 & *2.3 \\ 

63 & HD\,125823 & a\,Cen & B7\,IIIp &  &19*2 & 4.0*0.2 & 3.2*0.1\,^{p} & 3.6 & 4.7 & 8.812 & 15 & >*1.3 \\ 

64 & HD\,36526 & V1099\,Ori & B8\,Vp &  &16*3 & 4.0*0.3 & 2.5*0.3\,^{p} & 2.4 & 2.0 & 1.5405 &  & >*10 \\ 

\hline
\end{longtable}
}
\end{landscape}
\twocolumn


\newpage
\onecolumn
		{\small \begin{longtable}{l l l l l l l l l} 
\caption{\label{tab:ref}List of references for propeties retrieved from the literature with superscript letters indicating the type of parameter. Superscript letters in column (2) indicate information we inferred from MiMeS observations or other archival data, as indicated in the text.}\\
\hline
\mcc{ID} & \mcc{Star} & \mcc{Ref.} & \mcc{ID} & \mcc{Star} & \mcc{Ref.} \\
\mcc{(1)} & \mcc{(2)} & \mcc{(3)} & \mcc{(1)} & \mcc{(2)} & \mcc{(3)} \\
\hline
\endfirsthead
\caption{Continued}\\
\hline
\mcc{ID} & \mcc{Star} & \mcc{Ref.} & \mcc{ID} & \mcc{Star} & \mcc{Ref.} \\
\mcc{(1)} & \mcc{(2)} & \mcc{(3)} & \mcc{(1)} & \mcc{(2)} & \mcc{(3)} \\
\hline
\endhead
\hline
\multicolumn{6}{r}{Following on the next page} \\
\endfoot
\hline
\multicolumn{6}{l}{$^s$ Stellar parameters, $^r$ Rotational parameters, $^b$ Magnetic field parameters} \\
\multicolumn{6}{l}{$^a$ H$\alpha$ proxy, $^u$ UV proxy, $^x$ X-ray proxy} \\
\endlastfoot
\nopagebreak[0]1 & HD\,148937\,  & \protect{\citet{2012MNRAS.419.2459W}$^{srba}$} & 32 & HD\,149277\,$^{ra}$ & \protect{\citet{2006A&A...450..777B}$^{b}$} \\ 
\nopagebreak[4] & & \protect{A. Fullterton (priv. com.)$^{u}$} &  & & \protect{\citet{2007A&A...470..685L}$^{s}$} \\ 
\nopagebreak[2] & & \protect{\citet{2012ApJ...746..142N}$^{x}$} & 33 & HD\,184927\,  & \protect{\citet{1997A&A...320..172W}$^{r}$} \\ 
\nopagebreak[2]2 & CPD\,-28\,2561\,  & \protect{Barba et al. (MiMeS in prep)$^{srba}$} &  & & \protect{Yakunin et al. (MiMeS in prep)$^{sbau}$} \\ 
\nopagebreak[0]3 & HD\,37022\,  & \protect{\citet{2006A&A...448..351S}$^{s}$} & 34 & HD\,37776\,  & \protect{\citet{2007A&A...470..685L}$^{s}$} \\ 
\nopagebreak[4] & & \protect{\citet{2008A&A...487..323S}$^{ra}$} &  & & \protect{\citet{2011ApJ...726...24K}$^{b}$} \\ 
\nopagebreak[4] & & \protect{\citet{2006A&A...451..195W}$^{b}$} &  & & \protect{\citet{2011A&A...534L...5M}$^{r}$} \\ 
\nopagebreak[4] & & \protect{\citet{1994ApJ...425L..29W}$^{u}$} &  & & \protect{Shultz et al. (MiMeS in prep)$^{a}$} \\ 
\nopagebreak[4] & & \protect{\citet{2005ApJS..160..557S}$^{x}$} &  & & \protect{\citet{1990ApJ...365..665S}$^{u}$} \\ 
\nopagebreak[0]4 & HD\,191612\,  & \protect{\citet{2011MNRAS.416.3160W}$^{srba}$} & 35 & HD\,136504\,$^{a}$ & \protect{\citet{2005A&A...440..249U}$^{sr}$} \\ 
\nopagebreak[4] & & \protect{A. Fullerton (priv. com.)$^{u}$} &  & & \protect{\citet{2012ApJ...750....2S}$^{b}$} \\ 
\nopagebreak[4] & & \protect{\citet{2007MNRAS.375..145N}$^{x}$} &  & & \protect{\citet{2009AN....330..317H}$^{b}$} \\ 
\nopagebreak[0]5 & NGC\,1624-2\,  & \protect{\citet{2012MNRAS.425.1278W}$^{srbax}$} & 36 & HD\,156424\,  & \protect{E. Alecian (MiMeS in prep)$^{srba}$} \\ 
\nopagebreak[0]6 & HD\,47129\,  & \protect{\citet{2008A&A...489..713L}$^{srau}$} & 37 & HD\,156324\,  & \protect{E. Alecian (MiMeS in prep)$^{srba}$} \\ 
\nopagebreak[2] & & \protect{\cite{Grunhut2012_plaskett}$^{b}$} & 38 & HD\,121743\,  & \protect{\citet{1990AJ....100.1994W}$^{s}$} \\ 
\nopagebreak[2]7 & HD\,108\,  & \protect{\citet{2010MNRAS.407.1423M}$^{srb}$} &  & & \protect{E. Alecian (MiMeS in prep)$^{rba}$} \\ 
\nopagebreak[4] & & \protect{\citet{2012MNRAS.422.2314M}$^{au}$} &  & & \protect{\citet{1992ApJS...81..795G}$^{x}$} \\ 
\nopagebreak[2] & & \protect{\citet{2006MNRAS.370.1623L}$^{x}$} & 39 & HD\,3360\,  & \protect{\citet{2003A&A...406.1019N}$^{srbau}$} \\ 
\nopagebreak[2]8 & ALS\,15218\,  & \protect{\citet{2011ApJS..194....5G}$^{s}$} &  & & \protect{\citet{2011MNRAS.416.1456O}$^{x}$} \\ 
\nopagebreak[2] & & \protect{\citet{2012MNRAS.423.3413N}$^{rba}$} & 40 & HD\,186205\,  & \protect{\citet{2000CoSka..30...12Z}$^{r}$} \\ 
\nopagebreak[4] & & \protect{\citet{2011ApJS..194....7N}$^{x}$} &  & & \protect{J. Grunhut (MiMeS priv. com.)$^{sba}$} \\ 
\nopagebreak[0]9 & HD\,57682\,  & \protect{‘\citet{2009MNRAS.400L..94G}$^{sbaux}$} & 41 & HD\,67621\,  & \protect{Alecian et al. (MiMeS in prep)$^{srba}$} \\ 
\nopagebreak[2] & & \protect{\citet{2012arXiv1207.6988G}$^{r}$} & 42 & HD\,200775\,  & \protect{\citet{2008MNRAS.385..391A}$^{srb}$} \\ 
\nopagebreak[2]10 & HD\,37742\,  & \protect{‘\citet{2008MNRAS.389...75B}$^{srba}$} &  & & \protect{\citet{2005ApJ...618..360H}$^{x}$} \\ 
\nopagebreak[2] & & \protect{\citet{1996A&AS..116..257K}$^{u}$} & 43 & HD\,35912\,  & \protect{\citet{2010A&A...510A..22S}$^{sa}$} \\ 
\nopagebreak[4] & & \protect{\citet{2008A&A...478..513R}$^{x}$} &  & & \protect{\citet{2005A&A...430.1143B}$^{rb}$} \\ 
\nopagebreak[0]11 & HD\,149438\,$^{a}$  & \protect{‘\cite{2006A&A...448..351S}$^{s}$} & 44 & HD\,66522\,  & \protect{\citet{1997A&A...324..949Z}$^{s}$} \\ 
\nopagebreak[4] & & \protect{\citet{2006MNRAS.370..629D}$^{rbu}$} &  & & \protect{\citet{1997A&A...325.1125L}$^{s}$} \\ 
\nopagebreak[4] & & \protect{\citet{2003A&A...398..203M}$^{x}$} &  & & \protect{E. Alecian (MiMeS in prep)$^{rba}$} \\ 
\nopagebreak[0]12 & HD\,37061\,$^{u}$ & \protect{‘\citet{2011A&A...530A..57S}$^{sra}$} & 45 & HD\,182180\,  & \protect{\citet{RiviniusHR7355_sub}$^{srba}$} \\ 
\nopagebreak[2] & & \protect{\citet{2008MNRAS.387L..23P}$^{b}$} & 46 & HD\,55522\,$^{a}$ & \protect{\citet{2004A&A...413..273B}$^{sr}$} \\ 
\nopagebreak[4] & & \protect{\citet{2005ApJS..160..557S}$^{x}$} &  & & \protect{\citet{2007AN....328...41B}$^{b}$} \\ 
\nopagebreak[0]13 & HD\,63425\,  & \protect{\citet{2011MNRAS.412L..45P}$^{srbau}$} & 47 & HD\,142184\,  & \protect{\citet{2012MNRAS.419.1610G}$^{srba}$} \\ 
\nopagebreak[2]14 & HD\,66665\,  & \protect{‘\citet{2011MNRAS.412L..45P}$^{srbau}$} &  & & \protect{\citet{2011MNRAS.416.1456O}$^{x}$} \\ 
\nopagebreak[0]15 & HD\,46328\,  & \protect{\citet{2011IAUS..272..180F}$^{srbau}$} & 48 & HD\,58260\,  & \protect{\citet{1989ApJ...346..459B}$^{s}$} \\ 
\nopagebreak[4] & & \protect{\citet{2011MNRAS.416.1456O}$^{x}$} &  & & \protect{\citet{2007A&A...468..263C}$^{sb}$} \\ 
\nopagebreak[2]16 & ALS\,8988\,  & \protect{‘\citet{2008A&A...481L..99A}$^{srb}$} &  & & \protect{\citet{1979A&AS...35..313P}$^{ra}$} \\ 
\nopagebreak[4] & & \protect{\citet{2008ApJ...675..464W}$^{x}$} &  & & \protect{\citet{1990ApJ...365..665S}$^{u}$} \\ 
\nopagebreak[0]17 & HD\,47777\,  & \protect{E. Alecian (priv. com.)$^{srb}$} & 49 & HD\,36485\,  & \protect{\citet{2010MNRAS.401.2739L}$^{srba}$} \\ 
\nopagebreak[4] & & \protect{\citet{2009A&A...506.1055N}$^{x}$} &  & & \protect{\citet{1990ApJ...365..665S}$^{u}$} \\ 
\nopagebreak[0]18 & HD\,205021\,  & \protect{‘\citet{2001MNRAS.326.1265D}$^{srbu}$} & 50 & HD\,208057\,  & \protect{‘\citet{2001A&A...378..861C}$^{s}$} \\ 
\nopagebreak[4] & & \protect{\citet{2010A&A...515A..74L}$^{s}$} &  & & \protect{\citet{2009IAUS..259..393H}$^{rbau}$} \\ 
\nopagebreak[2] & & \protect{\citet{2008MNRAS.387..759C}$^{a}$} & 51 & HD\,306795\,  & \protect{‘\citet{2008ApJ...686.1269M}$^{srb}$} \\ 
\nopagebreak[4] & & \protect{\citet{2009A&A...495..217F}$^{x}$} &  & & \protect{\citet{2008ApJ...672..590M}$^{a}$} \\ 
\nopagebreak[0]19 & ALS\,15211\,  & \protect{\citet{2011ApJS..194....5G}$^{s}$} & 52 & HD\,25558\,$^{b}$ & \protect{‘\citet{2010A&A...515A..74L}$^{sr}$} \\ 
\nopagebreak[2] & & \protect{\citet{2012MNRAS.423.3413N}$^{b}$} & 53 & HD\,35298\,$^{a}$ & \protect{‘\citet{2007A&A...470..685L}$^{s}$} \\ 
\nopagebreak[4] & & \protect{\cite{2011ApJS..194....7N}$^{x}$} &  & & \protect{\citet{2005A&A...430.1143B}$^{rb}$} \\ 
\nopagebreak[2]20 & HD\,122451\,$^{a}$ & \protect{‘\citet{2006A&A...455..259A} $^{s}$} &  & & \protect{\citet{2011AN....332..974Y}$^{b}$} \\ 
\nopagebreak[2] & & \protect{\citet{2011A&A...536L...6A} $^{rb}$} & 54 & HD\,130807\,  & \protect{‘\citet{2011A&A...536L...6A}$^{srb}$} \\ 
\nopagebreak[2] & & \protect{H. Henrichs (priv. com.)$^{u}$} & 55 & HD\,142990\,  & \protect{‘\citet{2007A&A...468..263C}$^{s}$} \\ 
\nopagebreak[4] & & \protect{\cite{2005A&A...437..599R}$^{x}$} &  & & \protect{\citet{2005A&A...430.1143B}$^{rb}$} \\ 
\nopagebreak[2]21 & HD\,127381\,  & \protect{\citet{henrichs_siglupi}$^{srbau}$} &  & & \protect{\citet{2004A&A...421..203S}$^{au}$} \\ 
\nopagebreak[0]22 & ALS\,3694\,  & \protect{\citet{2006A&A...450..777B}$^{b}$} & 56 & HD\,37058\,  & \protect{\citet{2007AstBu..62..319G}$^{s}$} \\ 
\nopagebreak[4] & & \protect{\citet{2007A&A...470..685L} $^{s}$} &  & & \protect{\cite{1979A&AS...35..313P}$^{r}$} \\ 
\nopagebreak[4] & & \protect{\citet{2006ApJ...648..580H} $^{r}$} &  & & \protect{\citet{2005A&A...430.1143B}$^{rb}$} \\ 
\nopagebreak[2]23 & HD\,163472\,  & \protect{\citet{2003A&A...411..565N}$^{su}$} &  & & \protect{\citet{2004AJ....128..787R}$^{x}$} \\ 
\nopagebreak[2] & & \protect{\citet{2012A&A...537A.148N}$^{rb}$} & 57 & HD\,35502\,  & \protect{‘\citet{2007A&A...470..685L}$^{s}$} \\ 
\nopagebreak[4] & & \protect{C. Neiner (priv. com.)$^{a}$} &  & & \protect{\citet{2008AstBu..63..139R}$^{b}$} \\ 
\nopagebreak[4] & & \protect{\citet{2011MNRAS.416.1456O}$^{x}$} &  & & \protect{Bohlender et al. (in prep)$^{rba}$} \\ 
\nopagebreak[2]24 & HD\,96446\,  & \protect{\citet{Neiner2012_96446}$^{srba}$} &  & & \protect{\citet{1992ApJS...81..795G}$^{x}$} \\ 
\nopagebreak[2] & & \protect{\citet{1990ApJ...365..665S}$^{u}$} & 58 & HD\,176582\,  & \protect{‘\citet{2011AJ....141..169B}$^{srba}$} \\ 
\nopagebreak[0]25 & HD\,66765\,  & \protect{\citet{2007A&A...468..263C}$^{s}$} & 59 & HD\,189775\,  & \protect{‘\citet{2002MNRAS.333....9L}$^{s}$} \\ 
\nopagebreak[4] & & \protect{Alecian et al. (MiMeS in prep)$^{rba}$} &  & & \protect{Bohlender et al. (priv. com.)$^{rb}$} \\ 
\nopagebreak[0]26 & HD\,64740\,  & \protect{\citet{1990ApJ...358..274B}$^{sr}$} & 60 & HD\,61556\,  & \protect{‘\citet{2003IBVS.5397....1R}$^{r}$} \\ 
\nopagebreak[4] & & \protect{\citet{1990ApJ...365..665S}$^{u}$} &  & & \protect{Shultz et al. (in prep)$^{sba}$} \\ 
\nopagebreak[2] & & \protect{Peralta et al. (MiMeS in prep)$^{ba}$} & 61 & HD\,175362\,  & \protect{\citet{1997A&A...320..257L}$^{s}$} \\ 
\nopagebreak[4] & & \protect{\citet{1994ApJ...420..387D}$^{x}$} &  & & \protect{\citet{2005A&A...430.1143B}$^{rb}$} \\ 
\nopagebreak[2]27 & ALS\,15956\,  & \protect{\citet{2006A&A...450..777B}$^{sb}$} &  & & \protect{\cite{2004A&A...421..203S}$^{au}$} \\ 
\nopagebreak[4] & & \protect{\citet{2011ApJS..194....7N}$^{x}$} &  & & \protect{\citet{1992ApJS...81..795G}$^{x}$} \\ 
\nopagebreak[0]28 & ALS\,9522\,  & \protect{\citet{2008A&A...481L..99A}$^{srba}$} & 62 & HD\,105382\,  & \protect{\citet{2001A&A...366..121B}$^{sra}$} \\ 
\nopagebreak[4] & & \protect{\citet{2012ApJ...753..117G}$^{x}$} &  & & \protect{\citet{2011A&A...536L...6A}$^{b}$} \\ 
\nopagebreak[0]29 & HD\,36982\,$^{u}$ & \protect{\citet{2004ApJ...601..979W}$^{r}$} & 63 & HD\,125823\,  & \protect{\citet{2010A&A...520A..44B}$^{srb}$} \\ 
\nopagebreak[2] & & \protect{\citet{2012MNRAS.420..773P}$^{sba}$} & 64 & HD\,36526\,  & \protect{\citet{2007A&A...470..685L}$^{s}$} \\ 
\nopagebreak[4] & & \protect{\citet{2005ApJS..160..557S}$^{x}$} &  & & \protect{\citet{2005A&A...430.1143B}$^{r}$} \\ 
\nopagebreak[2]30 & HD\,37017\,$^{a}$ & \protect{\citet{1998A&A...337..183B}$^{sr}$} &  & & \protect{\citet{2008AstBu..63..139R}$^{b}$} \\ 
\nopagebreak[4] & & \protect{\citet{1987ApJ...323..325B}$^{b}$} &  & &  \\ 
\nopagebreak[4] & & \protect{\citet{1990ApJ...365..665S}$^{u}$} &  & &  \\ 
\nopagebreak[4] & & \protect{\citet{2011MNRAS.416.1456O}$^{x}$} &  & &  \\ 
\nopagebreak[2]31 & HD\,37479\,  & \protect{\citet{1989A&A...224...57H}$^{s}$} &  & &  \\ 
\nopagebreak[4] & & \protect{\citet{2010ApJ...714L.318T}$^{r}$} &  & &  \\ 
\nopagebreak[4] & & \protect{\citet{2012MNRAS.419..959O}$^{ba}$} &  & &  \\ 
\nopagebreak[4] & & \protect{\citet{1990ApJ...365..665S}$^{u}$} &  & &  \\ 
\nopagebreak[4] & & \protect{\citet{2004A&A...421..715S}$^{x}$} &  & &  \\ 
\hline
\end{longtable}
}
\twocolumn

\subsection{Sample selection}
\label{sec:sample}

Magnetic fields in hot stars can be detected through the circular polarisation induced in spectral lines by the Zeeman effect, using various types of instruments. The bulk of cooler magnetic ApBp stars were generally detected with first-generation photo-polarimeters, measuring for example the degree of polarisation in the wings of a Balmer line \citep[e.g.][]{1980ApJS...42..421B}. 

However, apart from a few strongly magnetic He-strong stars such as \sigorie, the bulk of hot magnetic OB stars were detected with second generation instruments, such as the low resolution ($R\simeq$ a few thousands) spectropolarimetry optics used in FORS 1 and 2 (VLT) and the high resolution ($R\simeq$ a few tens of thousands) spectropolarimeters MUSICOS, ESPaDOnS, Narval and HARPSpol at the TBL, CFHT, TBL and ESO-3.6m, respectively. These two classes of instruments differ in that low resolution spectropolarimeters are only sensitive to the disk-integrated, brightness-weighted longitudinal field component, whereas high-resolution instruments can probe field configurations through the rotationally induced Doppler shifts within the resolved line profiles \citep[see][]{2009ARA&A..47..333D,2010arXiv1010.2248P}.

We use the existing compilations of ApBp stars \citep[e.g.][]{2005A&A...430.1143B,2007A&A...470..685L,2008AstBu..63..139R} as well as an exhaustive review of the literature to identify hot stars with confirmed field detections, which we complement with new detections from the MiMeS project. 

Some concerns have recently been raised about claimed magnetic detections (usually near the $3\sigma$ level) obtained with the FORS instruments that were not reproduced with other high-resolution instruments \citep[see][]{2009MNRAS.398.1505S,2012ApJ...750....2S}. \citet{2012A&A...538A.129B} performed an in-depth study of the complete set of FORS circular polarisation measurements in the ESO archive, exploring the effect of various data reduction procedures and carefully considering all known sources of uncertainties. Using their new prescription for FORS data analysis, most of the claimed marginal detections were found to have very low significance, in agreement with the results from high-resolution instruments. They also provided updated longitudinal field values and new magnetic detection statuses for stars that were reported magnetic in the literature at the $<6\sigma$ level. We therefore base our selection on these new detection statuses for stars that were only detected with the FORS instruments. 

It is worth noting that stars with chemical abundance peculiarities can have effective temperatures that do not reflect their spectral types, as the latter is determined from spectral morphology. 
In particular He-strong/weak stars, which form the majority of the cooler part of our sample, are identified by their unusually strong/weak He lines, lines that are the basic means to classify B-type stars.
Given that photometric/spectral effective temperature determinations are not always readily available, it is therefore difficult to assess the completeness of our sample at the low temperature boundary. 
We therefore consider all magnetic stars with spectral type B5 and earlier, as well as additional stars of later spectral type known to have effective temperatures greater than 16\,kK. We believe the sample at these temperatures (and above) to be a substantially complete representation of the currently known hot magnetic stars.

Although we consider a detailed review of the large sample of stars evaluated for inclusion in Table \ref{tab:stars} beyond the scope of this work, two noteworthy objects require a brief mention. The first of these is the Be star $\omega$\,Ori, reported to be magnetic by \citet{2003A&A...409..275N} based on MuSiCoS longitudinal field measurements. Recently, \citet{neiner_omegaori} have retracted this claim based on new ESPaDOnS and Narval measurements. The second is $\zeta$\,Ori\,A, reported to be magnetic by \citet{2008MNRAS.389...75B}. While no single observation of this star yields a significant magnetic detection, overall we consider the evidence presented by \citet{2008MNRAS.389...75B} to be sufficiently compelling that we retain this star in our list. Note that $\zeta$\,Ori\,A occupies a unique position in the magnetic confinement-rotation diagram (see \S\,\ref{sec:theo}).

\subsection{Physical parameters}
\label{sec:physical}

Effective temperatures and surface gravities (columns 6 and 7 of Table \ref{tab:stars}) were retrieved from the literature.
An $s$ superscript in column (6) indicates stellar parameters that were determined by modern spectral modelling, with NTLE model atmospheres such as \textsc{cmfgen}, \textsc{tlusty} or \textsc{fastwind} for the hotter stars, or such as LTE \textsc{atlas} models with the polarised radiative transfer code \textsc{zeeman} for the cooler stars \citep{1998ApJ...496..407H,2003ApJS..146..417L,2005A&A...435..669P,1979ApJS...40....1K,1988ApJ...326..967L,2001A&A...374..265W}.
For the other stars, temperatures and gravities were generally derived from photometry combined with spectral type calibrations. 
Some details are given in the notes of Appendix \ref{sec:notes} in cases where significant discrepancies were found in the literature values or when we had to estimate $\logg$ from the luminosity class.

When modern spectral modelling is available, we use the literature value for the luminosity, radius and mass (columns 8, 9 and 10). 
The luminosity is generally obtained through a distance estimate and photometry, and the spectroscopic mass is derived from the surface gravity and radius, unless a better estimate is available from a binary orbit. 

For most of the remaining stars, marked with superscript $p$ or $c$ in column (8), we derive the luminosity from photometry (see \S \ref{sec:lum}) using tabulated bolometric corrections, or using the spectral energy distribution (SED) fitting code \textsc{Chorizos} \citep{2004PASP..116..859M} for stars with sufficient photometric data. 

\begin{figure*}
	\begin{center}
	\includegraphics[width=168mm]{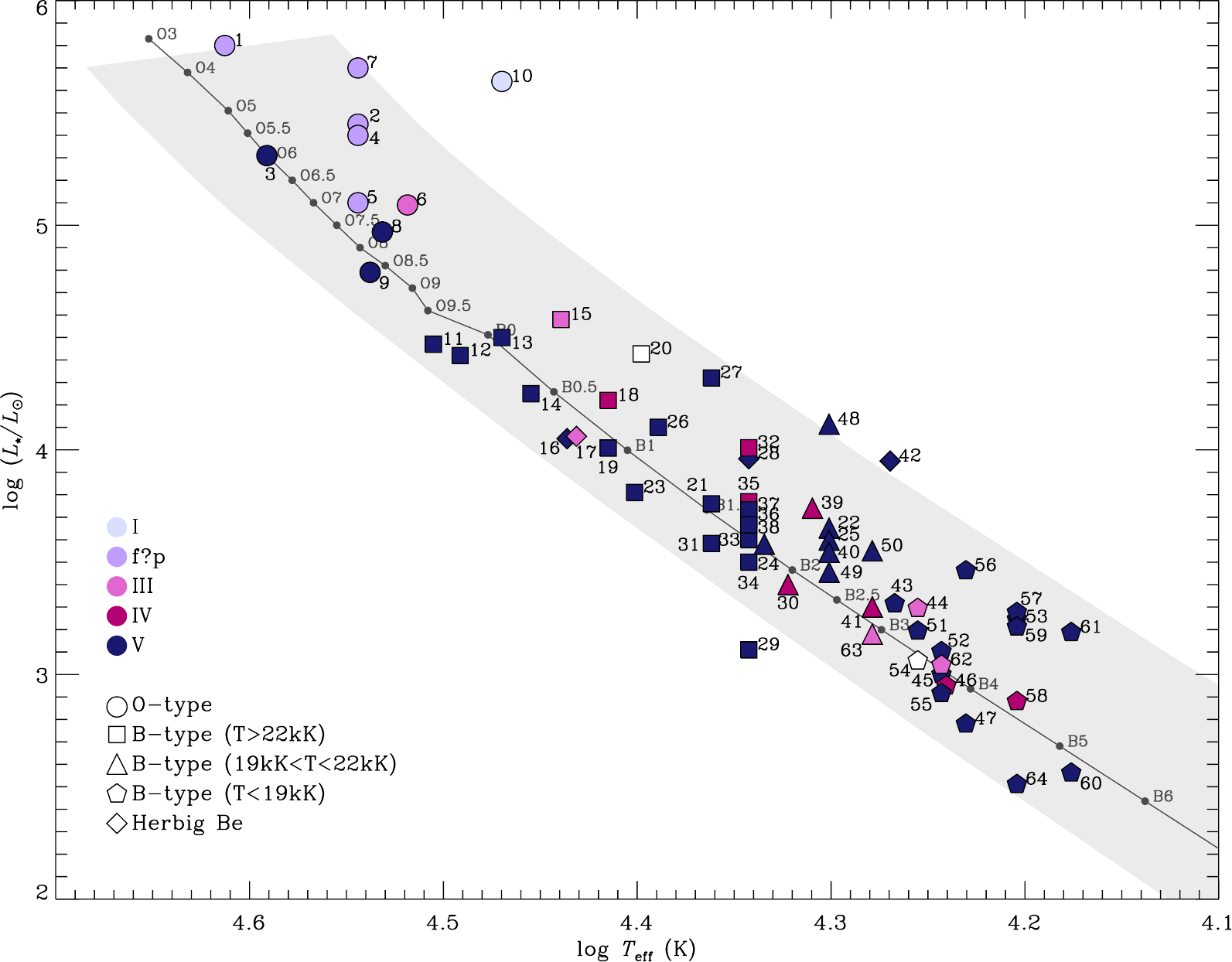}
	\caption{\label{fig:HR} Location of the magnetic stars in the HR diagram. The labels refer to ID sequence number listed in column (1) of Table \ref{tab:stars}. The various symbol shapes represent effective temperature ranges and colours denote luminosity classes, as indicated in the legend. The shaded region shows the main sequence, from ZAMS to TAMS \citep[from the galactic evolutionary tracks of][]{2011A&A...530A.115B}. The grey line shows the mid-way main sequence with spectral types calibrations from \citet{2005A&A...436.1049M} for O-type stars and \citet{1987A&A...177..217D} for B-type stars. }
	\end{center}
\end{figure*}

In Figure \ref{fig:HR}, we locate the magnetic OB stars on the HR diagram. The symbol shapes represent the O-type stars (circles), B-type stars hotter than 22\,kK (squares), those between 22\,kK and 19\,kK (triangles) and those that are cooler than 19\,kK (pentagons), and known Herbig Be stars (HeBe; diamonds). The luminosity classes are colour coded. 
The labels refer to the identification numbers in column (1) of Table \ref{tab:stars}. 

The position of the spectral types, from the calibrations of \citet{2005A&A...436.1049M} for the O-type stars and \citet{1987A&A...177..217D} for the B-type stars, are indicated on the dark grey line that runs approximately mid-way between the zero-age main sequence and the terminal-age main sequence; the main sequence itself is shown by the light grey shaded area \citep[from the galactic evolutionary tracks of][]{2011A&A...530A.115B}.

\subsubsection{Luminosity derivation}
\label{sec:lum}

For each star in our sample without modern spectral modelling, Table \ref{tab:phot} gathers visual magnitudes and colours (columns 3-5) in the Johnson $UBV$ system, from the compilations of \citet{2006yCat.2168....0M} and \citet{2005yCat.5125....0R}\footnote{As no Johnson $UBV$ measurements are available for HD\,61556 (ID\,60) and HD\,306795 (ID\,51), we use Str\"omgren photometry from \citet{1998A&AS..129..431H} with the transformation given by \citet{1990PASP..102.1331T}.}. 
We also provide $RJHK$ magnitudes (columns 6-9) from the NOMAD catalog \citep{2005yCat.1297....0Z}, which will be used below for SED fitting with \textsc{Chorizos}.

\begin{table*}
	\begin{center}
	\caption{\label{tab:phot}Photometry of magnetic stars without modern spectral modelling (\S\,\ref{sec:lum}).}
		{\begin{tabular}{ c l  D{.}{.}{2,2} D{.}{.}{2,2} D{*}{\,\pm\,}{-1} D{.}{.}{2,2} D{.}{.}{2,2} D{.}{.}{2,2} D{.}{.}{2,2} }
\hline
\mcc{ID} & \mcc{Star} & \mcc{$V$} & \mcc{($B-V$)} & \mcc{$(U-B)$} & \mcc{$R$} & \mcc{$J$} & \mcc{$H$} & \mcc{$K$} \\
\mcc{ } & \mcc{ } & \mcc{$\pm0.01$\,mag} & \mcc{$\pm0.01$\,mag} & \mcc{mag} & \mcc{$\pm0.02$\,mag} & \mcc{$\pm0.02$\,mag} & \mcc{$\pm0.02$\,mag} & \mcc{$\pm0.02$\,mag} \\
\mcc{(1)} & \mcc{(2)} & \mcc{(3)} & \mcc{(4)} & \mcc{(5)} & \mcc{(6)} & \mcc{(7)} & \mcc{(8)} & \mcc{(9)} \\
\hline
20 & HD\,122451\,$^\dagger$ & 0.61 & -0.24 &  &  & & & \\

22 & ALS\,3694 & 10.38 & 0.33 & -0.55*0.02 & 10.26 & 9.62 & 9.49 & 9.47 \\ 

25 & HD\,66765 & 6.62 & -0.16 & -0.8*0.2 & 6.69 & 6.89 & 7.01 & 7.01 \\ 

27 & ALS\,15956 & 10.80 & 0.05 & -0.69*0.02 & 10.70 & 10.62 & 10.64 & 10.66 \\ 

29 & HD\,36982 & 8.45 & 0.13 & -0.57*0.02 & 8.42 & 7.74 & 7.64 & 7.47 \\ 

31 & HD\,37479 & 6.66 & -0.18 & -0.87*0.01 & 6.40 & 6.97 & 6.95 & 6.95 \\ 

32 & HD\,149277\,$^\dagger$ & 8.38 & 0.04 & -0.658*0.009 & 8.43 & 8.26 & 8.26 & 8.28 \\ 

35 & HD\,136504\,$^\dagger$ & 3.37 & -0.18 & -0.74*0.02 & 3.44 & 3.99^* & 3.93^* & 4.13^* \\ 

36 & HD\,156424 & 8.72 & 0.06 &  &  & & & \\

37 & HD\,156324 & 8.75 & 0.10 &  &  & & & \\

38 & HD\,121743 & 3.82 & -0.22 &  &  & & & \\

40 & HD\,186205\,$^\dagger$ & 8.53 & 0.05 & -0.60*0.02 & 8.50 & 8.40 & 8.48 & 8.48 \\ 

41 & HD\,67621 & 6.33 & -0.20 & -0.802*0.004 & 6.41 & 6.73 & 6.86 & 6.84 \\ 

43 & HD\,35912 & 6.41 & -0.18 &  &  & & & \\

44 & HD\,66522 & 7.20 & 0.05 & -0.624*0.005 & 7.19 & 7.03 & 7.02 & 7.01 \\ 

48 & HD\,58260\,$^\dagger$ & 6.74 & -0.13 & -0.756*0.006 & 6.79 & 6.98 & 7.11 & 7.12 \\ 

49 & HD\,36485\,$^\dagger$ & 6.85 & -0.16 & -0.72*0.01 & 6.92 & 7.17 & 7.23 & 7.28 \\ 

50 & HD\,208057\,$^\dagger$ & 5.07 & -0.17 & -0.7*0.1 & 5.16 & 5.39 & 5.49 & 5.54 \\ 

51 & HD\,306795 & 10.62 & 0.02 &  &  & & & \\

52 & HD\,25558\,$^\dagger$ & 5.32 & -0.08 & -0.57*0.01 & 5.35 & 5.48 & 5.58 & 5.53 \\ 

53 & HD\,35298\,$^\dagger$ & 7.89 & -0.14 & -0.6*0.2 & 7.94 & 8.12 & 8.17 & 8.27 \\ 

54 & HD\,130807 & 4.32 & -0.15 & -0.62*0.01 & 4.38 & 4.78^* & 4.72 & 4.73 \\ 

55 & HD\,142990\,$^\dagger$ & 5.43 & -0.09 & -0.653*0.003 & 5.46 & 5.58 & 5.67 & 5.65 \\ 

56 & HD\,37058\,$^\dagger$ & 7.30 & -0.13 & -0.79*0.02 & 7.23 & 7.60 & 7.73 & 7.75 \\ 

57 & HD\,35502\,$^\dagger$ & 7.35 & -0.04 & -0.54*0.04 & 7.32 & 7.39 & 7.42 & 7.43 \\ 

59 & HD\,189775 & 6.14 & -0.19 & -0.66*0.02 & 6.20 & 6.46 & 6.53 & 6.64 \\ 

60 & HD\,61556\,$^\dagger$ & 4.78 & -0.14 &  &  & & & \\

61 & HD\,175362\,$^\dagger$ & 5.37 & -0.15 & -0.7*0.1 & 5.43 & 5.62 & 5.66 & 5.68 \\ 

62 & HD\,105382\,$^\dagger$ & 4.46 & -0.16 & -0.68*0.01 & 4.51 & 5.12 & 4.95 & 4.87 \\ 

63 & HD\,125823 & 4.37 & -0.19 &  &  & & & \\

64 & HD\,36526 & 8.31 & -0.11 &  &  & & & \\

\hline
\multicolumn{9}{l}{$^*$ Because of their brightness, these stars have uncertainties of $\pm0.24$\,mag according to the 2mass specifications.}\\
\end{tabular}
}
	\end{center}
\end{table*}

For all these stars, we derive the luminosity using bolometric corrections ($BC$) and extinction ($A_V$) evaluated from the intrinsic colour $(B-V)_0$. The results are compiled in Table \ref{tab:BC}. 
The distance modulus ($DM$; column 4) is estimated using either Hipparcos parallax measurements or a distance estimate from an association with a stellar cluster.
The Hipparcos distances are corrected for Lutz-Kelker-type effects \citep{1973PASP...85..573L} using the technique described by \citeauthor{2001AJ....121.2737M} \citeyearpar{2001AJ....121.2737M,2005ESASP.576..179M} updated to the new reduction of the Hipparcos data \citep{2007A&A...474..653V} by \citet{2008arXiv0804.2553M}.
\begin{table*}
	\begin{center}
	\caption{\label{tab:BC}Luminosity determination based on bolometric correction and extinction from intrinsic colours (\S\,\ref{sec:lum}).}
		{\begin{tabular}{l l D{*}{\,\pm\,}{4,4} D{*}{\,\pm\,}{4,4} D{.}{.}{2,2} D{.}{.}{2,2} D{.}{.}{2,2} D{*}{\,\pm\,}{4,4} D{*}{\,\pm\,}{4,4} D{*}{\,\pm\,}{3,3}}
\hline
\mcc{ID} & \mcc{Star} & \mcc{$\log g$} & \mcc{$DM$} & \mcc{$BC$} & \mcc{$(B-V)_0$} & \mcc{$A_V$} & \mcc{$M_V$} & \mcc{$M_\mathrm{bol}$} & \mcc{$\lum$} \\
\mcc{ } & \mcc{ } & \mcc{cgs} & \mcc{mag} & \mcc{$\pm0.2$\,mag} & \mcc{$\pm0.02$\,mag} & \mcc{$\pm0.25$\,mag} & \mcc{mag} & \mcc{mag} & \mcc{} \\
\mcc{(1)} & \mcc{(2)} & \mcc{(3)} & \mcc{(4)} & \mcc{(5)} & \mcc{(6)} & \mcc{(7)} & \mcc{(8)} & \mcc{(9)} & \mcc{(10)} \\
\hline
20 & HD\,122451\,$^\dagger$ & 3.5*0.4 &  & -2.48 &   &   & -3.8*0.5 ^{a} & -6.3*0.5 & 4.4*0.2 \\ 

22 & ALS\,3694 & 4.0*0.4 & 11.2*0.6\,^{c} & -1.98 & -0.18 & 1.57 & -2.4*0.7 & -4.4*0.8 & 3.7*0.3 \\ 

25 & HD\,66765 & 4.0*0.2 & 8.4*0.4 & -1.98 & -0.18 & 0.06 & -1.9*0.5 & -3.8*0.6 & 3.4*0.2 \\ 

27 & ALS\,15956 & 4.0*0.4 & 12.8*0.6\,^{c} & -2.32 & -0.20 & 0.78 & -2.8*0.7 & -5.1*0.7 & 3.9*0.3 \\ 

29 & HD\,36982 & 4.0*0.2 & 8.3*0.2\,^{c} & -2.21 & -0.19 & 1.01 & -0.8*0.3 & -3.0*0.4 & 3.1*0.2 \\ 

31 & HD\,37479 & 4.0*0.2 & 8.5*0.4\,^{c} & -2.31 & -0.20 & 0.07 & -1.9*0.5 & -4.2*0.5 & 3.6*0.2 \\ 

32 & HD\,149277\,$^\dagger$ & 4.0*0.4 & 10.7*0.6\,^{c} & -2.21 & -0.19 & 0.72 & -3.1*0.7 & -5.3*0.7 & 4.0*0.3 \\ 

35 & HD\,136504\,$^\dagger$ & 4.0*0.2 & 6.1*0.3 & -2.21 & -0.19 & 0.04 & -2.7*0.4 & -5.0*0.4 & 3.9*0.2 \\ 

36 & HD\,156424 & 4.0*0.3 & 10*1\,^{c} & -2.21 & -0.19 & 0.78 & -2*1 & -4*1 & 3.7*0.4 \\ 

37 & HD\,156324 & 4.0*0.3 & 10*1\,^{c} & -2.21 & -0.19 & 0.92 & -2*1 & -5*1 & 3.7*0.4 \\ 

38 & HD\,121743 & 4.0*0.3 & 6.03*0.07 & -2.21 & -0.19 & 0.00 & -2.2*0.3 & -4.4*0.4 & 3.7*0.2 \\ 

40 & HD\,186205\,$^\dagger$ & 4.0*0.2 & 14*1 & -1.98 & -0.18 & 0.70 & -6*1 & -8*1 & 5.1*0.5 \\ 

41 & HD\,67621 & 4.0*0.3 & 8.0*0.2 & -1.85 & -0.17 & 0.00 & -1.7*0.3 & -3.5*0.5 & 3.3*0.2 \\ 

43 & HD\,35912 & 4.0*0.1 & 8.2*0.7 & -1.78 & -0.16 & 0.00 & -1.8*0.7 & -3.5*0.7 & 3.3*0.3 \\ 

44 & HD\,66522 & 4.0*0.4 & 8.3*0.4 & -1.71 & -0.16 & 0.64 & -1.8*0.4 & -3.5*0.5 & 3.3*0.2 \\ 

48 & HD\,58260\,$^\dagger$ & 3.8*0.3 & 9.7*0.5 & -1.97 & -0.18 & 0.15 & -3.2*0.6 & -5.1*0.6 & 3.9*0.3 \\ 

49 & HD\,36485\,$^\dagger$ & 4.2*0.2 & 8.4*0.7\,^{c} & -1.99 & -0.18 & 0.06 & -1.6*0.7 & -3.6*0.8 & 3.3*0.3 \\ 

50 & HD\,208057\,$^\dagger$ & 3.9*0.2 & 6.4*0.1 & -1.85 & -0.17 & 0.00 & -1.4*0.3 & -3.2*0.5 & 3.2*0.2 \\ 

51 & HD\,306795 & 3.9*0.2 & 11.6*0.6\,^{c} & -1.71 & -0.16 & 0.55 & -1.5*0.7 & -3.2*0.7 & 3.2*0.3 \\ 

52 & HD\,25558\,$^\dagger$ & 4.0*0.2 & 6.5*0.1 & -1.64 & -0.15 & 0.21 & -1.4*0.3 & -3.0*0.4 & 3.1*0.2 \\ 

53 & HD\,35298\,$^\dagger$ & 4.0*0.3 & 12*1 & -1.41 & -0.14 & 0.00 & -4*1 & -5*1 & 3.9*0.5 \\ 

54 & HD\,130807 & 4.2*0.2 & 5.5*0.2 & -1.72 & -0.15 & 0.00 & -1.2*0.3 & -2.9*0.4 & 3.1*0.2 \\ 

55 & HD\,142990\,$^\dagger$ & 4.2*0.2 & 6.16*0.09 & -1.64 & -0.15 & 0.17 & -0.9*0.3 & -2.6*0.4 & 2.9*0.2 \\ 

56 & HD\,37058\,$^\dagger$ & 3.8*0.2 & 8.5*0.4\,^{c} & -1.71 & -0.16 & 0.09 & -1.3*0.5 & -3.0*0.6 & 3.1*0.2 \\ 

57 & HD\,35502\,$^\dagger$ & 4.0*0.3 & 9*2 & -1.41 & -0.14 & 0.30 & -2*2 & -4*2 & 3.3*0.8 \\ 

59 & HD\,189775 & 4.0*0.3 & 6.9*0.1 & -1.41 & -0.14 & 0.00 & -0.8*0.3 & -2.2*0.4 & 2.8*0.1 \\ 

60 & HD\,61556\,$^\dagger$ & 4.0*0.3 & 5.2*0.2 & -1.24 & -0.12 & 0.00 & -0.4*0.3 & -1.7*0.4 & 2.6*0.1 \\ 

61 & HD\,175362\,$^\dagger$ & 4.0*0.3 & 5.60*0.08 & -1.24 & -0.12 & 0.00 & -0.2*0.3 & -1.5*0.4 & 2.5*0.1 \\ 

62 & HD\,105382\,$^\dagger$ & 4.0*0.2 & 5.7*0.2 & -1.64 & -0.15 & 0.00 & -1.2*0.3 & -2.9*0.4 & 3.0*0.2 \\ 

63 & HD\,125823 & 4.0*0.2 & 5.73*0.06 & -1.85 & -0.17 & 0.00 & -1.4*0.3 & -3.2*0.4 & 3.2*0.1 \\ 

64 & HD\,36526 & 4.0*0.3 & 8.4*0.7\,^{c} & -1.41 & -0.14 & 0.08 & -0.1*0.7 & -1.5*0.8 & 2.5*0.3 \\ 

\hline
\multicolumn{10}{l}{$^a$ From the SB2 analysis of \protect{\citet{2006A&A...455..259A}}.} \\
\multicolumn{10}{l}{$^c$ Distance estimates from associations with stellar clusters (Hipparcos otherwise).}\\
\end{tabular}
}
	\end{center}
\end{table*}

The theoretical $BC$ and $(B-V)_0$ (columns 5 and 6) are determined from a smooth interpolation of the grids provided by \citet{2005A&A...436.1049M}, and \citet{2006A&A...457..637M} for the O-type stars and \citet{2007ApJS..169...83L} for the B-type stars. 
We use an extinction $R_\mathrm{V}=3.1$ to derive the extinction $A_V=R_V\,E(B-V)$ (column 7). 
The absolute visual magnitude ($M_V=V-A_V-DM$), the bolometric magnitude ($M_\mathrm{bol}=M_V+BC$) and the luminosity [$\lum=(M_{\mathrm{bol},\odot}-M_\mathrm{bol})/2.5$] are given in columns 8 to 10.

With a typical uncertainty of 2\,000\,K in $\teff$ and 0.3\,dex in $\logg$, we estimate an uncertainty of 0.2 and 0.02\,mag in $BC$ and $(B-V)_0$, respectively. Given the wide range of $R_\mathrm{V}$ often encountered in the literature for OB stars, we adopt a conservative error in $A_V$ of 0.25\,mag. 
In most cases, $BC$, $A_\mathrm{V}$ and $DM$ contribute equally to the uncertainty, leading to 0.2-0.3\,dex for the luminosity.  In five cases (ID: 36, 37, 40, 53 and 57) the luminosity error estimate from the bolometric correction method is more than 0.4\,dex, given the large uncertainty in distance.

\begin{table*}
	\begin{center}
	\caption{\label{tab:lumCho}Luminosity determination based on SED fitting with \textsc{Chorizos} (\S\,\ref{sec:lum}).}
		{\begin{tabular}{l l D{*}{\,\pm\,}{4,4} D{*}{\,\pm\,}{4,4} D{*}{\,\pm\,}{4,4} D{*}{\,\pm\,}{4,4} D{*}{\,\pm\,}{4,4}}
\hline
\mcc{ID} & \mcc{Star} & \mcc{$DM$} & \mcc{$A_V$} & \mcc{$\logg$} & \mcc{$\ms$} & \mcc{$\lum$} \\
\mcc{ } & \mcc{ } & \mcc{mag} & \mcc{mag} & \mcc{cgs} & \mcc{$\msol$} & \mcc{} \\
\mcc{(1)} & \mcc{(2)} & \mcc{(3)} & \mcc{(4)} & \mcc{(5)} & \mcc{(6)} & \mcc{(7)} \\
\hline
25 & HD\,66765 & 8.6*0.5 & 0.21*0.02 & 3.9*0.2 & 7.5*0.6 & 3.6*0.2 \\ 

27 & ALS\,15956 & 13.6*0.6 & 0.92*0.02 & 3.6*0.2 & 11*1 & 4.3*0.2 \\ 

35 & HD\,136504\,$^\dagger$ & 5.6*0.5 & 0.19*0.04 & 4.0*0.2 & 8.6*0.7 & 3.8*0.2 \\ 

40 & HD\,186205\,$^\dagger$ & 9.9*0.5 & 0.75*0.02 & 4.0*0.2 & 7.4*0.6 & 3.5*0.2 \\ 

48 & HD\,58260\,$^\dagger$ & 10.0*0.5 & 0.24*0.01 & 3.5*0.2 & 9*1 & 4.1*0.2 \\ 

49 & HD\,36485\,$^\dagger$ & 8.6*0.4 & 0.19*0.02 & 4.0*0.1 & 7.1*0.4 & 3.5*0.1 \\ 

50 & HD\,208057\,$^\dagger$ & 7.2*0.5 & 0.12*0.02 & 3.9*0.2 & 7.1*0.6 & 3.6*0.2 \\ 

53 & HD\,35298\,$^\dagger$ & 9.7*0.6 & 0.12*0.02 & 3.8*0.2 & 5.6*0.5 & 3.2*0.2 \\ 

54 & HD\,130807 & 5.3*0.3 & 0.13*0.01 & 4.1*0.1 & 5.7*0.2 & 3.1*0.1 \\ 

56 & HD\,37058\,$^\dagger$ & 9.4*0.5 & 0.07*0.03 & 3.8*0.2 & 6.6*0.5 & 3.5*0.2 \\ 

57 & HD\,35502\,$^\dagger$ & 8.9*0.6 & 0.40*0.01 & 3.8*0.2 & 5.7*0.5 & 3.3*0.2 \\ 

59 & HD\,189775 & 7.9*0.5 & 0.00*0.02 & 3.8*0.2 & 5.5*0.5 & 3.2*0.2 \\ 

61 & HD\,175362\,$^\dagger$ & 7.2*0.4 & 0.08*0.01 & 3.7*0.2 & 5.3*0.4 & 3.2*0.1 \\ 

\hline
\end{tabular}
}
	\end{center}
\end{table*}

For the stars with a complete set of $UBVRJHK$ photometry, we perform SED fitting using the Bayesian (spectro)photometric code \textsc{Chorizos}. The results are presented in Table \ref{tab:lumCho}.  
In the latest \textsc{Chorizos} version, the user can select distance to be an independent parameter, by applying atmosphere models (\textsc{tlusty} for OB stars) calibrated in luminosity with the help of Geneva stellar evolution tracks (excluding rotational effects).  
The parameters of such models are the logarithmic distance (column 3), the extinction (here fixed at $R_{5495}=3.1$\footnote{The extinction law is defined by the monochromatic quantity $R_{5495} \equiv A_{5495}/E(4405-5495)$ instead of a band-integrated one such as $R_{V} \equiv A_{V}/E(B-V)$, because the former depends only on the properties of the dust while the latter also depends on the input SED and the amount of dust present along the line of sight. See Ma\'{\i}z Apell\'aniz (2012) for details.}), the reddening (transformed to $A_\mathrm{V}$ in column 4), the effective temperature (here fixed to the literature estimate) and the luminosity class. 
The distance prior probability range has been left relatively wide around the Hipparcos or cluster-estimated value and the luminosity class prior probability was based on the gravity estimates used for the bolometric correction approach (column 3 of table \ref{tab:BC}) with an interval of 1.5$\sigma$.
From these fitted values, we can derive an estimate of the surface gravity (column 5), the evolutionary mass 
(column 6) and the luminosity (column 7).

Good fits to the photometry are achieved for the 13 stars displayed in Table \ref{tab:lumCho}, leading to better estimates of their luminosity (especially for the two stars with the largest uncertainty with the bolometric correction approach). 
Poorer fits were obtained for the remaining 9 stars with complete photometry.  
Incompatibility between optical and near-IR photometry, probably due to near-IR excess, could be a possible cause of the discrepancy. 
Therefore, for the stars with good fits, we use the luminosity, gravity and mass derived from \textsc{Chorizos}. 
For the remaining stars, we opt for the bolometric correction luminosity determination.

\subsection{Rotational and magnetic parameters}
\label{sec:rb}

Monitoring of the disk-integrated longitudinal field variations provides a natural and direct way to determine rotational periods for magnetic stars \citep[in the context of the Oblique Rotator Model; e.g.][]{1950MNRAS.110..395S}. Photometric and spectral variability associated with the magnetic field also provide a convenient and easy way to determine periods, even though some ambiguity can exist between e.g. short rotational periods and long pulsation periods.

In Table \ref{tab:stars}, column (11) gives the rotational period in days. When no period is available we use the measured $\vsini$ (column 12) as a lower limit to the equatorial velocity. In two cases (ALS\,15211, ID\,19; ALS\,15956, ID\,27), no $\vsini$ measurements are available, due to a lack of high-resolution spectra.
These stars would be prime candidates for further monitoring.

In four cases (HD\,96446,  HD\,136504, HD\,58260 and HD\,37058; ID 24, 35, 48 and 56) more than one period is reported in the literature. In these cases, we use the longest period for a lower limit on the equatorial velocity. We provide the magnetospheric calculations for the alternative periods in the notes of Appendix \ref{sec:notes}.

Column (13) gives the estimated polar strength ($B_p$) of the surface dipole in kilogauss. When only longitudinal magnetic measurements are available, we use a value of three times the strongest longitudinal field measurement (corresponding to a conservative limb-darkening coefficient\footnote{In the case of a dipolar field, the dipole strength can be expressed as $\bp\geq 4\frac{15-5\epsilon}{15+\epsilon}|\mbl |_\mathrm{max}$, where $\epsilon$ is the limb-darkening coefficient and $|\mbl|_\mathrm{max}$ is the maximum of the disk-integrated longitudinal field variation \citep{1967ApJ...150..547P}.} of 0.6), setting a lower limit on the dipolar field strength. A superscript $m$ in column (13) indicates stars that are known to possess a magnetic field with a significant contribution from multipole components higher than a simple dipole.

\section{Two-parameter classification of magnetospheres}
\label{sec:theo}

\subsection{Alfv\'en radius $\ra$ vs.\ Kepler co-rotation radius $\rk$}
\label{sec:mcp}

The high luminosity of massive stars drives powerful, high-speed stellar winds.
MHD  simulation studies \cite[e.g.][]{2002ApJ...576..413U,2008MNRAS.385...97U}
show that the overall net effect of a large-scale, dipole magnetic field in diverting such a wind can be well
characterised by a single
\textit{wind magnetic confinement parameter},
\begin{equation}
\eta_{\ast} \equiv \frac {B_{eq}^2 \, \rs^2} {\mdot \, \vinf}
\, ,
\label{eq:esdef}
\end{equation}
where $\beq = \bp/2$ is the field strength at the magnetic equatorial surface radius $\rs$,
and $\mdot$ and $\vinf$ are the fiducial mass-loss rate and terminal speed that the star \textit{would have} in the \textit{absence} of any magnetic field.

This confinement parameter sets the scaling for the ratio of the magnetic to wind kinetic energy density.
For a dipole field, the  $r^{-6}$ radial decline of magnetic energy density is much steeper than the $r^{-2}$ decline of the wind's mass and energy density;
this means the wind always dominates beyond the \textit{Alfv\'en radius} $\ra$ \citep{2008MNRAS.385...97U}, 
given by the approximate general scaling, 
\begin{equation}
    \frac{\ra}{\rs} \approx 0.3 + \left ( \etas + 0.25 \right)^{1/4} 
    \, .
    \label{eq:radef}
\end{equation}
Magnetic loops extending above $\ra$ are drawn open by the wind, while those with an apex below $\ra$ remain closed.
Indeed, the trapping of wind upflow from opposite footpoints of closed magnetic loops leads to strong collisions that may form X-ray emitting, magnetically confined wind shocks \citep[MCWS][see \S\ref{sec:xray}]{1997ApJ...485L..29B,1997A&A...323..121B}.
In models with negligible rotation, the post-shock material eventually cools and falls back onto the star, leading to a relatively complex, dynamic pattern of infall and wind outflow \citep[see e.g. lower row of Figure 9 of][also Figure\,\ref{fig:cartoon}]{2008MNRAS.385...97U}.

For the simple 2D axisymmetric case of a magnetic dipole that is aligned with a star's rotation axis,
\citet{2008MNRAS.385...97U}
extended these MHD simulation studies to explore the additional effect of stellar rotation.
They found it convenient to cast results in terms of the ratio of the rotation speed $\vrot$ to orbital speed $\vcrit$ at the equatorial surface radius $\rs$,
\begin{equation}
W \equiv  \frac{\vrot}{\vcrit} = \frac{\omega\rs}{\sqrt{G\ms/\rs}} 
\, ,
\label{eq:wdef}
\end{equation}
where the latter equality expresses this ratio in terms of the angular rotation frequency $\omega$, 
with $\ms$ the  stellar mass.
To avoid the complications associated with a rotationally distorted, oblate stellar surface, \citet{2008MNRAS.385...97U}  restricted their simulations to cases with $W \le 0.5$.
But if we associate $\rs$ with the \textit{actual equatorial} radius for the given rotation rate $\omega$ , then even for more rapid,  near-critical rotation, $W$ simply compares the star's equatorial rotation speed to the speed $\vcrit$ needed to reach Keplerian orbit near this equatorial surface\footnote{
For critical rotation ($W=1$), $\rs = 3 \rp /2$, where $\rp$ is the \textit{fixed} polar radius.
In terms of the associated critical rotational frequency 
$\omega_\mathrm{crit}  \equiv \sqrt{8GM/27 \rp^3}$, 
one can alternatively define a critical rotation ratio
 $\Omega\equiv\omega/\omega_\mathrm{crit}$. 
 We then find $W=\Omega (2\rs / 3\rp)$, with
  $\rs / \rp =  3 \cos[(\cos^{-1}[\Omega]+\pi)/3]/\Omega$ \citep{1966ApJ...146..152C}.
}.

In a magnetic star, torques from the magnetic field on any wind outflow can maintain rigid-body co-rotation up to roughly the  
Alfv\'en radius, so that the azimuthal speed of the confined wind plasma \textit{increases} with radius as $v_\phi = \vrot r/\rs$.
The outward centrifugal force from such rigid-body rotation will balance the inward force of gravity at the
\textit{Kepler corotation radius},
\begin{equation}
\rk \equiv  \left(\frac{GM}{\omega^2}\right)^{1/3} = W^{-2/3} \rs
    \, .
    \label{eq:rkdef}
\end{equation}
Together the two parameters $\etas$ and $W$ thus define the relative locations of the 
Alfv\'en and Kepler radii with respect to the equatorial radius.

\subsection{Dynamical vs.\ Centrifugal Magnetospheres}
\label{sec:dmcm}

For the simple case of  field-aligned rotation, \citet{2008MNRAS.385...97U} carried out an extensive MHD simulation parameter study varying both $W$ and $\etas$.
For $\etas < 1$, the field exerts only a modest perturbation on the wind;
but for $\etas > 1$,  outflow near the magnetic equator is trapped  within the Alfv\'{e}n radius by closed magnetic loops, forming a wind-fed circumstellar \textit{magnetosphere}. 
It was found that the dynamical evolution of this trapped magnetospheric material depends crucially on the rotation parameter $W$, and specifically on the relative magnitude of the associated Kepler vs.\ the Alfv\'{e}n radii.

\begin{figure}
\begin{center}
	\includegraphics[width=50mm]{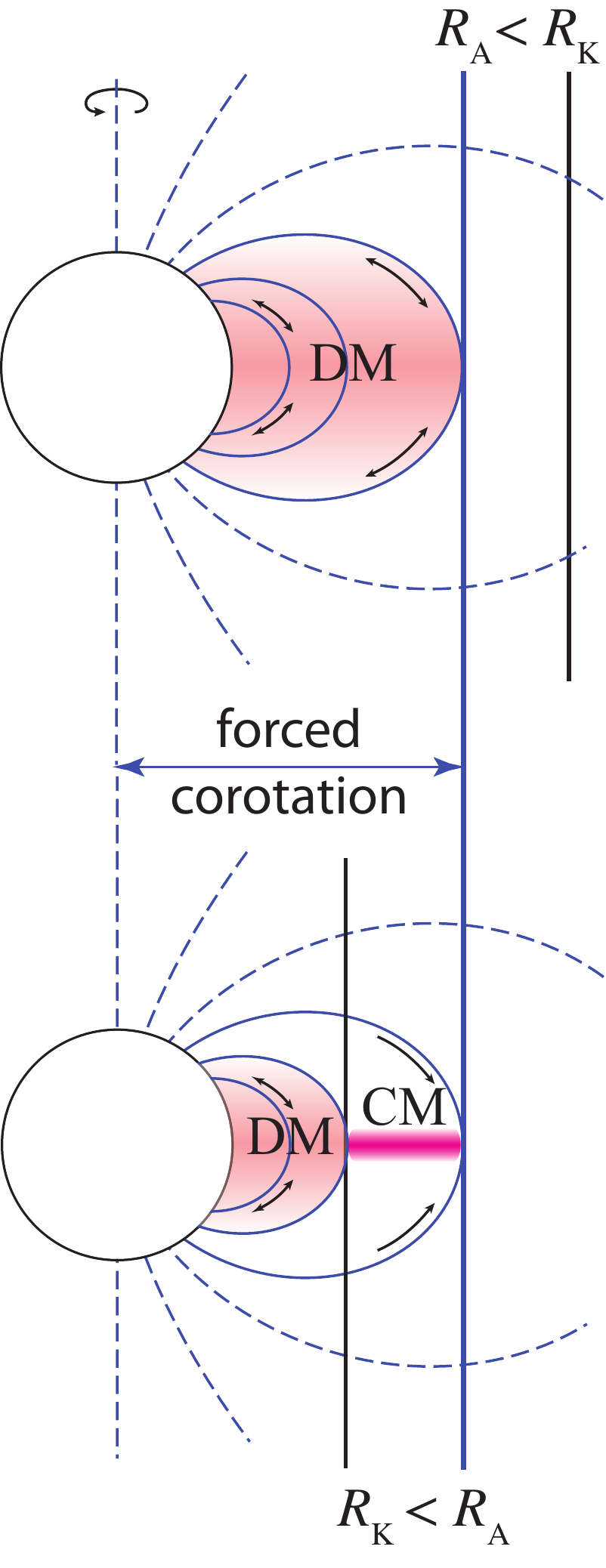}
\caption{\label{fig:cartoon}
Sketch of the regimes for dynamical vs.\ centrifugal magnetospheres (DM vs. CM).\
The top panel illustrates the case of a slowly rotating star with Kepler radius beyond the Alfv\'{e}n radius ($\rk > \ra$);
the lack of centrifugal support means that trapped material falls back to the star on a dynamical timescale, forming a DM, with colour illustrating the rough time-averaged distribution of density.
The lower panel is for a star with more rapid rotation and $\rk < \ra$, leading then to a region between these radii where a net outward centrifugal force against gravity is balanced by the magnetic tension of closed loops; this allows material to build up to  the much higher density of CM.	
}
\end{center}
\end{figure}

In a simplified, schematic form, Figure \ref{fig:cartoon} here illustrates that, depending on the relative positions of $\rk$ vs.\ $\ra$, regions of trapped equatorial material can be alternatively characterised as forming a \textit{dynamical} vs. \textit{centrifugal} magnetosphere (DM vs. CM).
 As sketched in the upper panel of Figure \ref{fig:cartoon}, for slowly rotating stars with $\ra < \rk $, material trapped in closed magnetic loops falls back to the star on a dynamical timescale, forming a DM 
 \citep{2012MNRAS.423L..21S}. 
In contrast, the lower panel shows that, for the more rapidly rotating case with $\ra > \rk$, material caught in the region between $\ra$ and $\rk$ is centrifugally supported against infall, and so builds up to a much denser CM (for a given fiducial mass-loss rate).
 Even for such rapid rotators, the inner regions below $\rk$ again have the infall of a DM, but the plasma density, and thus any circumstellar emission, is much lower than that of the CM region.

 Indeed, since the much longer confinement time allows material to accumulate to high density even if the feeding by the wind mass flux is weak, such a CM can exhibit rotationally modulated line emission even in the relatively low-luminosity, but strongly magnetic Bp stars, so long as the stellar rotation is sufficient to give $ \rk < \ra$
\citep{2005MNRAS.357..251T,2005ApJ...630L..81T}.
For slowly rotating magnetic stars with a DM, accumulating sufficiently high density plasma for line emission requires a much stronger wind to overcome the dynamical timescale leakage of infall back onto the star. For the luminous, slowly rotating magnetic O-type star HD\,191612 (ID\,4), 
\cite{2012MNRAS.423L..21S} showed that the emission from its wind-fed DM matches its observed H$\alpha$ emission quite well.

The transition from stars with a pure DM to those with a CM occurs near $\rk = \ra$; 
from equations (\ref{eq:radef}) and (\ref{eq:rkdef}), the associated transition value $W_\mathrm{t}$ for the rotation fraction is
\begin{equation}
W_\mathrm{t} =  \left [  0.3 + \left ( \etas + 0.25 \right)^{1/4} \right]^{-3/2} 
\, ,
\end{equation}
which in the strong confinement limit, $\etas \gg 1$,  simply requires $W_\mathrm{t}  \approx \etas^{-3/8}$.

\begin{figure*}
\begin{center}

	\includegraphics[width=168mm]{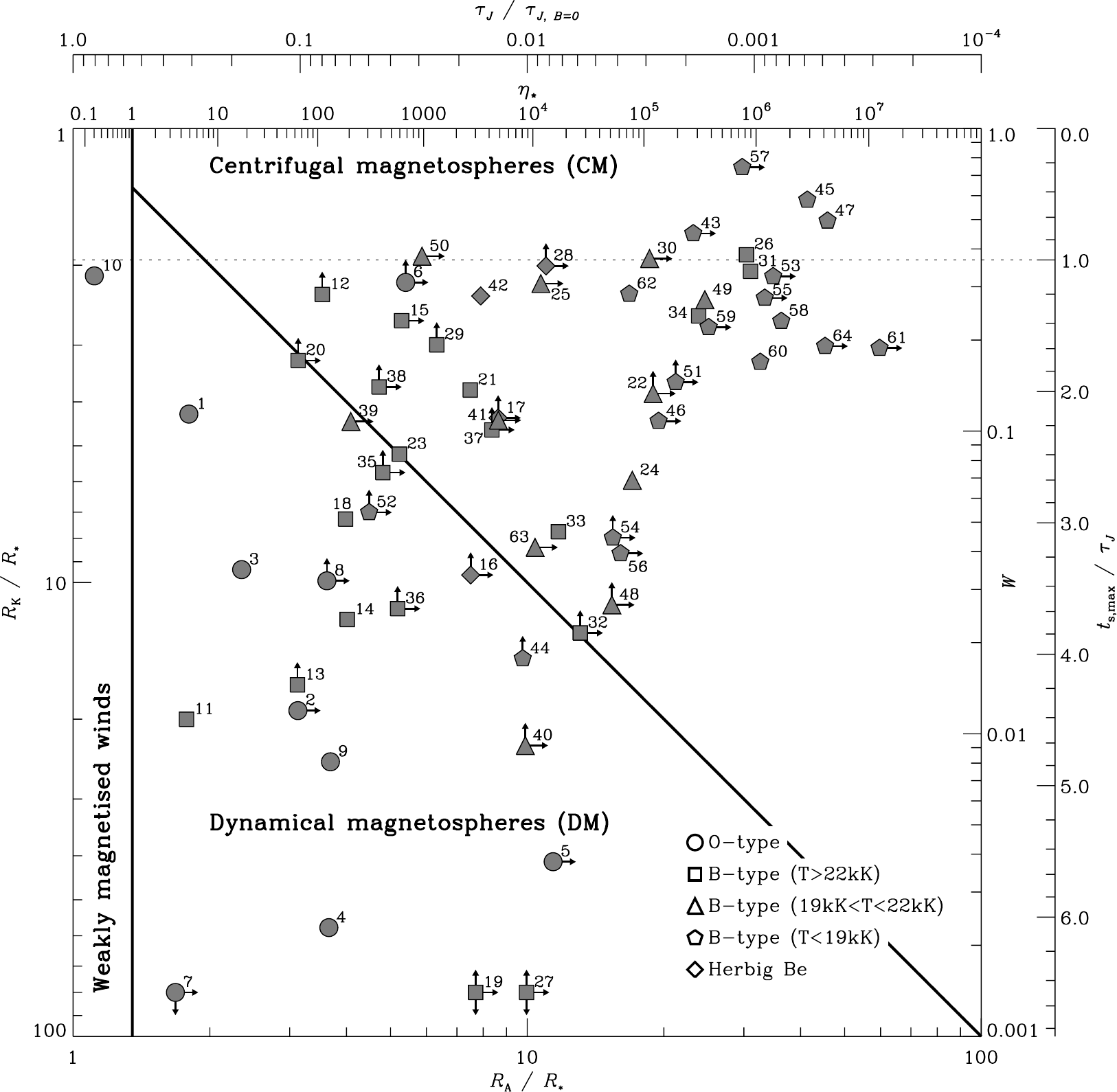}
	\caption{\label{fig:etaW}	
	Location of magnetic massive stars in the magnetic confinement-rotation diagram; a log-log plot with Kepler radius $\rk$ increasing downward and Alfv\'{e}n radius $\ra$  increasing to the right.
	The right and upper axes give respectively the corresponding rotation fraction $W$ and magnetic confinement parameter $\etas$. The solid lines separates the magnetosphere domains of weakly magnetised winds (with $\etas<1$), dynamical magnetospheres with $\ra<\rk$ (DM), and centrifugal magnetospheres with $\ra>\rk$ (CM), as defined in \S\ref{sec:dmcm}. 
	The additional upper and right axes give respectively the corresponding spindown timescale $\tauj$ (normalised by the value in a non-magnetised wind) and maximum spindown age $\tsmax$ (normalised by the spindown time and therefore the number of spindown e-folds), as defined in \S\,\ref{sec:spindown}. Stars above the dashed line have a maximum spindown age less than one spindown time.
	As in Figure \ref{fig:HR}, the symbol shapes denote spectral type, and the numbers correspond to the ID in column (1) of Table \ref{tab:stars}. The three downward arrows indicate two stars (ALS\,15211, ID\,19; ALS\,15956, ID\,27) for which no $\vsini$ measurement is available (e.g. $W>0$), and HD\,108 (ID\,7) for which $\rk\sim500\rs$. }

\end{center}
\end{figure*}

Figure \ref{fig:etaW} plots our sample of magnetic stars in the magnetic confinement-rotation diagram, a log-log plane with $\rk/\rs$ increasing downward on the ordinate vs.\ $\ra/\rs$ increasing to the right along the  abscissa.
As detailed in the next subsection (\S\,\ref{sec: mparams}),
the placement of the individual stars depends on inference of the relevant parameters that set the magnetic confinement $\etas$ (noted on the top axis) and rotation fraction $W$ (on the right axis).
The vertical line at $\etas = 1$ ($\ra/\rs \approx 1.3$) separates  weakly magnetised winds at the far left from the broad domain of stars with significant magnetospheres, with the diagonal line separating the stars with a CM to the upper right from those with just a DM to the lower left.
As detailed in \S\,\ref{sec:spindown},
the additional upper and right axes refer to associated stellar spindown properties, namely the stellar spindown timescale ($\tauj$) and the maximum spindown age ($\tsmax$), respectively.
Stars above the horizontal dotted line have a maximum spindown age $\tsmax$  that is less than one spindown time $\tauj$.

\subsection{Calculation of magnetospheric parameters}
\label{sec: mparams}

In this section, we determine the magnetospheric parameters described in \S\,\ref{sec:mcp} for all the stars in the sample.
Table \ref{tab:magneto} compiles our calculations of $\eta_\star$, $\ra/\rs$, $W$ and $\rk/\rs$ (equations 1 to 4) in columns (4) to (7).


\afterpage{\small \onecolumn \begin{landscape}  \begin{longtable}{l l c D{.}{.}{3,5} D{.}{.}{5,5} D{.}{.}{3,5} D{.}{.}{5,5} D{.}{.}{5,5} D{.}{.}{5,5} D{.}{.}{5,5} c c D{.}{.}{3,3} }
\caption{\label{tab:magneto}List of magnetospheric parameters, magnetic spindown properties, and H$\alpha$, UV and X-ray proxies.}\\
\hline
\mcc{ID} & \mcc{Star} & \mcc{Remark} & \mcc{$W$} & \mcc{$\rk/\rs$} & \mcc{$\etas$} & \mcc{$\ra/\rs$} & \mcc{$\tauj$} & \mcc{$\tsmax$} & \mcc{$\log(\ra/\rk)$} & \mcc{H$\alpha$} & \mcc{UV} & \mcc{$\lxls$} \\
\mcc{ } & \mcc{ } & \mcc{ } & \mcc{} & \mcc{} & \mcc{} & \mcc{} & \mcc{Myr} & \mcc{Myr} & \mcc{} & \mcc{ } & \mcc{ } & \mcc{} \\
\mcc{(1)} & \mcc{(2)} & \mcc{(3)} & \mcc{(4)} & \mcc{(5)} & \mcc{(6)} & \mcc{(7)} & \mcc{(8)} & \mcc{(9)} & \mcc{(10)} & \mcc{(11)} & \mcc{(12)} & \mcc{(13)} \\
\hline
\endfirsthead
\caption{Continued}\\
\hline
\mcc{ID} & \mcc{Star} & \mcc{Remark} & \mcc{$W$} & \mcc{$\rk/\rs$} & \mcc{$\etas$} & \mcc{$\ra/\rs$} & \mcc{$\tauj$} & \mcc{$\tsmax$} & \mcc{$\log(\ra/\rk)$} & \mcc{H$\alpha$} & \mcc{UV} & \mcc{$\lxls$} \\
\mcc{ } & \mcc{ } & \mcc{ } & \mcc{} & \mcc{} & \mcc{} & \mcc{} & \mcc{Myr} & \mcc{Myr} & \mcc{} & \mcc{ } & \mcc{ } & \mcc{} \\
\mcc{(1)} & \mcc{(2)} & \mcc{(3)} & \mcc{(4)} & \mcc{(5)} & \mcc{(6)} & \mcc{(7)} & \mcc{(8)} & \mcc{(9)} & \mcc{(10)} & \mcc{(11)} & \mcc{(12)} & \mcc{(13)} \\
\hline
\endhead
\hline
\multicolumn{13}{r}{Following on the next page} \\
\endfoot
\hline
\multicolumn{13}{l}{$^\dagger$ Notes in Appendix.} \\
\endlastfoot
1 & HD\,148937& &1.1\,\mbox{e-}1 & 4.3 & 4.9 & 1.8 & 0.82 &1.8 &-0.37 & var em & var & -6.6 \\ 

2 & CPD\,-28\,2561& &1.2\,\mbox{e-}2 & 19 & >6.4\,\mbox{e}1 & >3.1 & <0.92 &<4.1 &>-0.79 & var em &  &   \\ 

3 & HD\,37022\,$^\dagger$&SB1 &3.5\,\mbox{e-}2 & 9.4 & 1.8\,\mbox{e}1 & 2.4 & 3.1 &10 &-0.60 & var em & per & -5.6 \\ 

4 & HD\,191612\,$^\dagger$&SB2 &2.3\,\mbox{e-}3 & 57 & 1.3\,\mbox{e}2 & 3.7 & 0.38 &2.3 &-1.20 & var em & var & -6.0 \\ 

5 & NGC\,1624-2& &3.8\,\mbox{e-}3 & 41 & >1.5\,\mbox{e}4 & >11 & <0.24 &<1.4 &>-0.56 & var em &  & -6.6 \\ 

6 & HD\,47129\,$^\dagger$&SB2 &>3.1\,\mbox{e-}1 & <2.2 & >6.8\,\mbox{e}2 & >5.4 & <4.1 &<4.8 &>0.39 & em & var & -5.6 \\ 

7 & HD\,108& &8.3\,\mbox{e-}5 & 526 & >3.5 & >1.7 & <0.90 &<8.5 &>-2.50 & var em & var & -6.2 \\ 

8 & ALS\,15218\,$^\dagger$& &>3.2\,\mbox{e-}2 & <9.9 & >1.2\,\mbox{e}2 & >3.6 & <3.1 &<10 &>-0.44 & em &  & -6.3 \\ 

9 & HD\,57682& &8.1\,\mbox{e-}3 & 24 & 1.3\,\mbox{e}2 & 3.7 & 2.2 &10 &-0.83 & var em & var & -6.3 \\ 

10 & HD\,37742\,$^\dagger$&SB2 &3.3\,\mbox{e-}1 & 2.1 & 2.0\,\mbox{e-}1 & 1.1 & 3.9 &4.4 &-0.28 & var em & abs & -6.7 \\ 

11 & HD\,149438& &1.1\,\mbox{e-}2 & 20 & 4.6 & 1.8 & 21 &97 &-1.05 & var abs & per & -6.3 \\ 

12 & HD\,37061\,$^\dagger$&SB2 &>2.8\,\mbox{e-}1 & <2.3 & 1.1\,\mbox{e}2 & 3.5 & 27 &<35 &>0.18 & abs & stab abs & -6.9 \\ 

13 & HD\,63425& &>1.5\,\mbox{e-}2 & <16 & 6.3\,\mbox{e}1 & 3.1 & 22 &<93 &>-0.73 & stab abs & var &   \\ 

14 & HD\,66665& &2.4\,\mbox{e-}2 & 12 & 1.9\,\mbox{e}2 & 4.0 & 12 &46 &-0.48 & stab abs & var &   \\ 

15 & HD\,46328&$\beta$\,Cep &2.3\,\mbox{e-}1 & 2.7 & >6.2\,\mbox{e}2 & >5.3 & <1.6 &<2.3 &>0.30 & em & em & -6.8 \\ 

16 & ALS\,8988&HeBe &>3.3\,\mbox{e-}2 & <9.6 & >2.7\,\mbox{e}3 & >7.5 & <23 &<78 &>-0.11 & HeBe &  & -8.8 \\ 

17 & HD\,47777&HeBe &>1.1\,\mbox{e-}1 & <4.3 & >4.8\,\mbox{e}3 & >8.6 & <8.9 &<19 &>0.30 & HeBe &  & -6.9 \\ 

18 & HD\,205021\,$^\dagger$&SB2,$\beta$\,Cep &5.1\,\mbox{e-}2 & 7.3 & 1.8\,\mbox{e}2 & 4.0 & 43 &128 &-0.26 & abs & per & -7.2 \\ 

19 & ALS\,15211\,$^\dagger$& &  &  &>3.0\,\mbox{e}3 & >7.7 & <17 &  & >-2.29 &  &  & -7.8 \\ 

20 & HD\,122451\,$^\dagger$&SB2,$\beta$\,Cep &>1.7\,\mbox{e-}1 & <3.2 & >6.5\,\mbox{e}1 & >3.1 & <14 &<26 &>-0.02 & abs & stab abs & -7.3 \\ 

21 & HD\,127381& &1.4\,\mbox{e-}1 & 3.8 & 2.7\,\mbox{e}3 & 7.5 & 127 &254 &0.30 & stab abs & var &   \\ 

22 & ALS\,3694& &>1.3\,\mbox{e-}1 & <3.8 & >1.2\,\mbox{e}5 & >18 & <3.0 &<6.1 &>0.69 &  &  &   \\ 

23 & HD\,163472&$\beta$\,Cep &8.4\,\mbox{e-}2 & 5.2 & 5.9\,\mbox{e}2 & 5.2 & 157 &390 &0.00 & abs & per & -8.6 \\ 

24 & HD\,96446\,$^\dagger$& &6.9\,\mbox{e-}2 & 6.0 & 7.9\,\mbox{e}4 & 17 & 2.2 &5.8 &0.46 & stab abs & em &   \\ 

25 & HD\,66765& &3.1\,\mbox{e-}1 & 2.2 & >1.2\,\mbox{e}4 & >10 & <4.8 &<5.6 &>0.69 & stab abs &  &   \\ 

26 & HD\,64740& &3.8\,\mbox{e-}1 & 1.9 & 8.2\,\mbox{e}5 & 30 & 1.7 &1.6 &1.20 & var em & per & -7.3 \\ 

27 & ALS\,15956& &  &  &>8.8\,\mbox{e}3 & >9 & <7.4 &  & >-2.14 &  &  & -6.8 \\ 

28 & ALS\,9522&HeBe &>3.5\,\mbox{e-}1 & <2.0 & >1.3\,\mbox{e}4 & >11 & <1.3 &<1.4 &>0.74 & HeBe &  & -7.3 \\ 

29 & HD\,36982& &>1.9\,\mbox{e-}1 & <3.0 & 1.3\,\mbox{e}3 & 6.3 & 8.6 &<14 &>0.32 & stab abs & stab abs & -7.8 \\ 

30 & HD\,37017\,$^\dagger$&SB2 &3.7\,\mbox{e-}1 & 1.9 & >1.1\,\mbox{e}5 & >18 & <3.7 &<3.6 &>0.98 & var em & per & -7.8 \\ 

31 & HD\,37479& &3.4\,\mbox{e-}1 & 2.1 & 8.9\,\mbox{e}5 & 31 & 4.6 &5.0 &1.18 & var em & per & -5.9 \\ 

32 & HD\,149277\,$^\dagger$&SB2 &>2.2\,\mbox{e-}2 & <12 & >2.7\,\mbox{e}4 & >13 & <2.6 &<10 &>0.01 & abs &  &   \\ 

33 & HD\,184927& &4.7\,\mbox{e-}2 & 7.7 & 1.7\,\mbox{e}4 & 11 & 1.7 &5.2 &0.18 & stab abs & per &   \\ 

34 & HD\,37776\,$^\dagger$& &2.4\,\mbox{e-}1 & 2.6 & 3.1\,\mbox{e}5 & 23 & 0.68 &0.97 &0.96 & var em & per &   \\ 

35 & HD\,136504\,$^\dagger$&SB2,$\beta$\,Cep &>7.3\,\mbox{e-}2 & <5.7 & >4.2\,\mbox{e}2 & >4.8 & <12 &<32 &>-0.08 & abs & abs &   \\ 

36 & HD\,156424& &>2.6\,\mbox{e-}2 & <11 & >5.7\,\mbox{e}2 & >5.2 & <15 &<55 &>-0.34 & var em &  &   \\ 

37 & HD\,156324& &>1.0\,\mbox{e-}1 & <4.6 & >4.2\,\mbox{e}3 & >8.4 & <5.9 &<13 &>0.26 & var em &  &   \\ 

38 & HD\,121743&$\beta$\,Cep &>1.4\,\mbox{e-}1 & <3.7 & >3.8\,\mbox{e}2 & >4.7 & <18 &<35 &>0.10 & abs & abs & -7.2 \\ 

39 & HD\,3360&SPB &1.1\,\mbox{e-}1 & 4.4 & >2.1\,\mbox{e}2 & >4.1 & <19 &<43 &>-0.03 & abs & per & -7.8 \\ 

40 & HD\,186205\,$^\dagger$& &>9.1\,\mbox{e-}3 & <22 & >8.5\,\mbox{e}3 & >9.9 & <6.9 &<32 &>-0.36 & stab abs &  &   \\ 

41 & HD\,67621& &1.1\,\mbox{e-}1 & 4.4 & >4.9\,\mbox{e}3 & >8.6 & <22 &<49 &>0.29 & stab abs & abs &   \\ 

42 & HD\,200775\,$^\dagger$&SB2, HeBe &2.8\,\mbox{e-}1 & 2.3 & 3.3\,\mbox{e}3 & 7.9 & 3.6 &4.6 &0.53 & HeBe &  & -6.3 \\ 

43 & HD\,35912& &4.5\,\mbox{e-}1 & 1.7 & >2.8\,\mbox{e}5 & >23 & <4.1 &<3.3 &>1.13 & abs &  &   \\ 

44 & HD\,66522& &>1.8\,\mbox{e-}2 & <14 & 8.1\,\mbox{e}3 & 9.8 & 30 &<124 &>-0.18 & abs &  &   \\ 

45 & HD\,182180& &5.8\,\mbox{e-}1 & 1.4 & 2.9\,\mbox{e}6 & 41 & 4.5 &2.5 &1.46 & var em & abs &   \\ 

46 & HD\,55522& &1.1\,\mbox{e-}1 & 4.4 & >1.4\,\mbox{e}5 & >19 & <21 &<48 &>0.64 & abs &  &   \\ 

47 & HD\,142184& &5.0\,\mbox{e-}1 & 1.6 & 4.3\,\mbox{e}6 & 45 & 9.5 &6.7 &1.46 & var em &  & -6.7 \\ 

48 & HD\,58260\,$^\dagger$& &>2.7\,\mbox{e-}2 & <11 & >5.2\,\mbox{e}4 & >15 & <0.29 &<1.1 &>0.14 & abs & em &   \\ 

49 & HD\,36485\,$^\dagger$&SB2 &2.7\,\mbox{e-}1 & 2.4 & 3.5\,\mbox{e}5 & 24 & 1.6 &2.1 &1.01 & var em & var &   \\ 

50 & HD\,208057\,$^\dagger$&SPB &3.8\,\mbox{e-}1 & 1.9 & >9.6\,\mbox{e}2 & >5.9 & <18 &<18 &>0.49 & abs & abs &   \\ 

51 & HD\,306795& &>1.5\,\mbox{e-}1 & <3.6 & >1.9\,\mbox{e}5 & >21 & <2.8 &<5.5 &>0.77 & abs &  &   \\ 

52 & HD\,25558\,$^\dagger$&SPB &>5.4\,\mbox{e-}2 & <7.0 & >3.1\,\mbox{e}2 & >4.5 & <185 &<542 &>-0.19 &  &  &   \\ 

53 & HD\,35298\,$^\dagger$& &3.2\,\mbox{e-}1 & 2.1 & >1.4\,\mbox{e}6 & >34 & <1.7 &<1.9 &>1.22 & stab abs &  &   \\ 

54 & HD\,130807& &>4.4\,\mbox{e-}2 & <8.0 & >5.3\,\mbox{e}4 & >15 & <20 &<64 &>0.29 &  &  &   \\ 

55 & HD\,142990\,$^\dagger$& &2.8\,\mbox{e-}1 & 2.4 & >1.2\,\mbox{e}6 & >33 & <9.3 &<12 &>1.15 & var em & per &   \\ 

56 & HD\,37058\,$^\dagger$& &3.9\,\mbox{e-}2 & 8.6 & >6.2\,\mbox{e}4 & >16 & <3.7 &<11 &>0.27 &  &  &   \\ 

57 & HD\,35502\,$^\dagger$&SB2 &7.4\,\mbox{e-}1 & 1.2 & >7.5\,\mbox{e}5 & >29 & <2.1 &<0.61 &>1.39 & var em &  & -6.0 \\ 

58 & HD\,176582& &2.3\,\mbox{e-}1 & 2.7 & 1.7\,\mbox{e}6 & 36 & 6.7 &9.9 &1.14 & var em &  &   \\ 

59 & HD\,189775& &2.2\,\mbox{e-}1 & 2.7 & >3.8\,\mbox{e}5 & >25 & <3.8 &<5.7 &>0.96 &  &  &   \\ 

60 & HD\,61556\,$^\dagger$& &1.7\,\mbox{e-}1 & 3.3 & 1.1\,\mbox{e}6 & 32 & 14 &26 &1.00 & stab abs &  &   \\ 

61 & HD\,175362\,$^\dagger$& &1.9\,\mbox{e-}1 & 3.0 & >1.2\,\mbox{e}7 & >59 & <0.74 &<1.2 &>1.29 & stab abs & var & -6.3 \\ 

62 & HD\,105382\,$^\dagger$& &2.8\,\mbox{e-}1 & 2.3 & 7.4\,\mbox{e}4 & 16 & 13 &16 &0.86 & abs &  & -6.8 \\ 

63 & HD\,125823& &4.1\,\mbox{e-}2 & 8.4 & >1.0\,\mbox{e}4 & >10 & <14 &<47 &>0.09 & var abs & per &   \\ 

64 & HD\,36526& &1.9\,\mbox{e-}1 & 3.0 & >4.1\,\mbox{e}6 & >45 & <3.9 &<6.4 &>1.18 &  &  &   \\ 

\hline
\end{longtable}
  \end{landscape} \twocolumn }


\subsubsection{Wind momentum}
\label{sec:mv}

To compute $\etas$ and $\ra$ from equations \ref{eq:esdef} and \ref{eq:radef}, we need, in addition to the stellar radius and surface magnetic field, estimates of the wind mass-loss rate $\mdot $ and terminal speed $\vinf$. Simulation models define the confinement in terms of wind properties a star \textit{would} have if it had \textit{no} magnetic field. 
Therefore instead of making empirical estimates of the wind properties of each magnetic star 
\citep[which are in any case difficult to obtain, see][]{2012MNRAS.423L..21S,2012arXiv1207.6988G}, we derive theoretical values based on inferred stellar parameters applied to radiation line-driven wind theory.

Following standard theory, we take  the wind terminal speed $\vinf$ to scale with the star's effective surface escape speed,
\begin{equation}
	\label{eq:vesc}
	\vesc \equiv  \left(\frac{2G\ms(1-\gae)}{\rs} \right)^{1/2},
\end{equation}
where $\gae \equiv \kappa_\mathrm{e} L/4 \pi G\ms c$ is the Eddington parameter for electron scattering opacity $\kappa_\mathrm{e}$.
For the order-unity ratio $\vinf/\vesc$, we use the factors recommended by \citet{2000A&A...362..295V,2001A&A...369..574V} \citep[based on the empirical study of][]{1995ApJ...455..269L}, which declines abruptly from 2.6 to 1.3 from the hot to cool side of the so-called  ``bi-stability''  jump at $\teff \approx 25,000$\,K (see below).

For mass-loss rates, we also use the recipe given by \citet{2000A&A...362..295V,2001A&A...369..574V}, assuming solar metallicity for all stars.
This predicts an associated strong mass-loss increase of nearly an order of magnitude from the hot to cool side of this bi-stability jump,
because iron recombination makes available more efficient driving-lines and so produces an increase in the line force.
But note that, whereas the expected decrease in $\vinf$ over this bi-stability jump is empirically quite well established, this predicted increase in $\mdot$ is not yet observationally confirmed \citep[e.g.][]{2008A&A...478..823M}.

For comparison, we therefore also compute mass-loss rates based on the standard (finite-disk-corrected) \citet[][hearafter CAK]{1975ApJ...195..157C} scaling.
Using the notation from \citet{1995ApJ...454..410G},  this can be written in the form,
\begin{equation}
	\label{eq:cak_mdot}
	\dot{M}_{B=0,\mathrm{CAK}}
	=
	\frac{1}{(1+\alf)^{1/\alf}}
	\frac{\alf}{1-\alf}
	\frac{\ls}{c^2}
	\left( \frac{\qbar\gae}{1-\gae} \right)^{-1+1/\alf},
\end{equation}
where we adopt $\qbar = 1000$ and $\alf \approx 0.55$ for the full sample, to represent the normalisation and effective power-exponent of the line opacity distribution, 
where the latter has been adjusted to account for ionisation effects \citep{2000A&AS..141...23P}, and is in good agreement with the observationally inferred value for non-magnetic O-type stars \citep{2004A&A...415..349R}.

\begin{figure*}
\begin{center}
	\includegraphics[width=84mm]{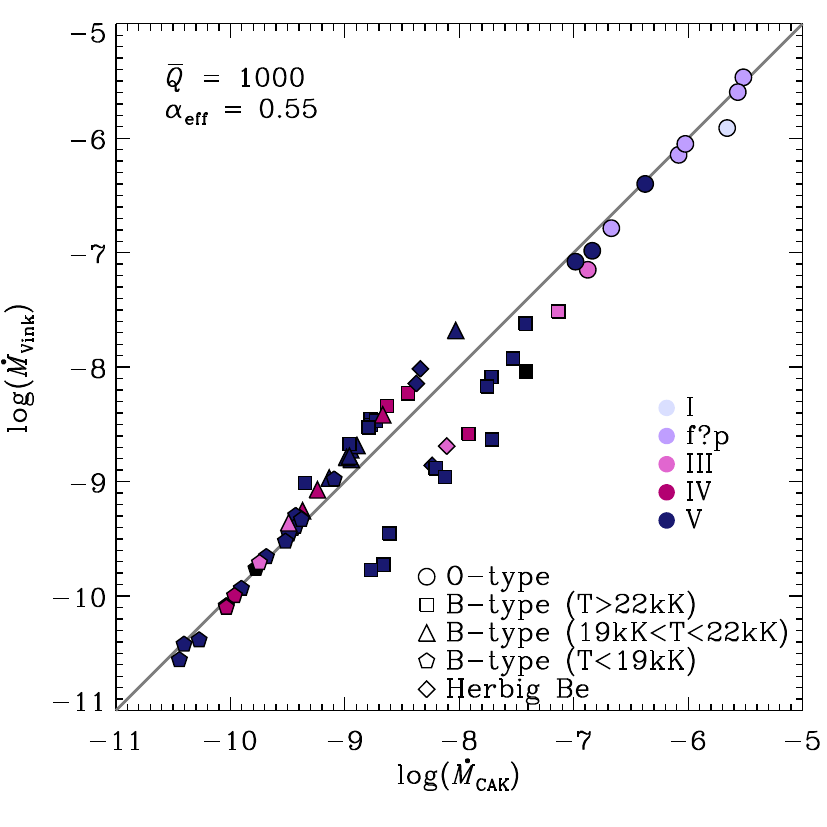}
	\includegraphics[width=84mm]{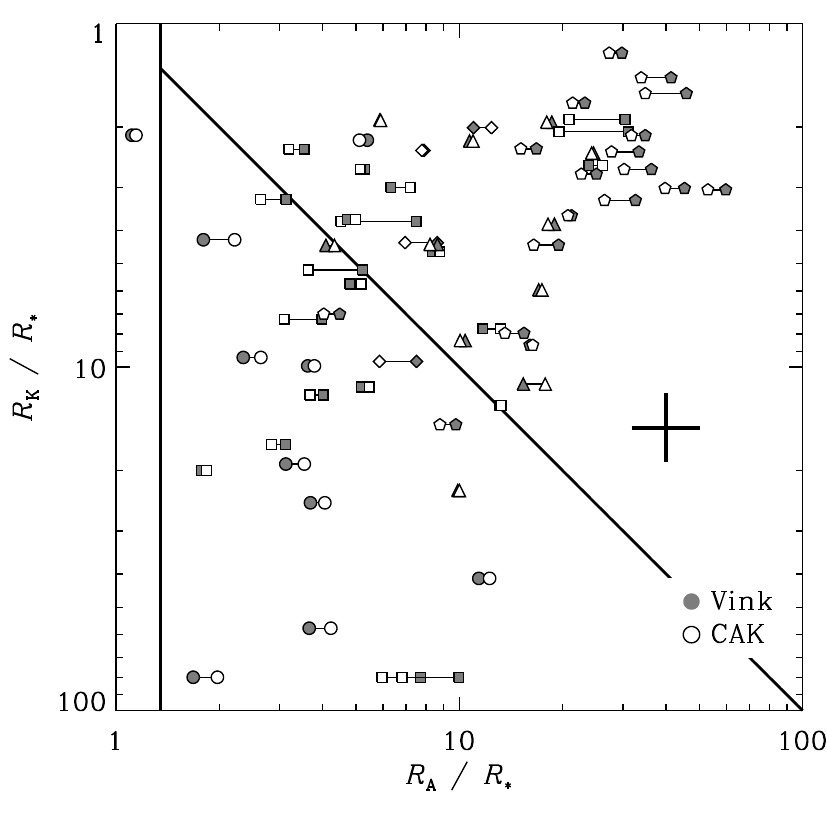}
   \caption{\label{fig:mdot} \textit{Left:} Comparison between the
mass-loss rate calculations by \citet{2000A&A...362..295V,2001A&A...369..574V} and the CAK scaling law described in \S\,\ref{sec:mv}. Note the large change in mass-loss over the bi-stability jump in the \citeauthor{2000A&A...362..295V} rates.  
\textit{Right:} Shift in Alv\'en radius from switching between the \citeauthor{2000A&A...362..295V} rates (filled symbols) and the CAK scaling law (empty symbols). 
The error bar in the lower right represents uncertainty in $\rk$ and $\ra$ estimated from the propagation of typical uncertainty in the stellar parameters \textit{only} ($\sim25$\,percent). The uncertainty in $\ra$ is in fact dominated by systematics in the mass-loss rate determinations. However, the relative position in the magnetic confinement-rotation diagram, and hence the magnetospheric classification, is not too sensitive to these systematics, as described in \S\,\ref{sec:mv}. }
\end{center}
\end{figure*}

The left panel of Figure \ref{fig:mdot}  compares the two mass-loss rate values for our full sample of magnetic massive stars, while
the right panel illustrates the shift in the confinement-rotation diagram resulting from switching between the two scalings.

For the hotter O-type stars the CAK scaling agrees quite well with the \citeauthor{2001A&A...369..574V} recipe, and, because of the weak $\ra \sim \mdot^{-1/4}$ dependence, this translates to a negligible shift in the confinement-rotation diagram. 
Of course, this comparison only reflects uncertainties due to different theoretical mass-loss descriptions. But recent multi-wavelength spectroscopic studies aiming to derive mass-loss rates in the O-type star domain typically yield rates that deviate from the \citeauthor{2001A&A...369..574V} prescription only by factors of $\sim$2-3, if small-scale wind inhomogeneities (``clumping'') are adequately accounted for \citep{2011A&A...528A..64S,2011A&A...535A..32N,2012arXiv1205.3075B} 
And as illustrated by Figure\,\ref{fig:mdot}, such discrepancies barely affect stellar positions in the confinement-rotation diagram.

For B-type stars, however, mass-loss rate differences are generally much larger, an order or magnitude or more near the bi-stability jump. Empirical mass-loss determinations for B-type dwarfs (which comprise most of our magnetic sample) are difficult at most, but studies of B-type supergiants have found a decrease in wind momentum compared to the theoretical \citeauthor{2001A&A...369..574V} predictions for the complete low-temperature region \citep{2008A&A...478..823M}. 
Further deviations from theoretical wind momentum of similar magnitude have also been observed for some late O-type stars with so-called weak winds \citep[for an overview see][]{2008A&ARv..16..209P}. 

Even with the weak $\mdot^{-1/4}$ scaling, the shift in $\ra$ associated with these large deviations can approach 0.3\,dex. 
For other quantities, such as the stellar spindown time, which scales as $\ra^2 \sim 1/\sqrt{\mdot}$, there can be a substantial change, by a factor of several,  for different mass-loss values near and below the bi-stability region, as discussed further in \S\,\ref{sec:spindown}.

In summary, these relatively large systematic differences in the adopted mass-loss rate will be a important concern for performing detailed modelling of magnetosphere signatures for individual stars.
However, it can be seen from Figure \ref{fig:mdot} (right) that despite these large differences, the overall appearance of the rotation-confinement diagram is not much affected and the basic, qualitative classification results presented here are quite robust against errors in the wind parameters.
To maintain a uniform standard, all the presented magnetosphere parameter values in Table \ref{tab:magneto} are based on the  \citeauthor{2001A&A...369..574V} scalings.

\subsubsection{Rotational oblateness}
\label{sec:oblate}

As mentioned in \S\,\ref{sec:mcp}, calculation of the Alfv\'en and Kepler radii requires the actual equatorial radius of the  star, in principle accounting for any rotationally induced oblateness. In practice, a 15 percent oblateness requires $\Omega\approx 0.8$, equivalent to $W=0.6$, and so except for the most rapid rotators, the difference between the polar and equatorial radii is generally much smaller than the uncertainty in the radius determination.
In our sample, only three stars have a period short enough for oblateness to become potentially significant.
For HD\,182180 (ID\,45) and HD\,142184 (ID\,47), we use the equatorial radii derived by \citet{RiviniusHR7355_sub} and \citet{2012MNRAS.419.1610G} from spectral analysis including the oblateness. 
As no such analysis is available for HD\,35502 (ID\,57), we use the radius derived from the Stefan-Boltzmann equation.

For simplicity, we ignore the effect of gravity darkening on the wind driving from the stellar surface.
For aligned rotators, the wind feeding the equatorial magnetosphere originates from mid-latitudes, where gravity darkening is weaker.
For non-aligned rotators, the magnetosphere will have a complex 3D structure that requires detailed modelling for each case.
But in general terms, the maximum density occurs near $\rk$ along the line defined by the \textit{intersection} between the magnetic and rotational equators \citep{2005MNRAS.357..251T,2005ApJ...630L..81T}.
In this context, the relative confinement and centrifugal support of the such magnetospheres should be well characterised by the Alfv\'{e}n and Kepler radius relative to the star's \textit{equatorial} radius, accounting for any rotational oblateness.

\subsubsection{Uncertainties  in Kepler and Alfv\'en radii}
\label{sec:error}

Let us now explore the effect of stellar parameter uncertainty on the position of the stars in the confinement-rotation diagram. 
As the radius and mass of the stars are generally derived from $\teff$, $\lum$ and $\logg$, we propagate the uncertainty on these quantities, assuming they are independent.

\begin{figure}
\begin{center}
	\includegraphics[width=84mm]{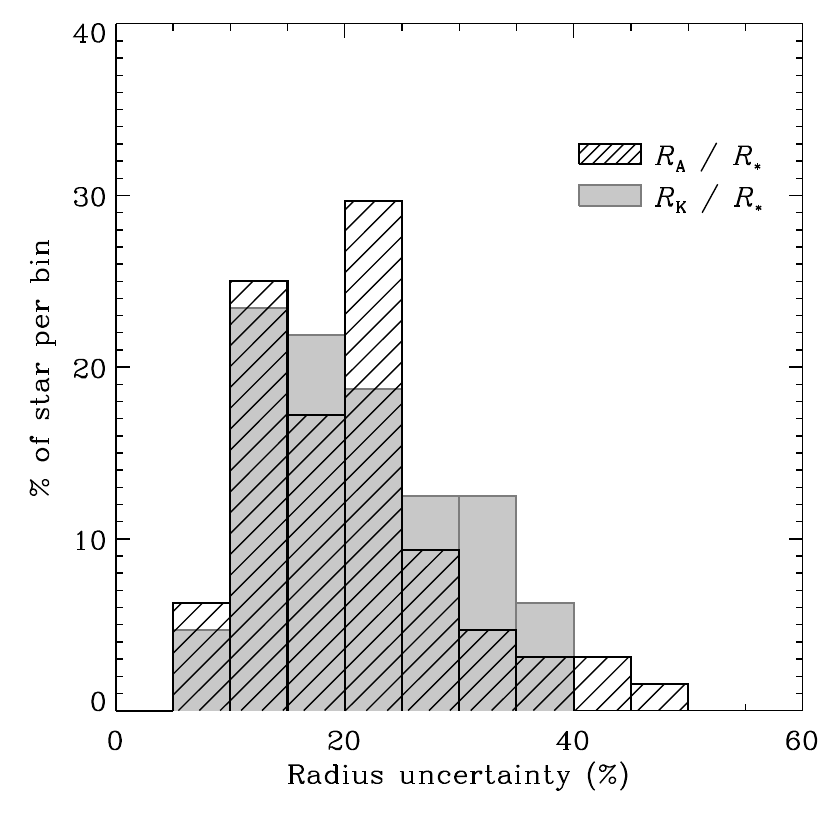}
	\caption{\label{fig:error} Distribution of uncertainties in our determination of the Kepler radius(shaded) and Alv\'en radius  (hatched), estimated by propagating the uncertainty on the stellar parameters ($\teff$, $\lum$ and $\logg$). }
\end{center}
\end{figure}

The Kepler radius can be expressed as,
\begin{equation}
	\label{eq:e_rk}
	\frac{\rk}{\rs} \propto \frac{g^{1/3}\teff^{3/2}}{L^{1/6}\omega^{2/3}}.
\end{equation}
In general, the rotational periods are accurate at 1-2 percent, so their uncertainty can be neglected. 
The quantities $\teff$, $\logg$ and $\lum$ have typical uncertainties of 10 percent, 0.2\,dex and 0.25\,dex, respectively. From equation \ref{eq:e_rk}, they contribute 15, 15 and 10 percent uncertainties to $\rk$, for a total uncertainty of 23 percent.
Figure \ref{fig:error} shows a histogram of the Kepler radius uncertainty distribution (grey shade) for our sample, confirming that the mean uncertainty is around 25 percent. A corresponding vertical error bar is shown in the confinement-rotation diagram in Figure \ref{fig:mdot} (right). 

For the error propagation in Alfv\'en radius, we follow the CAK scaling (for fixed $\alf=0.55$), for the dependence of wind momentum on stellar parameters (equations \ref{eq:vesc} and \ref{eq:cak_mdot}),

\begin{equation}
	\mdot\vinf \propto \frac{\teff^{2.24}\,\ls^{1.25}}{g^{0.31}}.
\end{equation}
With the typical uncertainties quoted above, $\teff$, $\logg$ and $\lum$ contribute respectively 22, 14 and 72 percent uncertainties to the wind momentum, for a total uncertainty of 76 percent. This uncertainty from stellar parameters is much smaller than that associated with the systematics discussed in \S\,\ref{sec:mv}. 
As such, we estimate the total uncertainty in $\ra$ through the scaling, 
\begin{equation}
	\frac{\ra}{\rs} \propto \frac{\bp^2\,\ls^{1/4}}{\teff\, (\mdot\vinf)^{1/4}}.
\end{equation}
As a large fraction of our sample has only lower limits on the dipole field strength, let us ignore for now its contribution to the uncertainty. Again with the uncertainties quoted above, $\teff$ and $\lum$ contribute to 10 and 14 percent uncertainties, whereas $\mdot\vinf$ contributes 20 percent uncertainty, for a total uncertainty in $\ra$ of 26 percent. Figure \ref{fig:error} presents the uncertainty distribution attributed to the stellar parameters for $\ra$ (hashed histogram) with a mean value again confirming the above estimate. 
One can see that the corresponding horizontal error bar in the confinement-rotation diagram of Figure \ref{fig:mdot} (right) is smaller than the systematic uncertainty associated with the theoretical mass-loss rate determination. 
 
 Moreover, since the dipolar field strength is generally constrained with an accuracy of $\sim30$ percent, it would only contribute a 15 percent  uncertainty to $\ra$, again much smaller than the systematic uncertainty from the wind momentum.

\subsection{Spindown time and spindown age}
\label{sec:spindown}

Let us next turn to considering the rotational evolution for our sample of magnetic massive stars.
The angular momentum loss rate from a magnetised wind can be written in terms of the mass-loss rate, the Alfv\'{e}n radius $\ra$, and the stellar rotation frequency $\omega = \vrot/\rs$ \citep{1967ApJ...148..217W,2009MNRAS.392.1022U},
\beqn
{\dot J} \approx \frac{2}{3} \mdot \omega \ra^2 \, .
\label{jdotdef}
\eeqn
The associated timescale for magnetic-wind-induced spindown of the stellar angular momentum $J = I \omega$ can then be written in the form
\begin{align}
\tauj \equiv \frac{J}{\dot J}  \approx \frac{3}{2} \, \frac{f \ms \rs^2 \omega}{\mdot \ra^2 \omega}
&= \frac{3}{2} \, f \tau_M  \left ( \frac{\rs}{\ra} \right )^2\, , \label{taujdef} \\
&= \tau_{J,B=0}  \left ( \frac{\rs}{\ra} \right )^2\, , \nonumber 
\end{align}
where $\tau_M \equiv \ms/\mdot$ is a characteristic mass-loss timescale,
and $\tau_{J,B=0}$ defines the spindown time in the case of no magnetic field (i.e. $\ra=\rs$).
The star's moment of inertia $I= f \ms \rs^2$ can be evaluated from the radius of gyration $\beta=f^{1/2}$ tabulated from internal structure models such as \citet{2004A&A...424..919C}.
If we assume for simplicity a fixed radius $\rs$ and moment of inertia factor $f \approx 0.1$, as well as a constant angular momentum loss rate ${\dot J}$, 
then the stellar rotational period $P$ will simply increase exponentially with age $t$ from its initial value $P_o$,
\beqn
P(t) = P_o e^{t/\tauj} \, .
\label{eq:poft}
\eeqn
We can then use equation \ref{eq:poft} to define a star's spindown \textit{age}, $t_s$, in terms of the spindown time $\tauj$, and its inferred present-day critical rotation fraction $W = P_\mathrm{orb}/P$ relative to its initial rotation fraction $W_o$ at age $t=0$,
\beqn
\frac{\ts}{\tauj} = \ln{W_o} - \ln{W}
\, .
\label{tstauj}
\eeqn
Taking the initial rotation to be critical, $W_o =1$, yields a simple upper limit to the spindown age,
\beqn
\tsmax = \tauj ~  \ln (1/W)
\, .
\label{tsmax}
\eeqn
If the initial rotation is subcritical, $W_o < 1$, then the actual spindown age is shorter by a time $\Delta \ts =  \tauj \, \ln W_o$.

As noted previously, the extra axes in Figure \ref{fig:etaW} give the spindown timescale $\tauj$ normalised by the value in a non-magnetised wind (i.e. by how much the magnetic braking enhances the stellar spindown) along the top, and the maximum spindown age $\tsmax$ normalised by the spindown time (i.e. the number of spindown e-folds) along the right.
For each of the individual magnetic OB stars, columns (8) and (9) of Table \ref{tab:magneto} also list estimated values  for respectively 
the spindown time $\tauj$ and the maximum spindown age $\tsmax$ in Myr.
Future studies can thereby compare $\tsmax$ with other indicators of stellar age, for example from stellar evolution tracks or cluster association. To the extent that such independent age estimates are available, then within the limits of the stated assumptions of constancy in $\rs$, $f$, and ${\dot J}$, a comparison with this spindown age could be used to estimate an initial rotation fraction $W_o$.

More immediately, note that among the full magnetic sample, many of the most slowly rotating stars are O-type stars.
The high luminosities of these stars drive strong stellar winds that lead to a rapid angular momentum mass loss and thus very short spindown times.
These characteristics help to explain their very slow rotation relative to many of the B-type targets. 
Except for Plaskett's star (ID\,6), which has likely been spun up by binary interaction and show CM-type emission at high velocity  \citep{Grunhut2012_plaskett}, all the rapidly rotating stars near the top of Figure \ref{fig:etaW} are lower luminosity B-type stars with weaker winds;  for magnetic B-type stars, the spindown time is thus generally longer than for the magnetic O-type stars, typically several Myrs.

Indeed,  extended photometric monitoring of the strongly magnetic B-type star $\sigma$\,Ori\,E (ID\,31) has provided a direct measurement of the change in rotation period, yielding  a spindown time of 1.34\,Myr \citep{2010ApJ...714L.318T,Townsend2012_MOST}. This is remarkably close to the spindown time of 1.4\,Myr   \textit{predicted} previously by the scaling developed from MHD simulations \citep{2009MNRAS.392.1022U}, but such very close agreement was likely fortuitous given the uncertainties in the mass-loss rate and stellar parameters. Indeed, the \citet{2000A&A...362..295V,2001A&A...369..574V} mass-loss rate we use here is roughly a factor 10 smaller than the CAK mass-loss rate assumed by \citet{2009MNRAS.392.1022U}, leading to a factor $\sim \sqrt{10}$ longer estimate for the spindown time, 4.6\,Myr. This emphasises that our listed values spindown time and age are only estimates accurate to within a factor of 3 or so\footnote{A change of period has also been measured for HD\,37776 \citep[ID\,34,][]{2011A&A...534L...5M} with a spindown timescale of 0.37\,Myr, comparable to our $\tauj$ of 0.68\,Myr. However, as discussed by \citet{2009MNRAS.392.1022U}, the complex field geometry of HD\,37776 \citep{2011ApJ...726...24K} will have a potentially strong impact on the angular momentum evolution of this star.}.

\section{H$\alpha$ as a magnetospheric proxy}
\label{sec:disc}

We now explore how magnetospheric H$\alpha$ emission characteristics correlate with their position in the magnetic confinement-rotation diagram. 

\subsection{Identification of H$\alpha$ magnetospheric emission}
\label{sec:ha_obs}

In Table \ref{tab:magneto}, column (11) indicates the emission (em) versus absorption (abs) nature of the H$\alpha$ line, and, if enough observations ($\ga 5$) are available, whether the profile is stable (stab) or variable (var).
We flag Herbig Be stars (HeBe) because of the difficulty in disentangling magnetospheric emission from the emission produced by the accretion disks characteristic to this class of pre-main sequence stars. 
Similarly in the case of spectroscopic binaries, slowly pulsating B-stars and $\beta$\,Cep stars (indicated in column 3), variation in the absorption line profile may have non-magnetic origins, but emission can generally be attributed to a magnetosphere. 

\begin{figure*}
\begin{center}
	\includegraphics[width=84mm]{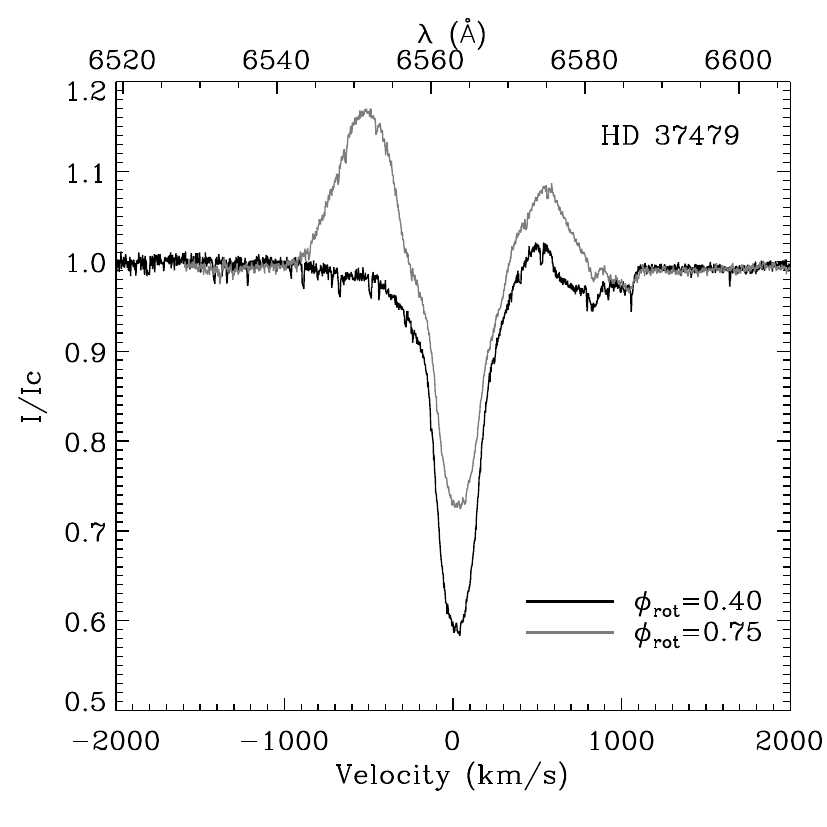}
	\includegraphics[width=84mm]{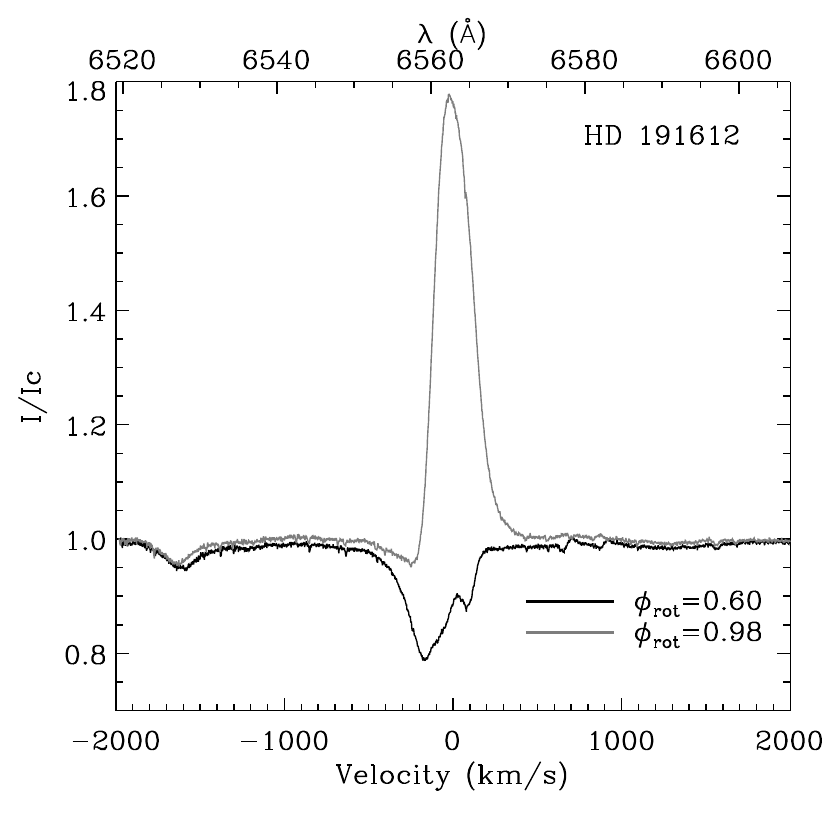}
	\caption{\label{fig:pro_ha} Example of H$\alpha$ line profiles, from MiMeS observations, for a centrifugal magnetosphere (left, \sigorie, ID\,31) and a dynamical magnetosphere (right, HD\,191612, ID\,4) and minimum (black) and maximum (grey) emission. Note how the extended emission wings are located at large velocities (outside $\vsini$) for the CM, whereas the central emission is localised in a narrow range of velocities (inside $\vsini$) in the line core for the DM.}
\end{center}
\end{figure*}

This emission can have various distinguishing characteristics: (1) a central absorption core with broad emission wings that extend well beyond the photospheric $\vsini$ (Fig. \ref{fig:pro_ha}, left), (2) strong, narrow emission that overwhelms the photospheric absorption profile (Fig. \ref{fig:pro_ha}, right, grey profile), (3) weak overlying emission that only partially fills the underlying absorption (Fig. \ref{fig:pro_ha}, right, black profile). 

For type (3), there can be confusion with the line-filling effect of a non-magnetised stellar wind; a clear identification requires monitoring for rotational modulation\footnote{Modulated variations in the core of absorption lines could also have other origins, for example changes in the photospheric structure due to large helium abundance inhomogeities on the surface of chemically peculiar stars  \citep[e.g. a\,Cen; ID\,63,][]{2010A&A...520A..44B} }. 
For type (1) and (2), a single observation can suffice to identify a magnetospheric origin. 

For type (1) the extended emission wings suggest plasma held in extended rigid-body rotation around the star, presumably by the stellar magnetic field. These correspond to centrifugal magnetospheres, as described by \citet{2005MNRAS.357..251T}. Multiple occurrences of this type of emission can be found in \citet{2011AJ....141..169B}, \citet{2012MNRAS.419..959O}, \citet{Grunhut2012_plaskett}, as well as the references listed in Table \ref{tab:ref}.

For type (2), the narrow central emission suggests that the trapped plasma is kept at low velocities, without much broadening from rotation or from a high-speed outflow like in a non-magnetic stellar wind \citep{2012MNRAS.423L..21S}. 
These correspond to dynamical magnetospheres. Example of such emission can be found in \citet{2007MNRAS.381..433H}, \citet{2012arXiv1207.6988G}, \citet{2012MNRAS.425.1278W}, as well as the references listed in Table \ref{tab:ref}.

\subsection{H$\alpha$ emission in the classification diagram}
\label{sec:ha_rark}

\begin{figure*}
\begin{center}

	\includegraphics[width=84mm]{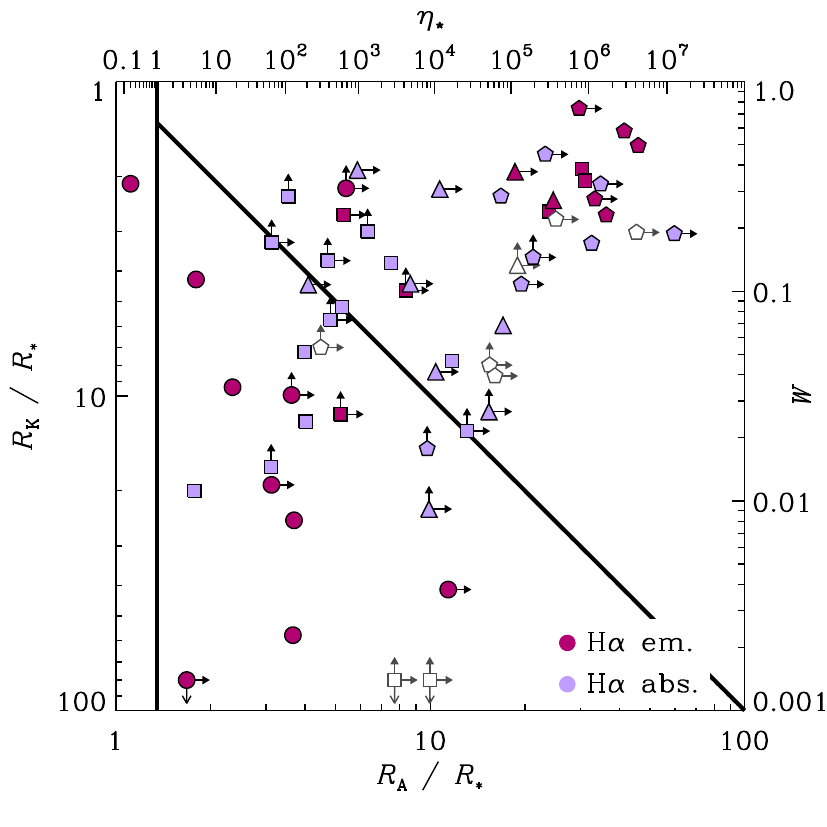}
	\includegraphics[width=84mm]{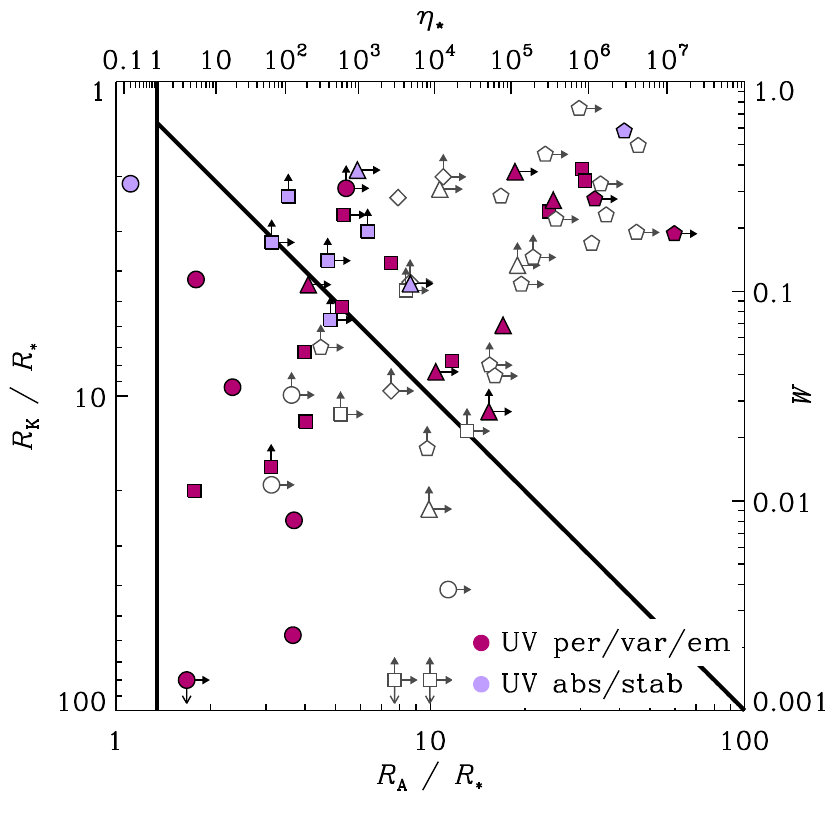}
	\caption{\label{fig:rark_ha} Location of magnetic massive stars in the magnetic confinement-rotation diagram (see Figure \ref{fig:etaW}). 
	The symbols are coloured to mark the presence (dark pink) or absence (light purple) of a magnetospheric signature in H$\alpha$ emission (left) and UV resonance line (right), as described in Table \ref{tab:magneto}, and in \S \ref{sec:ha_obs} and \S \ref{sec:uv} respectively. The symbols are empty when no information is available. 
 }

\end{center}
\end{figure*}

Figure \ref{fig:rark_ha} (left) again plots stars in the confinement-rotation diagram, 
with symbols now coloured to mark the presence (dark pink) or absence (light purple) of magnetospheric H$\alpha$ emission. Herbig stars are omitted here because of their intrinsic emission not associated with magnetic fields. 
While stars with and without emission are found throughout the diagram, note that in the DM region with $\ra < \rk$, all the emission occurs, (with just one exception, HD\,156424; ID\,36) in O-type stars, for which the large luminosity suggests the wind feeding of the DM is strong enough to build up sufficient density for emission within the dynamical infall timescale.
The same strong wind mass loss that feeds the DM emission means that they have relatively strong angular momentum loss that spins down the stars to their observed slow rotation rates near the bottom of the confinement-rotation diagram. 

Conversely, in the CM region with $\ra > \rk$, the emission (again with one exception, Plaskett's star, ID\,6) occurs in B-type stars, for which the lower luminosity and wind feeding requires the longer retention timescale of a CM to build up sufficient density for emission.
In fact, all the non-emitting stars in this region are also B-type, indicating that a CM is a necessary but not sufficient condition for emission for such low luminosity stars with relatively weak wind mass loss. Most of the B-type stars with emission are in the extreme upper right of the diagram, with both strong confinement and rapid rotation. Their wide separation from the $\ra=\rk$  line implies a large radial extent for their CM.

Overall, this link between H$\alpha$ emission and location in the magnetic confinement-rotation diagram provides a useful categorisation that connects the rotation, mass loss, and circumstellar emission properties of massive star magnetospheres.

\subsection{Magnetospheric vs stellar properties}
\label{sec:ha_param}

\begin{figure*}
	\begin{center}
	\includegraphics[width=84mm]{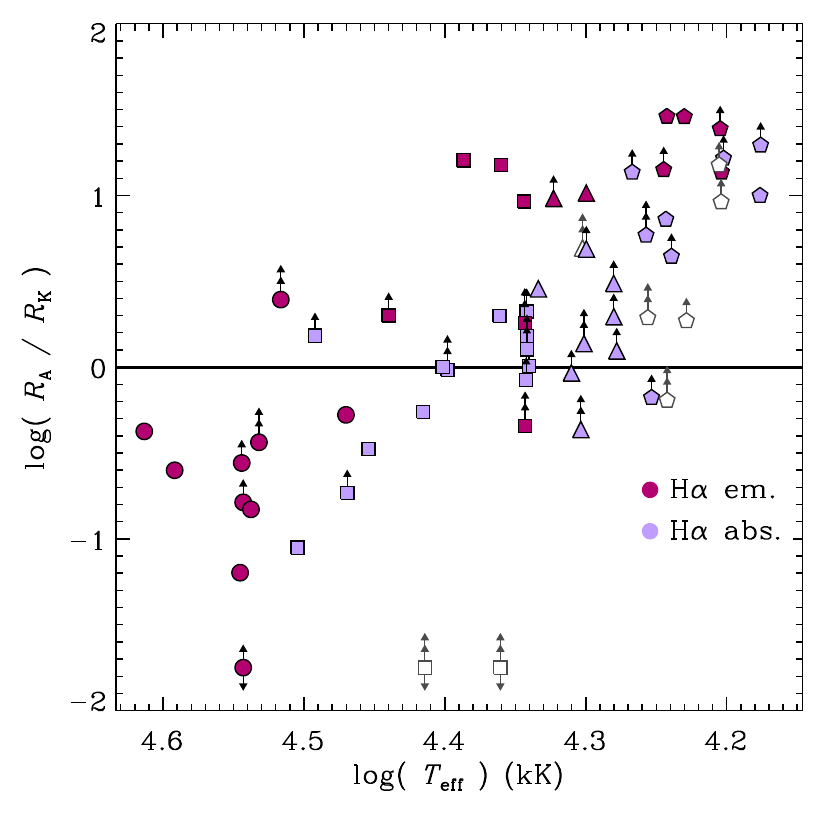}
	\includegraphics[width=84mm]{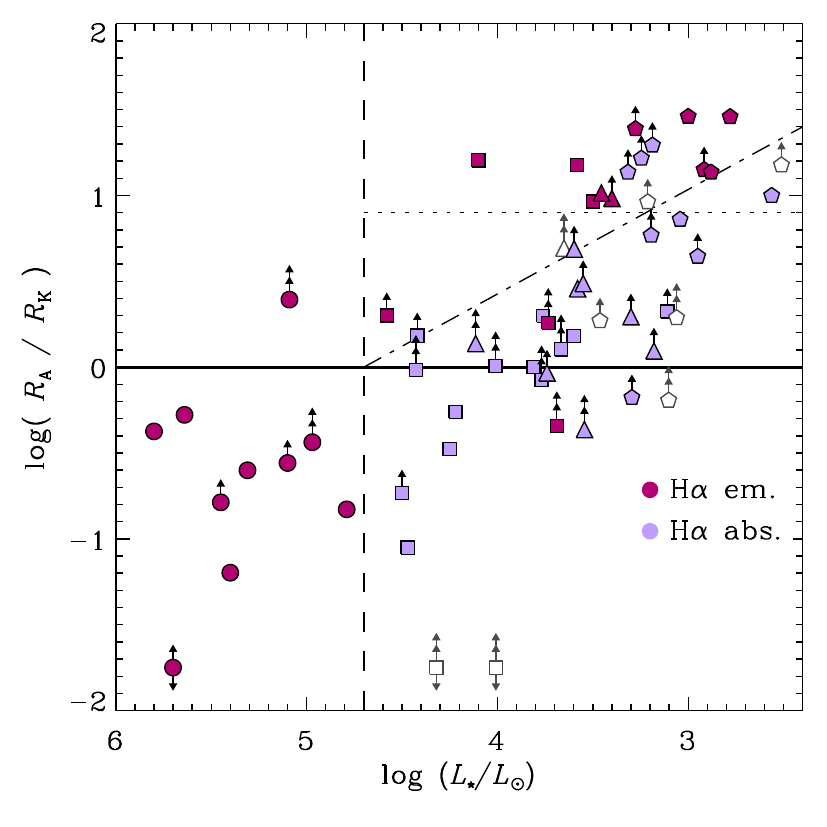}
	\caption{\label{fig:par_ha}
	Location of magnetic massive stars in a $\log$-$\log$ plot of $\ra/\rk$ vs. the effective temperature (left) and the luminosity (right). The symbols are coloured to mark the presence (dark pink) or absence (light purple) of magnetospheric H$\alpha$ emission, as described in Table \ref{tab:magneto} and \S\ref{sec:ha_obs}. The symbols are empty when no H$\alpha$ information is available. Single arrows indicate a limit on either $\ra$ or $\rk$, whereas double arrows mark stars for which both $\ra$ and $\rk$ are limits. 
	In the righthand diagram, the vertical dashed line represents the luminosity transition between O-type and B-type main sequence stars. The horizontal dotted line and the diagonal dot-dashed line are illustrative division of the CM domain according to potential mass leakage mechanisms (see discussion in \S\ref{sec:ha_param}). }
	\end{center}
\end{figure*}

\begin{figure*}
	\begin{center}
	\includegraphics[width=84mm]{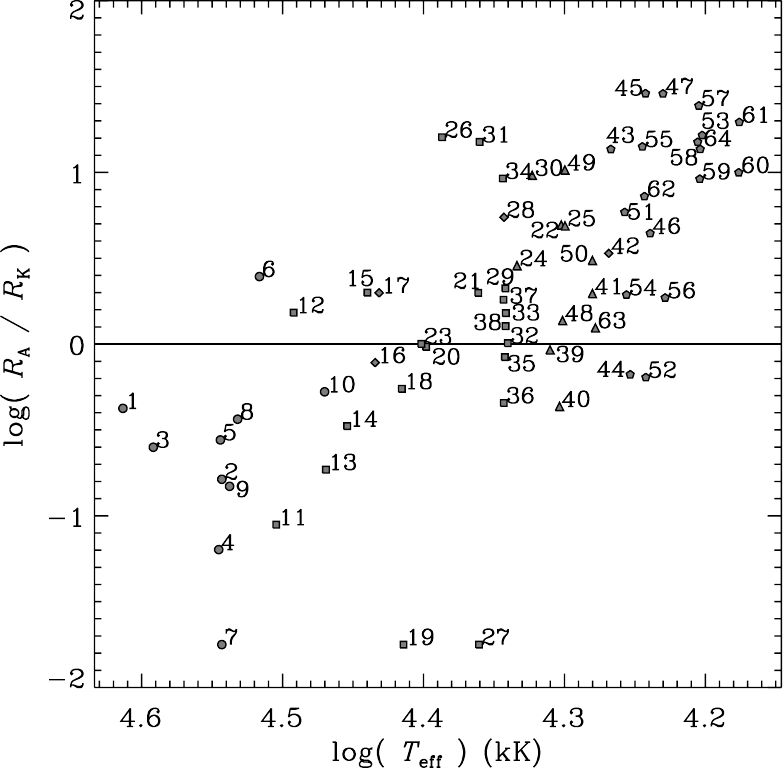}
	\includegraphics[width=84mm]{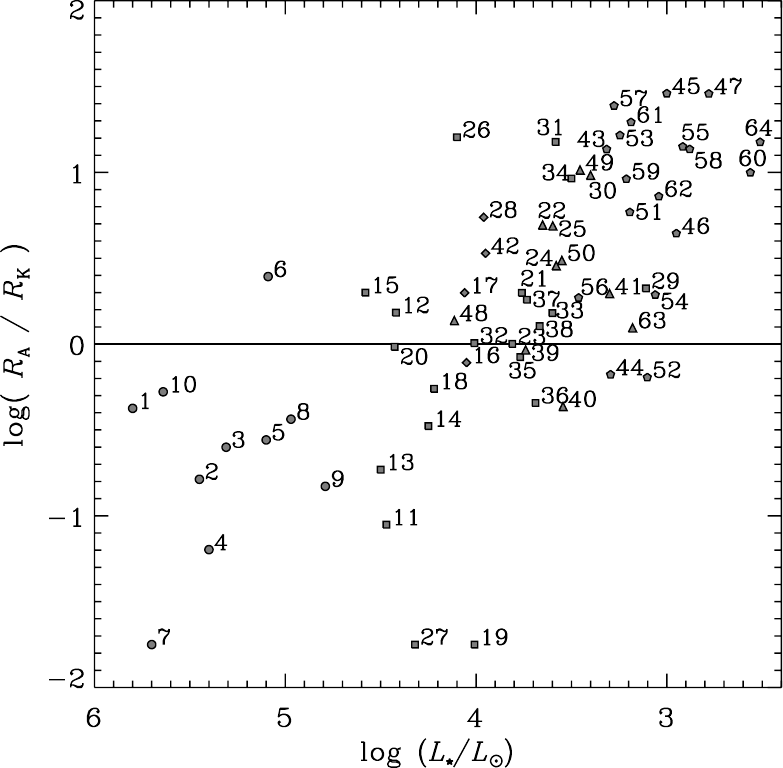}
	\caption{\label{fig:par_chart}
	Finding charts for the location of magnetic massive stars in a $\log$-$\log$ plot of $\ra/\rk$ vs. the effective temperature (left) and the luminosity (right). The label numbers correspond to the ID in column (1) of Table \ref{tab:stars}.}
	\end{center}
\end{figure*}

To explore further this categorisation, Figure \ref{fig:par_ha} plots (again for all the non-Herbig stars) the log of  the ratio $\ra/\rk$ (column 10 of Table \ref{tab:magneto}) vs. stellar effective temperature $\teff$ (left) or bolometric luminosity $\ls$ (right), with symbols again marking spectral type, coloured for the presence (dark pink) or absence (light purple) of magnetospheric H$\alpha$ emission (Figure \ref{fig:par_chart} provides a finding chart). 
The single upward arrows indicate stars that could be shifted upward due to either a higher polar field than the minimum inferred from the available longitudinal field measurements (increasing $\ra$), or a higher rotation rate than the minimum inferred from the measured $\vsini$ (decreasing $\rk$); 
the {\em double} upward arrows indicate stars for which {\em both} limits are at play.

The solid horizontal line at $\ra = \rk$ separates the domains for dynamical magnetospheres at the bottom from centrifugal magnetospheres at the top, with the distance above the line characterising the radial extent for the centrifugal support.
Each plot again shows that H$\alpha$ emission occurs both in O-type stars to the left, and in B-type stars to the (mostly upper) right;  but the demarcation is particularly distinct in the plot vs. bolometric luminosity.

 In that plot, the vertical dashed line corresponds roughly to the main sequence transition from O- to B-type stars \citep{2005A&A...436.1049M}.
The O-type stars to the left all have clear Balmer emission. Except for Plaskett's star (ID\,6), which has likely been spun up by binary interaction, they also are all relatively slow rotators with $\ra < \rk$ (DM).
Their high luminosity means they have strong stellar winds, implying both a rapid stellar spindown, as well as a sufficient magnetospheric density to give line emission in the short residence time for a dynamical magnetosphere.

For B-type stars to the right of this vertical  line, emission is most common in the most rapidly rotating stars above the horizontal dotted line at $\log (\ra/\rk ) = 0.9$.
The 4 stars above this line without detected emission are all relatively late type stars, with low luminosity and so likely a very low wind mass-loss rate to feed the expected CM.
The 3 stars {\em below} this line {\em with} detected emission have arrows indicating they could shift upward, with the two lower luminosity stars (HD\,156424 and HD\,156324; ID\,36, 37) having double arrows indicating potentially significant revision in both field strength and rotation rate.
Indeed, although the current MiMeS observations do not allow for clear determination of a rotation period, in both cases the nightly variation of longitudinal field measurements points toward periods of order a day. These stars are prime candidates for follow up observations.

The third, relatively high-luminosity B-type star ($\xi^1$\,CMa; ID\,15) has a well determined period 
and Kepler radius,
 but still has a single arrow from 
the limited polar field estimate.
Its current position --  just above the horizontal solid line,  and just to the right of the vertical dashed line -- makes it a particularly interesting test case for magnetospheric models, very near the transition from DM to CM, and from O-type stars to B-type stars mass loss.

Indeed, if the CM/B-type star occurrence of H$\alpha$ emission depends on a {\em combination} of the radial extent of the CM (set by $\ra/\rk$ and thus the vertical position in Figure \ref{fig:par_ha}) and on the mass-loss rate feeding the CM (set by the luminosity and thus the horizontal position in Figure \ref{fig:par_ha}), then we can identify a possible further division along the illustrative {\em diagonal} dot-dashed line in Figure \ref{fig:par_ha}. 
It would thus be of particular interest to clarify the position, and the emission properties, of stars  with current placements near this illustrative diagonal line. 

Establishing empirically whether the onset of emission is better delineated by the horizontal dotted line or the diagonal dot-dashed line has potentially important implications for our theoretical understanding of the magnetospheric mass budget.
The former would indicate that the CM mass depends mainly on the {\em capacity} for the magnetic field to confine centrifugally supported material, which eventually fills to a fixed level even if at the slow rate from a weak wind.
The latter would indicate a competing {\em leakage} from the CM that decreases with distance above the $\ra=\rk$ line.
To build up sufficient density for emission, stars near the $\ra=\rk$ line with high leakage require a high mass-loss rate and thus high luminosity, representing the left end of the diagonal; stars further above the line with lower leakage can fill their CM even with the weaker wind from a lower luminosity, representing the upper right end of the diagonal.

As a shorthand, we might identify these as the `capacity' vs. `leakage' models for determining the onset of  CM emission.  Hopefully, the classification and physical arguments here will motivate a concentrated observational program to clarify the position, and the emission properties, of the key stars for establishing this discrimination.

\section{Other magnetospheric proxies}
\label{sec:other}

\subsection{Ultraviolet variability}
\label{sec:uv}

In hot, massive stars, strong UV resonance lines like C\,\textsc{iv}\,$\lambda\lambda$\,1548, 1550, Si\,\textsc{iv}\,$\lambda\lambda$\,1393, 1403 and N\,\textsc{v}\,$\lambda\lambda$\,1239, 1243 are typically used as diagnostics of the stellar wind. In O-type stars with dense winds, the line profiles generally exhibit a characteristic P-Cygni profile showing red-side emission and blue-side absorption, with the blue edge of the latter marking the wind terminal speed.
In B-type stars with weaker winds, the emission is weak or absent, and the blue edge of the shallower absorption may not extend to the terminal speed.
Both types can exhibit intrinsic variability, but this is most common, distinctive and well-studied in O-type stars, 
for which it is generally
characterised by discrete absorption components (DACs) that start near line-centre and 
slowly propagate across the blue absorption trough \citep[e.g.][]{1989ApJS...69..527H,1996A&AS..116..257K}. 
These are likely representations of spiral-shaped density compressions, referred to as Corotating Interacting Regions
\citep[CIRs,][]{1986A&A...165..157M}, caused by faster moving streams overtaking slower moving streams, where the interacting interface between the two is shocked. The projected velocity in the line of sight progresses because of the stellar rotation, rather than because of the outflow itself \citep{1996ApJ...462..469C}

In magnetic OB stars these UV lines can be strongly affected by changes in velocity, density, and/or ionisation balance.
Indeed, it was recognised early on that \textit{periodic} variations of the resonance lines could point toward the presence of a rotating magnetosphere \cite[e.g.][]{1990ApJ...365..665S,1993npsp.conf..186H,1994ApJ...425L..29W}.
In contrast, the appearance of DACs is often found to be cyclical but never strictly periodic.
Moreover, unlike the blueward-propagating DACs, in magnetic OB stars  UV line variation occurs nearly coherently and  synchronously over the full velocity range of the profile \citep[e.g.][]{2012MNRAS.422.2314M,henrichs_siglupi}.
Thus, even in those stars without sufficient monitoring to clearly establish a period, one could use the morphological character of variations between two or more observations to flag the likely presence of a strong field \citep[e.g.][]{henrichs_siglupi}.

In B-type stars, UV profiles show only shallow (if any) blueward wind absorption and weak or absent redward emission. The appearance of strong redward emission along with filling in of the absorption \cite[see Figure 5 of][]{1990ApJ...365..665S} can likewise be used to flag the likely presence of a magnetosphere, even without multiple observations to show variability.

Column (12) of Table \ref{tab:magneto} gives a summary characterisation of UV resonance lines for the full sample of magnetic stars.
The listed UV signatures of a field include periodicity (per), profile variability with morphology similar to periodic stars (var), and, for B-type stars, distinct redward emission with missing blueward absorption (em).
Stars lacking a clear UV magnetic signature are those with only pure absorption (abs), and those with 5 or more observations showing stable absorption (stab abs).
When available, these characterisations are from the literature, as summarised in Table \ref{tab:ref}, and otherwise are based on visual inspection of IUE archive spectra.

In analogy with the organisation of H$\alpha$ signatures shown in the left panel of Figure \ref{fig:rark_ha}, the right panel again plots stars in the magnetic confinement-rotation diagram, 
but now with stars showing one or more UV signatures for a field marked in dark pink, and those with absorption profiles consistent with lack of a field marked in light purple.
Stars without UV observations (or IUE spectra with too low signal-to-noise ratio) have empty symbols. 

Note that, in contrast to H$\alpha$ emission, such UV magnetic signatures occur throughout the diagram,  and for all spectral types.  In particular,  B-type stars with weak winds  show a UV magnetic signature even in the slow-rotation, DM region, for which B-type star H$\alpha$ emission is not seen.
Thus UV variation seems to be a wide-spread phenomenon among magnetic OB stars, as long as some confinement is present, and therefore represents a relatively robust proxy of magnetism. In fact, a number of magnetic OB stars had been first identified as peculiar UV stars \citep[e.g.][]{1993npsp.conf..186H,2003A&A...406.1019N,henrichs_siglupi}. 

The few stars without signs of UV variability cluster at lower $\ra$, but many other stars show UV field signatures in the same region of the diagram. As suggested by \citet{1990ApJ...365..665S} for some of the Bp stars, the exact behaviour of the variability might be closely tied with the geometry of the magnetic field with respect to the observer. Thus detailed modelling of the UV line profiles for magnetic OB stars may help constrain the geometry of the magnetospheres and clarify the velocity and ionisation structure of the trapped material.

Indeed, UV resonance-line synthesis from MHD models shows clear P-Cygni absorption troughs that are modulated on the rotation phase. 
For relatively strong lines, such troughs are actually deeper when viewing the magnetosphere pole-on than equator-on \citep{2008cihw.conf..125U}. 
This somewhat counterintuitive effect occurs because the overdense material around the magnetic equator is characterised by very low velocities, whereas the outflow above the pole more closely resembles that of a normal, non-magnetic wind. 
Thus the absorption column above the pole covers a much wider velocity range, leading to wider and deeper troughs.

However, further calculations also suggest that the phase variability of such UV lines is quite sensitive to the actual strength of the line itself, and so may depend critically on the stellar mass-loss rate as well as on the magnetospheric ionisation state (Sundqvist et al. in prep). 
The strong UV lines in  HD\,108 (ID\,7) indeed seem to display the characteristic variability described above \citep{2012MNRAS.422.2314M}, but 
those in \oriunc\ (ID\,3) show effectively the opposite behaviour \citep{2008cihw.conf..125U}.
Thus further modelling work is still needed to fully understand how magnetic fields affect the formation of UV lines of OB stars.

\subsection{X-rays}
\label{sec:xray}

Massive stars are generally X-ray bright due to the intrinsic instability of the line-driving mechanism for radiative stellar winds \citep{1997A&A...322..878F,2002A&A...381.1015R,2003A&A...406L...1D}, with a well known canonical value for early OB star X-ray luminosity, $\lx,$ of $\sim10^{-7.2}\ls$ \citep{1997A&A...322..167B,2011ApJS..194....7N,2011ApJS..194....5G}. 
The magnetically channeled wind shocks (MCWS scenario) associated with magnetic massive stars should also generate even stronger and harder X-ray emission, by the radiative cooling of the shock heated plasma in the magnetosphere \citep{1997ApJ...485L..29B}. 
For example, the X-rays from the O-type star \oriunc\ (ID\,3) are more luminous and harder than in typical O-type stars, and modulated by the rotational period. \citet{2005ApJ...628..986G} used 2D MHD simulations, including an explicit energy equation, to track the shock heated material and its radiative cooling, and were able to reproduce the X-ray properties of \oriunc,  including the star's elevated X-ray luminosity, high plasma temperature, rotational modulation, and narrow spectral lines. Therefore, it seems at first glance that luminous, hard and variable X-ray emission could be a proxy for magnetism in massive stars. 

However, these characteristics are not always present in magnetic massive stars.
For example, the B-type star $\tau$\,Sco (ID\,11) is X-ray luminous and indeed displays a hard X-ray spectrum \citep{2003A&A...398..203M}, but it does not show evidence of rotational modulation \citep{2010ApJ...721.1412I}. 
The B-type star NU\,Ori (ID\,12) does not show any significant variability over the duration of a $\sim10$\,d \textit{Chandra} observation \citep{2005ApJS..160..557S}, and has a soft spectrum.
Another prototypical magnetic O-type star is the Of?p star HD\,191612 (ID\,4) which is quite luminous, but has a rather soft spectrum \citep{2010A&A...520A..59N}.  \citet{2011MNRAS.416.1456O} recently examined a small subset of magnetic B stars and noted that they too have diverse X-ray properties, including a few that are not X-ray overluminous at all.

We present here a first attempt to cast the X-ray characteristics of our large sample of magnetic OB stars as a function of their magnetospheric properties, focusing on just the X-ray luminosity.
We perform a review of the literature to extract X-ray fluxes for the stars in our sample. Where possible, we use X-ray fluxes derived from pointed observations by modern X-ray observatories (\textit{Chandra} and \textit{XMM}) and reported in papers that carefully model the emission properties, correcting for interstellar absorption. A large majority of the O-type and very early B-type stars in our sample fall into this category. And for these stars, differences in the instrument bandpasses and uncertainties associated with the multi-temperature emission modelling and the ISM correction should lead to errors in the reported X-ray fluxes of less than a factor of two. We correct all of the literature X-ray luminosities for the distances adopted by the authors of each paper to derive an X-ray flux, and then recompute the X-ray luminosity using the distances we adopt for each star, which of course are consistent with the distances we use for the bolometric luminosity determinations. We then compute the X-ray efficiency ratio $\lxls$ (column 13 of Table \ref{tab:magneto}), so that even if better distance determinations are made for some of these objects in the future, their $\lxls$ values will not have to be adjusted.

For many of the later B-type stars, no X-ray measurements exist, and for others only survey data, primarily from \textit{ROSAT}, exists. The X-ray fluxes derived for these stars are more uncertain, primarily because of  the lack of detailed spectral modelling and in some cases the lack of detailed ISM absorption corrections. Additionally, the bandpass of \textit{ROSAT} is softer than that of either \textit{XMM} or \textit{Chandra}, further skewing comparisons between the derived X-ray luminosities. There are similar considerations for the small number of B-type stars for which only \textit{EINSTEIN} measurements exist \citep{1992ApJS...81..795G}. A more conservative estimate of the $\lxls$ uncertainties for these stars is required, with the overall error being probably up to 0.5 dex. Another factor potentially affecting our X-ray luminosity determinations is the contribution from unresolved binary companions \citep[e.g.][]{2012arXiv1205.3538P}. This phenomenon is more likely to be important for later B-type stars, observed with X-ray telescopes with poorer spatial resolution, and with lower intrinsic X-ray luminosities such that low-mass PMS companions could account for much of the observed X-ray emission for a given star.  However it is unlikely that all the X-ray bright magnetic B-type stars are affected by binarity. For example, \citet{2011ApJS..194....5G} have shown that the pre-main sequence population of the Carina Complex cannot explain all the X-ray emission of B-type stars and that some of them must be intrinsically X-ray bright. 

\begin{figure*}
	\begin{center}
	\includegraphics[width=84mm]{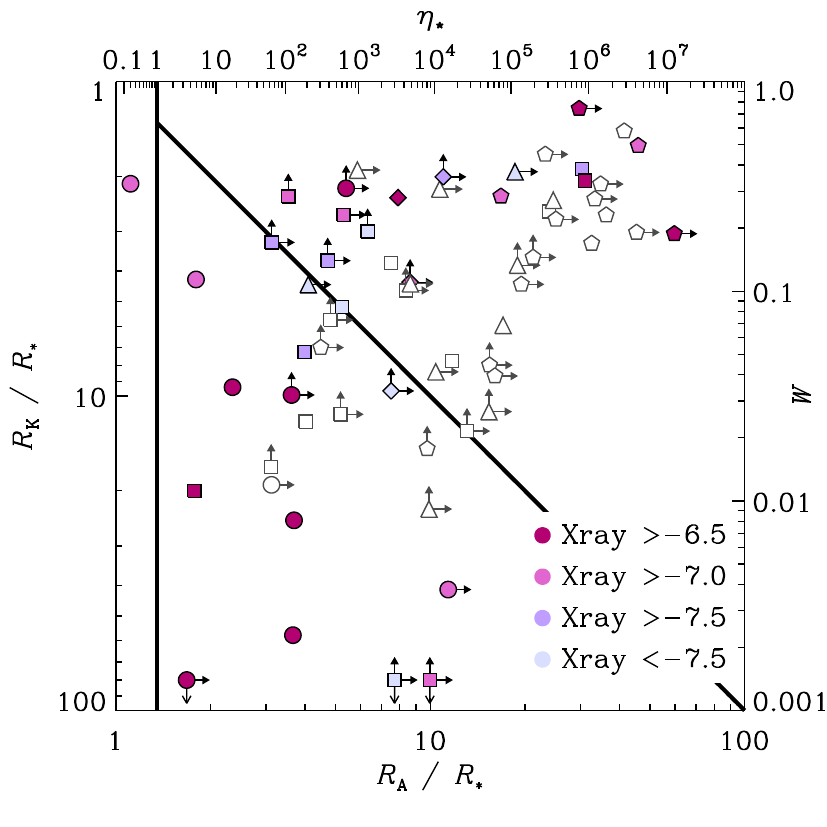}
	\includegraphics[width=84mm]{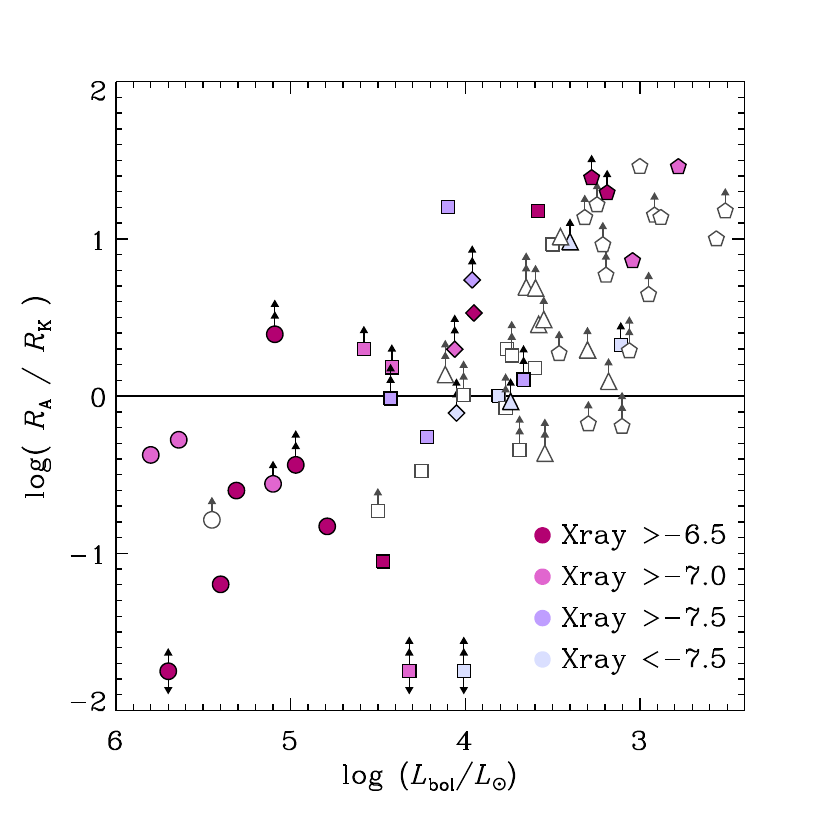}
	\caption{\label{fig:xrays}
	Location of magnetic massive stars in the magnetic confinement-rotation diagram (left) and in a log-log plot of $\ra/\rk$ vs. the luminosity (right). The stars are colour-coded according to their X-ray luminosity with respect to their bolometric luminosity ($\lxls$), in bins of 0.5\,dex. The dark pink shades represent stars with $\lxls$ greater than the canonical value of $-7$ for early OB stars and the light purple shades for stars below this value. The symbols are empty when no X-ray information is available.  }
	\end{center}
\end{figure*}

In Figure \ref{fig:xrays} (left), we plot the stars in the $\ra$-$\rk$ plane with a colour coding representing the X-ray efficiency ratio in bins of 0.5\,dex ([$>$$-6.5$], [$-6.5$,$-7.0$], [$-7.0$,$-7.5$] and [$<$$-7.5$]).
The dark pink shades are for stars with X-ray efficiency greater than the canonical value of $\lxls=-7.0$ for O-type stars. 
All the O-type stars show some level of overluminosity ($\lxls>-6.7$). 
Some of the B-type stars also show overluminosity. Most of them are located in the upper part of the CM region, although a few overluminous, very early B-type stars are located in the DM region. The right panel of Figure \ref{fig:xrays} presents a logarithmic plot of $\ra/\rk$ vs. the bolometric luminosity. One can see that enhanced X-ray emission generally occurs for the most luminous magnetic OB stars. 
The B-type stars with intermediate luminosities seem to have X-ray emission typical for their spectral type, although it has been shown that the $\lx$-$\ls$ relation breaks down at spectral type B2 and that the typical $\lx$ is $10^{-8}\ls$ or lower for later spectral types (Cohen et al. 1997), implying that \textit{any} mid B-type star in one of the three highest $\lxls$ bins is overluminous for its spectral subtype. 
Finally, some low luminosity stars with large $\ra/\rk$ (extended centrifugal magnetospheres) show enhanced X-ray emission.

A potential explanation for the X-ray emission enhancement in CMs is the centrifugal acceleration for fast rotators, which contributes by propelling the confined material up the magnetic loop leading to stronger shocks than what could be achieved by radiative acceleration alone like in a DM. 
However, Rigid-Field Hydrodynamical \citep[RFHD;][]{2007MNRAS.382..139T} simulations predict that the overall X-ray flux of stars in the CM region is also quite sensitive to the mass-loss rate. The distribution in temperature of the differential emission measure (DEM) is governed by both the pre-shock and post-shock characteristics of the magnetosphere. Both of these are affected by the wind properties, with the post-shock cooling length being longer for lower-density wind flows, leading to softer and weaker emission \citep{2011IAUS..272..194H}. 

Therefore, a complete survey of the X-ray properties of the magnetic OB stars would be highly desirable as X-ray emission could provide a different perspective on the structure and dynamics of magnetospheres, and the shock physics occurring in both DMs and CMs.
Future studies should include a consistent and uniform analysis of (1) X-ray plasma temperature distributions and (2) time-variability. 
Although a re-analysis of all the available X-ray observations is beyond the scope of the current paper, the work presented here can be used as a starting point for identifying interesting stars for which X-ray data already exists, as well as identifying stars with interesting positions in the magnetic confinement-rotation diagram for which acquiring X-ray observations should be a priority.
Another key development for understanding the trends identified above in the X-ray emission of magnetic massive stars would be more accurate and secure wind mass-loss rates for the B-type stars.

\section{Conclusions}
\label{sec:conclu}

The Magnetism in Massive Stars (MiMeS) project aims to study the scope and impact of stellar magnetism in massive stars using high resolution and high signal-to-noise ratio spectropolarimetric observations from Large Program time allocations. 
Within this context, the present study had two main goals: (1) To compile an exhaustive and well-documented list of confirmed magnetic, hot OB stars that are directly detected through the Zeeman effect, and; (2) To organise the stars in a way that accounts for both the strength of magnetic confinement of the stellar wind (through $\etas$ or $\ra$) and the dynamical role of stellar rotation (through $W$ or $\rk$).
Key results are:

\begin{itemize}

\item We have provided a compilation of relevant stellar parameters for our magnetic sample. We used the luminosity, mass, and radius obtained from modern spectral modelling from the literature, when available; otherwise, these were derived from a classical bolometric correction approach, and from SED fitting with the code \textsc{Chorizos} when a complete set of photometry was available. We have also compiled rotational periods and dipolar field strengths, as well as binarity and pulsation status. 

\item Using these parameters, we have placed the full sample of magnetic stars in a classification plane, the magnetic confinement-rotation diagram, characterising stellar rotation (as $\rk$ or $W$) vs. wind magnetic confinement (as $\ra$ or $\etas$).

\item We identified key domains within the magnetic rotation-confinement diagram,  representing weakly magnetised winds with $\etas \la 1$, or  dynamical magnetospheres (DM, with $\rs < \ra < \rk$) vs. centrifugal magnetospheres (CM, with $\rs < \rk < \ra$).

\item We have associated H$\alpha$ line emission characteristics with position in the confinement-rotation diagram. Slowly rotating O-type stars show DM magnetospheric emission, in contrast to B-type stars which generally only show CM magnetospheric emission.

\item In a plane plotting the ratio $\rk/\ra$ vs. stellar luminosity, we found a clear separation between O-type star DM emission and B-type stars for which appearance of CM emission requires higher $\rk/\ra$ for lower luminosity stars. This suggests that the CM leakage mechanism may depend on the degree of magnetic confinement.

\item We have also associated other magnetospheric proxies with position in the confinement-rotation diagram. UV resonance line variation occurs in all magnetosphere regimes and for stars of all temperatures; although detailed modelling will be needed in the future, UV spectroscopy seems an excellent proxy for identifying new magnetic OB stars. 
The earliest magnetic OB stars generally show X-ray overluminosity, as do the low-luminosity B-type stars with large centrifugal magnetosphere volumes (high $\ra/\rk$).  

\item We have calculated magnetic spindown timescales ($\tauj$) and inferred spindown ages ($t_{\rm s}$) for each star in our sample. O-type stars with strong winds have short spindown timescales and so mostly are slow rotators located in the DM region; B-type stars with weaker winds have longer spindown timescales, and thus extend well into the CM regime with rapid rotation.

\item Finally, we have identified stars which will be prime candidates for follow-up studies (with either unknown periods or only dipole field strength lower limits) that would lead to a more accurate placement on the confinement-rotation diagram, hence providing more clues to the answers of some of the questions raised in this paper. 

\end{itemize}

\section*{Acknowledgments}
  VP acknowledges support from Fonds qu\'eb\'ecois de la recherche sur la nature et les technologies.
SOP and JOS acknowledge support from NASA grant ATP NNX11AC40G.
GAW acknowledges support from the Natural Science and Engineering Research Council of Canada (NSERC) Discovery Grant program.
MEO acknowledges financial support from GA\,\v{C}R under grant number 209/11/1198. The Astronomical Institute Ond\v{r}ejov 
is supported by the project RVO:67985815.
EA thanks the Programme National de Physique Stellaire (PNPS) for its support.
RHDT  and Au-D acknowledge support from NASA grant NNX12AC72G.
   This research has made use of the SIMBAD database, operated at CDS, Strasbourg, France and the Canadian Astronomy Data Centre operated by the National Research Council of Canada.
The authors would like to thank the MiMeS observing team for their efforts and the anonymous referee for his comments.

\bibliographystyle{mn2e}
\bibliography{ipod}

\begin{thebibliography}{}

\bibitem[\protect\citeauthoryear{{Abt}, {Wang} \& {Cardona}}{{Abt}
  et~al.}{1991}]{1991ApJ...367..155A}
{Abt} H.~A.,  {Wang} R.,    {Cardona} O.,  1991, \apj, 367, 155

\bibitem[\protect\citeauthoryear{{Alecian}, {Catala}, {Wade}, {Donati},
  {Petit}, {Landstreet}, {B{\"o}hm}, {Bouret}, {Bagnulo}, {Folsom}, {Grunhut}
  \& {Silvester}}{{Alecian} et~al.}{2008a}]{2008MNRAS.385..391A}
{Alecian} E. et~al.,  2008a, \mnras, 385, 391

\bibitem[\protect\citeauthoryear{{Alecian}, {Kochukhov}, {Neiner}, {Wade}, {de
  Batz}, {Henrichs}, {Grunhut}, {Bouret}, {Briquet}, {Gagne}, {Naze}, {Oksala},
  {Rivinius}, {Townsend}, {Walborn}, {Weiss} \& {Mimes
  Collaboration}}{{Alecian} et~al.}{2011}]{2011A&A...536L...6A}
{Alecian} E. et~al., 2011, \aap, 536, L6

\bibitem[\protect\citeauthoryear{{Alecian}, {Wade}, {Catala}, {Bagnulo},
  {Boehm}, {Bohlender}, {Bouret}, {Donati}, {Folsom}, {Grunhut} \&
  {Landstreet}}{{Alecian} et~al.}{2008b}]{2008A&A...481L..99A}
{Alecian} E. et~al.,  2008b, \aap, 481, L99

\bibitem[\protect\citeauthoryear{{Ausseloos}, {Aerts}, {Lefever}, {Davis} \&
  {Harmanec}}{{Ausseloos} et~al.}{2006}]{2006A&A...455..259A}
{Ausseloos} M.,  {Aerts} C.,  {Lefever} K.,  {Davis} J.,    {Harmanec} P.,
  2006, \aap, 455, 259

\bibitem[\protect\citeauthoryear{{Babcock}}{{Babcock}}{1947}]{1947ApJ...105..105B}
{Babcock} H.~W.,  1947, \apj, 105, 105

\bibitem[\protect\citeauthoryear{{Babel} \& {Montmerle}}{{Babel} \&
  {Montmerle}}{1997a}]{1997ApJ...485L..29B}
{Babel} J.,  {Montmerle} T.,  1997a, \apjl, 485, L29

\bibitem[\protect\citeauthoryear{{Babel} \& {Montmerle}}{{Babel} \&
  {Montmerle}}{1997b}]{1997A&A...323..121B}
{Babel} J.,  {Montmerle} T.,  1997b, \aap, 323, 121

\bibitem[\protect\citeauthoryear{{Bagnulo}, {Landstreet}, {Fossati} \&
  {Kochukhov}}{{Bagnulo} et~al.}{2012}]{2012A&A...538A.129B}
{Bagnulo} S.,  {Landstreet} J.~D.,  {Fossati} L.,    {Kochukhov} O.,  2012,
  \aap, 538, A129

\bibitem[\protect\citeauthoryear{{Bagnulo}, {Landstreet}, {Mason}, {Andretta},
  {Silaj} \& {Wade}}{{Bagnulo} et~al.}{2006}]{2006A&A...450..777B}
{Bagnulo} S.,  {Landstreet} J.~D.,  {Mason} E.,  {Andretta} V.,  {Silaj} J.,
  {Wade} G.~A.,  2006, \aap, 450, 777

\bibitem[\protect\citeauthoryear{{Berghoefer}, {Schmitt}, {Danner} \&
  {Cassinelli}}{{Berghoefer} et~al.}{1997}]{1997A&A...322..167B}
{Berghoefer} T.~W.,  {Schmitt} J.~H.~M.~M.,  {Danner} R.,    {Cassinelli}
  J.~P.,  1997, \aap, 322, 167

\bibitem[\protect\citeauthoryear{{Bohlender}}{{Bohlender}}{1989}]{1989ApJ...346..459B}
{Bohlender} D.~A.,  1989, \apj, 346, 459

\bibitem[\protect\citeauthoryear{{Bohlender} \& {Landstreet}}{{Bohlender} \&
  {Landstreet}}{1990}]{1990ApJ...358..274B}
{Bohlender} D.~A.,  {Landstreet} J.~D.,  1990, \apj, 358, 274

\bibitem[\protect\citeauthoryear{{Bohlender}, {Landstreet}, {Brown} \&
  {Thompson}}{{Bohlender} et~al.}{1987}]{1987ApJ...323..325B}
{Bohlender} D.~A.,  {Landstreet} J.~D.,  {Brown} D.~N.,    {Thompson} I.~B.,
  1987, \apj, 323, 325

\bibitem[\protect\citeauthoryear{{Bohlender} \& {Monin}}{{Bohlender} \&
  {Monin}}{2011}]{2011AJ....141..169B}
{Bohlender} D.~A.,  {Monin} D.,  2011, \aj, 141, 169

\bibitem[\protect\citeauthoryear{{Bohlender}, {Rice} \& {Hechler}}{{Bohlender}
  et~al.}{2010}]{2010A&A...520A..44B}
{Bohlender} D.~A.,  {Rice} J.~B.,    {Hechler} P.,  2010, \aap, 520, A44

\bibitem[\protect\citeauthoryear{{Bolton}, {Harmanec}, {Lyons}, {Odell} \&
  {Pyper}}{{Bolton} et~al.}{1998}]{1998A&A...337..183B}
{Bolton} C.~T.,  {Harmanec} P.,  {Lyons} R.~W.,  {Odell} A.~P.,    {Pyper}
  D.~M.,  1998, \aap, 337, 183

\bibitem[\protect\citeauthoryear{{Borra}}{{Borra}}{1981}]{1981ApJ...249L..39B}
{Borra} E.~F.,  1981, \apjl, 249, L39

\bibitem[\protect\citeauthoryear{{Borra} \& {Landstreet}}{{Borra} \&
  {Landstreet}}{1980}]{1980ApJS...42..421B}
{Borra} E.~F.,  {Landstreet} J.~D.,  1980, \apjs, 42, 421

\bibitem[\protect\citeauthoryear{{Bouret}, {Donati}, {Martins}, {Escolano},
  {Marcolino}, {Lanz} \& {Howarth}}{{Bouret}
  et~al.}{2008}]{2008MNRAS.389...75B}
{Bouret} J.-C.,  {Donati} J.-F.,  {Martins} F.,  {Escolano} C.,  {Marcolino}
  W.,  {Lanz} T.,    {Howarth} I.~D.,  2008, \mnras, 389, 75

\bibitem[\protect\citeauthoryear{{Bouret}, {Hillier}, {Lanz} \&
  {Fullerton}}{{Bouret} et~al.}{2012}]{2012arXiv1205.3075B}
{Bouret} J.-C.,  {Hillier} D.~J.,  {Lanz} T.,    {Fullerton} A.~W.,  2012,
  \aap, 544, A67

\bibitem[\protect\citeauthoryear{{Briquet}, {Aerts} \& {De Cat}}{{Briquet}
  et~al.}{2001}]{2001A&A...366..121B}
{Briquet} M.,  {Aerts} C.,    {De Cat} P.,  2001, \aap, 366, 121

\bibitem[\protect\citeauthoryear{{Briquet}, {Aerts}, {L{\"u}ftinger}, {De Cat},
  {Piskunov} \& {Scuflaire}}{{Briquet} et~al.}{2004}]{2004A&A...413..273B}
{Briquet} M.,  {Aerts} C.,  {L{\"u}ftinger} T.,  {De Cat} P.,  {Piskunov}
  N.~E.,    {Scuflaire} R.,  2004, \aap, 413, 273

\bibitem[\protect\citeauthoryear{{Briquet}, {Hubrig}, {Sch{\"o}ller} \& {De
  Cat}}{{Briquet} et~al.}{2007}]{2007AN....328...41B}
{Briquet} M.,  {Hubrig} S.,  {Sch{\"o}ller} M.,    {De Cat} P.,  2007,
  Astronomische Nachrichten, 328, 41

\bibitem[\protect\citeauthoryear{{Brott}, {de Mink}, {Cantiello}, {Langer}, {de
  Koter}, {Evans}, {Hunter}, {Trundle} \& {Vink}}{{Brott}
  et~al.}{2011}]{2011A&A...530A.115B}
{Brott} I. et~al.,  2011, \aap,
  530, A115

\bibitem[\protect\citeauthoryear{{Bychkov}, {Bychkova} \& {Madej}}{{Bychkov}
  et~al.}{2005}]{2005A&A...430.1143B}
{Bychkov} V.~D.,  {Bychkova} L.~V.,    {Madej} J.,  2005, \aap, 430, 1143

\bibitem[\protect\citeauthoryear{{Castor}, {Abbott} \& {Klein}}{{Castor}
  et~al.}{1975}]{1975ApJ...195..157C}
{Castor} J.~I.,  {Abbott} D.~C.,    {Klein} R.~I.,  1975, \apj, 195, 157

\bibitem[\protect\citeauthoryear{{Catanzaro}}{{Catanzaro}}{2008}]{2008MNRAS.387..759C}
{Catanzaro} G.,  2008, \mnras, 387, 759

\bibitem[\protect\citeauthoryear{{Chauville}, {Zorec}, {Ballereau}, {Morrell},
  {Cidale} \& {Garcia}}{{Chauville} et~al.}{2001}]{2001A&A...378..861C}
{Chauville} J.,  {Zorec} J.,  {Ballereau} D.,  {Morrell} N.,  {Cidale} L.,
  {Garcia} A.,  2001, \aap, 378, 861

\bibitem[\protect\citeauthoryear{{Cidale}, {Arias}, {Torres}, {Zorec},
  {Fr{\'e}mat} \& {Cruzado}}{{Cidale} et~al.}{2007}]{2007A&A...468..263C}
{Cidale} L.~S.,  {Arias} M.~L.,  {Torres} A.~F.,  {Zorec} J.,  {Fr{\'e}mat} Y.,
     {Cruzado} A.,  2007, \aap, 468, 263

\bibitem[\protect\citeauthoryear{{Claret}}{{Claret}}{2004}]{2004A&A...424..919C}
{Claret} A.,  2004, \aap, 424, 919

\bibitem[\protect\citeauthoryear{{Collins} II \& {Harrington}}{{Collins} \&
  {Harrington}}{1966}]{1966ApJ...146..152C}
{Collins} II G.~W.,  {Harrington} J.~P.,  1966, \apj, 146, 152

\bibitem[\protect\citeauthoryear{{Cranmer} \& {Owocki}}{{Cranmer} \&
  {Owocki}}{1996}]{1996ApJ...462..469C}
{Cranmer} S.~R.,  {Owocki} S.~P.,  1996, \apj, 462, 469

\bibitem[\protect\citeauthoryear{{de Jager} \& {Nieuwenhuijzen}}{{de Jager} \&
  {Nieuwenhuijzen}}{1987}]{1987A&A...177..217D}
{de Jager} C.,  {Nieuwenhuijzen} H.,  1987, \aap, 177, 217

\bibitem[\protect\citeauthoryear{{Dessart} \& {Owocki}}{{Dessart} \&
  {Owocki}}{2003}]{2003A&A...406L...1D}
{Dessart} L.,  {Owocki} S.~P.,  2003, \aap, 406, L1

\bibitem[\protect\citeauthoryear{{Donati} \& {Landstreet}}{{Donati} \&
  {Landstreet}}{2009}]{2009ARA&A..47..333D}
{Donati} J.,  {Landstreet} J.~D.,  2009, \araa, 47, 333

\bibitem[\protect\citeauthoryear{{Donati}, {Babel}, {Harries}, {Howarth},
  {Petit} \& {Semel}}{{Donati} et~al.}{2002}]{2002MNRAS.333...55D}
{Donati} J.-F.,  {Babel} J.,  {Harries} T.~J.,  {Howarth} I.~D.,  {Petit} P.,
   {Semel} M.,  2002, \mnras, 333, 55

\bibitem[\protect\citeauthoryear{{Donati}, {Howarth}, {Bouret}, {Petit},
  {Catala} \& {Landstreet}}{{Donati} et~al.}{2006a}]{2006MNRAS.365L...6D}
{Donati} J.-F.,  {Howarth} I.~D.,  {Bouret} J.-C.,  {Petit} P.,  {Catala} C.,
   {Landstreet} J.,  2006a, \mnras, 365, L6

\bibitem[\protect\citeauthoryear{{Donati}, {Howarth}, {Jardine}, {Petit},
  {Catala}, {Landstreet}, {Bouret}, {Alecian}, {Barnes}, {Forveille}, {Paletou}
  \& {Manset}}{{Donati} et~al.}{2006b}]{2006MNRAS.370..629D}
{Donati} J.-F. et~al.,  2006b, \mnras, 370, 629

\bibitem[\protect\citeauthoryear{{Donati}, {Wade}, {Babel}, {Henrichs}, {de
  Jong} \& {Harries}}{{Donati} et~al.}{2001}]{2001MNRAS.326.1265D}
{Donati} J.-F.,  {Wade} G.~A.,  {Babel} J.,  {Henrichs} H.~f.,  {de Jong}
  J.~A.,    {Harries} T.~J.,  2001, \mnras, 326, 1265

\bibitem[\protect\citeauthoryear{{Drake}, {Linsky}, {Schmitt} \&
  {Rosso}}{{Drake} et~al.}{1994}]{1994ApJ...420..387D}
{Drake} S.~A.,  {Linsky} J.~L.,  {Schmitt} J.~H.~M.~M.,    {Rosso} C.,  1994,
  \apj, 420, 387

\bibitem[\protect\citeauthoryear{{Favata}, {Neiner}, {Testa}, {Hussain} \&
  {Sanz-Forcada}}{{Favata} et~al.}{2009}]{2009A&A...495..217F}
{Favata} F.,  {Neiner} C.,  {Testa} P.,  {Hussain} G.,    {Sanz-Forcada} J.,
  2009, \aap, 495, 217

\bibitem[\protect\citeauthoryear{{Feldmeier}, {Puls} \&
  {Pauldrach}}{{Feldmeier} et~al.}{1997}]{1997A&A...322..878F}
{Feldmeier} A.,  {Puls} J.,    {Pauldrach} A.~W.~A.,  1997, \aap, 322, 878

\bibitem[\protect\citeauthoryear{{Fourtune-Ravard}, {Wade}, {Marcolino},
  {Shultz}, {Grunhut}, {Henrichs} \& {Henrichs}}{{Fourtune-Ravard}
  et~al.}{2011}]{2011IAUS..272..180F}
{Fourtune-Ravard} C.,  {Wade} G.~A.,  {Marcolino} W.~L.~F.,  {Shultz} M.,
  {Grunhut} J.~H.,  {Henrichs} H.~F.,    {Henrichs} 2011, in {Neiner} C.,
  {Wade} G.,  {Meynet} G.,   {Peters} G.,  eds, IAU Symposium Vol.~272 Active OB Stars: Structure, Evolution, Mass
Loss, and Critical Limits, Cambridge Univ. Press, Cambridge, p. 180

\bibitem[\protect\citeauthoryear{{Gagn{\'e}}, {Fehon}, {Savoy}, {Cohen},
  {Townsley}, {Broos}, {Povich}, {Corcoran}, {Walborn}, {Remage Evans},
  {Moffat}, {Naz{\'e}} \& {Oskinova}}{{Gagn{\'e}}
  et~al.}{2011}]{2011ApJS..194....5G}
{Gagn{\'e}} M. et~al.,
  2011, \apjs, 194, 5

\bibitem[\protect\citeauthoryear{{Gagn{\'e}}, {Oksala}, {Cohen}, {Tonnesen},
  {ud-Doula}, {Owocki}, {Townsend} \& {MacFarlane}}{{Gagn{\'e}}
  et~al.}{2005}]{2005ApJ...628..986G}
{Gagn{\'e}} M.,  {Oksala} M.~E.,  {Cohen} D.~H.,  {Tonnesen} S.~K.,  {ud-Doula}
  A.,  {Owocki} S.~P.,  {Townsend} R.~H.~D.,    {MacFarlane} J.~J.,  2005,
  \apj, 628, 986

\bibitem[\protect\citeauthoryear{{Gayley}}{{Gayley}}{1995}]{1995ApJ...454..410G}
{Gayley} K.~G.,  1995, \apj, 454, 410

\bibitem[\protect\citeauthoryear{{Glagolevskij}, {Leushin} \&
  {Chountonov}}{{Glagolevskij} et~al.}{2007}]{2007AstBu..62..319G}
{Glagolevskij} Y.~V.,  {Leushin} V.~V.,    {Chountonov} G.~A.,  2007,
  Astrophysical Bulletin, 62, 319

\bibitem[\protect\citeauthoryear{{Grillo}, {Sciortino}, {Micela}, {Vaiana} \&
  {Harnden} Jr.}{{Grillo} et~al.}{1992}]{1992ApJS...81..795G}
{Grillo} F.,  {Sciortino} S.,  {Micela} G.,  {Vaiana} G.~S.,    {Harnden} Jr.
  F.~R.,  1992, \apjs, 81, 795

\bibitem[\protect\citeauthoryear{{Grunhut}, {Rivinius}, {Wade}, {Townsend},
  {Marcolino}, {Bohlender}, {Szeifert}, {Petit}, {Matthews}, {Rowe}, {Moffat},
  {Kallinger}, {Kuschnig}, {Guenther}, {Rucinski}, {Sasselov} \&
  {Weiss}}{{Grunhut} et~al.}{2012a}]{2012MNRAS.419.1610G}
{Grunhut} J.~H. et~al.,  2012a, \mnras, 419, 1610

\bibitem[\protect\citeauthoryear{{Grunhut}, {Wade}, {Leutenegger}, {Petit},
  {Rauw}, {Neiner}, {Martins}, {Cohen}, {Gagn{\'e}}, {Ignace}, {Mathis}, {de
  Mink}, {Moffat}, {Owocki}, {Shultz}, {Sundqvist} \& {the MiMeS
  Collaboration}}{{Grunhut} et~al.}{2012b}]{Grunhut2012_plaskett}
{Grunhut} J.~H. et~al., 2012b, \mnras,
  submitted (arXiv 1209.6326)

\bibitem[\protect\citeauthoryear{{Grunhut}, {Wade}, {Marcolino}, {Petit},
  {Henrichs}, {Cohen}, {Alecian}, {Bohlender}, {Bouret}, {Kochukhov}, {Neiner},
  {St-Louis} \& {Townsend}}{{Grunhut} et~al.}{2009}]{2009MNRAS.400L..94G}
{Grunhut} J.~H. et~al.,  2009,
  \mnras, 400, L94

\bibitem[\protect\citeauthoryear{{Grunhut}, {Wade}, {Sundqvist}, {ud-Doula},
  {Neiner}, {Ignace}, {Marcolino}, {Rivinius}, {Fullerton}, {Kaper},
  {Mauclaire}, {Buil}, {Garrel}, {Ribeiro}, {Ubaud} \& {the MiMeS
  Collaboration}}{{Grunhut} et~al.}{2012c}]{2012arXiv1207.6988G}
{Grunhut} J.~H. et~al., 2012c, MNRAS, in press (arXiv 1207.6988)

\bibitem[\protect\citeauthoryear{{Guarcello}, {Caramazza}, {Micela},
  {Sciortino}, {Drake} \& {Prisinzano}}{{Guarcello}
  et~al.}{2012}]{2012ApJ...753..117G}
{Guarcello} M.~G.,  {Caramazza} M.,  {Micela} G.,  {Sciortino} S.,  {Drake}
  J.~J.,    {Prisinzano} L.,  2012, \apj, 753, 117

\bibitem[\protect\citeauthoryear{{Hamaguchi}, {Yamauchi} \&
  {Koyama}}{{Hamaguchi} et~al.}{2005}]{2005ApJ...618..360H}
{Hamaguchi} K.,  {Yamauchi} S.,    {Koyama} K.,  2005, \apj, 618, 360

\bibitem[\protect\citeauthoryear{{Hauck} \& {Mermilliod}}{{Hauck} \&
  {Mermilliod}}{1998}]{1998A&AS..129..431H}
{Hauck} B.,  {Mermilliod} M.,  1998, \aaps, 129, 431

\bibitem[\protect\citeauthoryear{{Henrichs}, {Bauer}, {Hill}, {Kaper},
  {Nichols-Bohlin} \& {Veen}}{{Henrichs} et~al.}{1993}]{1993npsp.conf..186H}
{Henrichs} H.~F.,  {Bauer} F.,  {Hill} G.~M.,  {Kaper} L.,  {Nichols-Bohlin}
  J.~S.,    {Veen} P.~M.,  1993, in {Nemec} J.~M.,  {Matthews} J.~M.,  eds, IAU
  Colloq. 139: New Perspectives on Stellar Pulsation and Pulsating Variable
  Stars, Cambridge Univ. Press, Cambridge,
p.~186

\bibitem[\protect\citeauthoryear{{Henrichs}, {Kolenberg}, {Plaggenborg},
  {Marsden}, {Waite}, {Landstreet}, {Wade}, {Grunhut}, {Oksala} \& {Mimes
  Collaboration}}{{Henrichs} et~al.}{2012}]{henrichs_siglupi}
{Henrichs} H.~F. et~al., 2012, \aap, 545

\bibitem[\protect\citeauthoryear{{Henrichs}, {Neiner}, {Schnerr}, {Verdugo},
  {Alecian}, {Catala}, {Cochard}, {Guti{\'e}rrez}, {Huat}, {Silvester} \&
  {Thizy}}{{Henrichs} et~al.}{2009}]{2009IAUS..259..393H}
{Henrichs} H.~F. et~al.,  2009, in Strassmeier, K.~G., Kosovichev, A.~G. \& Beckmann J. ed, IAU Symposium Vol.~259, Cosmic Magnetic Fields:
from Planets, to Stars and Galaxies, Cambridge Univ. Press, Cambridge,
p. 393

\bibitem[\protect\citeauthoryear{{Hill}, {Townsend}, {Cohen} \&
  {Gagn{\'e}}}{{Hill} et~al.}{2011}]{2011IAUS..272..194H}
{Hill} N.~R.,  {Townsend} R.~H.~D.,  {Cohen} D.~H.,    {Gagn{\'e}} M.,  2011,
  in {C.~Neiner, G.~Wade, G.~Meynet, \& G.~Peters} ed., IAU Symposium Vol.~272,
 Active OB Stars: Structure, Evolution, Mass
Loss, and Critical Limits, Cambridge Univ. Press, Cambridge,
p. 194

\bibitem[\protect\citeauthoryear{{Hillier} \& {Miller}}{{Hillier} \&
  {Miller}}{1998}]{1998ApJ...496..407H}
{Hillier} D.~J.,  {Miller} D.~L.,  1998, \apj, 496, 407

\bibitem[\protect\citeauthoryear{{Howarth} \& {Prinja}}{{Howarth} \&
  {Prinja}}{1989}]{1989ApJS...69..527H}
{Howarth} I.~D.,  {Prinja} R.~K.,  1989, \apjs, 69, 527

\bibitem[\protect\citeauthoryear{{Howarth}, {Walborn}, {Lennon} \&
  {Puls}}{{Howarth} et~al.}{2007}]{2007MNRAS.381..433H}
{Howarth} I.~D.,  {Walborn} N.~R.,  {Lennon} D.~J.,    {Puls} J.,  2007,
  \mnras, 381, 433

\bibitem[\protect\citeauthoryear{{Huang} \& {Gies}}{{Huang} \&
  {Gies}}{2006}]{2006ApJ...648..580H}
{Huang} W.,  {Gies} D.~R.,  2006, \apj, 648, 580

\bibitem[\protect\citeauthoryear{{Hubrig}, {Briquet}, {De Cat}, {Sch{\"o}ller},
  {Morel} \& {Ilyin}}{{Hubrig} et~al.}{2009}]{2009AN....330..317H}
{Hubrig} S.,  {Briquet} M.,  {De Cat} P.,  {Sch{\"o}ller} M.,  {Morel} T.,
  {Ilyin} I.,  2009, Astronomische Nachrichten, 330, 317

\bibitem[\protect\citeauthoryear{{Hubrig}, {Briquet}, {Sch{\"o}ller}, {De Cat},
  {Mathys} \& {Aerts}}{{Hubrig} et~al.}{2006}]{2006MNRAS.369L..61H}
{Hubrig} S.,  {Briquet} M.,  {Sch{\"o}ller} M.,  {De Cat} P.,  {Mathys} G.,
  {Aerts} C.,  2006, \mnras, 369, L61

\bibitem[\protect\citeauthoryear{{Hummel}, {White}, {Elias} II, {Hajian} \&
  {Nordgren}}{{Hummel} et~al.}{2000}]{2000ApJ...540L..91H}
{Hummel} C.~A.,  {White} N.~M.,  {Elias} II N.~M.,  {Hajian} A.~R.,
  {Nordgren} T.~E.,  2000, \apjl, 540, L91

\bibitem[\protect\citeauthoryear{{Hunger}, {Heber} \& {Groote}}{{Hunger}
  et~al.}{1989}]{1989A&A...224...57H}
{Hunger} K.,  {Heber} U.,    {Groote} D.,  1989, \aap, 224, 57

\bibitem[\protect\citeauthoryear{{Ignace}, {Oskinova}, {Jardine}, {Cassinelli},
  {Cohen}, {Donati}, {Townsend} \& {ud-Doula}}{{Ignace}
  et~al.}{2010}]{2010ApJ...721.1412I}
{Ignace} R.,  {Oskinova} L.~M.,  {Jardine} M.,  {Cassinelli} J.~P.,  {Cohen}
  D.~H.,  {Donati} J.-F.,  {Townsend} R.~H.~D.,    {ud-Doula} A.,  2010, \apj,
  721, 1412

\bibitem[\protect\citeauthoryear{{Kaper}, {Henrichs}, {Nichols}, {Snoek},
  {Volten} \& {Zwarthoed}}{{Kaper} et~al.}{1996}]{1996A&AS..116..257K}
{Kaper} L.,  {Henrichs} H.~F.,  {Nichols} J.~S.,  {Snoek} L.~C.,  {Volten} H.,
    {Zwarthoed} G.~A.~A.,  1996, \aaps, 116, 257

\bibitem[\protect\citeauthoryear{{Kochukhov}, {Lundin}, {Romanyuk} \&
  {Kudryavtsev}}{{Kochukhov} et~al.}{2011}]{2011ApJ...726...24K}
{Kochukhov} O.,  {Lundin} A.,  {Romanyuk} I.,    {Kudryavtsev} D.,  2011, \apj,
  726, 24

\bibitem[\protect\citeauthoryear{{Kraus}, {Weigelt}, {Balega}, {Docobo},
  {Hofmann}, {Preibisch}, {Schertl}, {Tamazian}, {Driebe}, {Ohnaka}, {Petrov},
  {Sch{\"o}ller} \& {Smith}}{{Kraus} et~al.}{2009}]{2009A&A...497..195K}
{Kraus} S.,  {Weigelt} G.,  {Balega} Y.~Y.,  {Docobo} J.~A.,  {Hofmann} K.-H.,
  {Preibisch} T.,  {Schertl} D.,  {Tamazian} V.~S.,  {Driebe} T.,  {Ohnaka} K.,
   {Petrov} R.,  {Sch{\"o}ller} M.,    {Smith} M.,  2009, \aap, 497, 195

\bibitem[\protect\citeauthoryear{{Kurucz}}{{Kurucz}}{1979}]{1979ApJS...40....1K}
{Kurucz} R.~L.,  1979, \apjs, 40, 1

\bibitem[\protect\citeauthoryear{{Lamers}, {Snow} \& {Lindholm}}{{Lamers}
  et~al.}{1995}]{1995ApJ...455..269L}
{Lamers} H.~J.~G.~L.~M.,  {Snow} T.~P.,    {Lindholm} D.~M.,  1995, \apj, 455,
  269

\bibitem[\protect\citeauthoryear{{Landstreet}}{{Landstreet}}{1988}]{1988ApJ...326..967L}
{Landstreet} J.~D.,  1988, \apj, 326, 967

\bibitem[\protect\citeauthoryear{{Landstreet}, {Bagnulo}, {Andretta},
  {Fossati}, {Mason}, {Silaj} \& {Wade}}{{Landstreet}
  et~al.}{2007}]{2007A&A...470..685L}
{Landstreet} J.~D.,  {Bagnulo} S.,  {Andretta} V.,  {Fossati} L.,  {Mason} E.,
  {Silaj} J.,    {Wade} G.~A.,  2007, \aap, 470, 685

\bibitem[\protect\citeauthoryear{{Landstreet} \& {Borra}}{{Landstreet} \&
  {Borra}}{1978}]{1978ApJ...224L...5L}
{Landstreet} J.~D.,  {Borra} E.~F.,  1978, \apjl, 224, L5

\bibitem[\protect\citeauthoryear{{Lanz} \& {Hubeny}}{{Lanz} \&
  {Hubeny}}{2003}]{2003ApJS..146..417L}
{Lanz} T.,  {Hubeny} I.,  2003, \apjs, 146, 417

\bibitem[\protect\citeauthoryear{{Lanz} \& {Hubeny}}{{Lanz} \&
  {Hubeny}}{2007}]{2007ApJS..169...83L}
{Lanz} T.,  {Hubeny} I.,  2007, \apjs, 169, 83

\bibitem[\protect\citeauthoryear{{Lee} \& {Obrien}}{{Lee} \&
  {Obrien}}{1977}]{1977A&A....60..259L}
{Lee} P.,  {Obrien} A.,  1977, \aap, 60, 259

\bibitem[\protect\citeauthoryear{{Lefever}, {Puls}, {Morel}, {Aerts}, {Decin}
  \& {Briquet}}{{Lefever} et~al.}{2010}]{2010A&A...515A..74L}
{Lefever} K.,  {Puls} J.,  {Morel} T.,  {Aerts} C.,  {Decin} L.,    {Briquet}
  M.,  2010, \aap, 515, A74

\bibitem[\protect\citeauthoryear{{Leone}, {Bohlender}, {Bolton}, {Buemi},
  {Catanzaro}, {Hill} \& {Stift}}{{Leone} et~al.}{2010}]{2010MNRAS.401.2739L}
{Leone} F.,  {Bohlender} D.~A.,  {Bolton} C.~T.,  {Buemi} C.,  {Catanzaro} G.,
  {Hill} G.~M.,    {Stift} M.~J.,  2010, \mnras, 401, 2739

\bibitem[\protect\citeauthoryear{{Leone}, {Catalano} \& {Malaroda}}{{Leone}
  et~al.}{1997}]{1997A&A...325.1125L}
{Leone} F.,  {Catalano} F.~A.,    {Malaroda} S.,  1997, \aap, 325, 1125

\bibitem[\protect\citeauthoryear{{Leone} \& {Manfre}}{{Leone} \&
  {Manfre}}{1997}]{1997A&A...320..257L}
{Leone} F.,  {Manfre} M.,  1997, \aap, 320, 257

\bibitem[\protect\citeauthoryear{{Linder}, {Rauw}, {Martins}, {Sana}, {De
  Becker} \& {Gosset}}{{Linder} et~al.}{2008}]{2008A&A...489..713L}
{Linder} N.,  {Rauw} G.,  {Martins} F.,  {Sana} H.,  {De Becker} M.,
  {Gosset} E.,  2008, \aap, 489, 713

\bibitem[\protect\citeauthoryear{{Linder}, {Rauw}, {Pollock} \&
  {Stevens}}{{Linder} et~al.}{2006}]{2006MNRAS.370.1623L}
{Linder} N.,  {Rauw} G.,  {Pollock} A.~M.~T.,    {Stevens} I.~R.,  2006,
  \mnras, 370, 1623

\bibitem[\protect\citeauthoryear{{Linsky}, {Drake} \& {Bastian}}{{Linsky}
  et~al.}{1992}]{1992ApJ...393..341L}
{Linsky} J.~L.,  {Drake} S.~A.,    {Bastian} T.~S.,  1992, \apj, 393, 341

\bibitem[\protect\citeauthoryear{{Lutz} \& {Kelker}}{{Lutz} \&
  {Kelker}}{1973}]{1973PASP...85..573L}
{Lutz} T.~E.,  {Kelker} D.~H.,  1973, \pasp, 85, 573

\bibitem[\protect\citeauthoryear{{Lyubimkov}, {Rachkovskaya}, {Rostopchin} \&
  {Lambert}}{{Lyubimkov} et~al.}{2002}]{2002MNRAS.333....9L}
{Lyubimkov} L.~S.,  {Rachkovskaya} T.~M.,  {Rostopchin} S.~I.,    {Lambert}
  D.~L.,  2002, \mnras, 333, 9

\bibitem[\protect\citeauthoryear{{Ma{\'{\i}}z-Apell{\'a}niz}}{{Ma{\'{\i}}z-Apell{\'a}niz}}{2001}]{2001AJ....121.2737M}
{Ma{\'{\i}}z-Apell{\'a}niz} J.,  2001, \aj, 121, 2737

\bibitem[\protect\citeauthoryear{{Ma{\'{\i}}z-Apell{\'a}niz}}{{Ma{\'{\i}}z-Apell{\'a}niz}}{2004}]{2004PASP..116..859M}
{Ma{\'{\i}}z-Apell{\'a}niz} J.,  2004, \pasp, 116, 859

\bibitem[\protect\citeauthoryear{{Ma{\'{\i}}z Apell{\'a}niz}}{{Ma{\'{\i}}z
  Apell{\'a}niz}}{2005}]{2005ESASP.576..179M}
{Ma{\'{\i}}z Apell{\'a}niz} J.,  2005, in {Turon} C.,  {O'Flaherty} K.~S.,
  {Perryman} M.~A.~C.,  eds, The Three-Dimensional Universe with Gaia Vol.~576
  of ESA Special Publication,
p.~179

\bibitem[\protect\citeauthoryear{{Ma{\'{\i}}z Apell{\'a}niz}, {Alfaro} \&
  {Sota}}{{Ma{\'{\i}}z Apell{\'a}niz} et~al.}{2008}]{2008arXiv0804.2553M}
{Ma{\'{\i}}z Apell{\'a}niz} J.,  {Alfaro} E.~J.,    {Sota} A.,  2008, ArXiv 0804.2553

\bibitem[\protect\citeauthoryear{{Marcolino}, {Bouret}, {Walborn}, {Howarth},
  {Naz{\'e}}, {Fullerton}, {Wade}, {Hillier} \& {Herrero}}{{Marcolino}
  et~al.}{2012}]{2012MNRAS.422.2314M}
{Marcolino} W.~L.~F. et~al.,  2012, \mnras, 422, 2314

\bibitem[\protect\citeauthoryear{{Markova} \& {Puls}}{{Markova} \&
  {Puls}}{2008}]{2008A&A...478..823M}
{Markova} N.,  {Puls} J.,  2008, \aap, 478, 823

\bibitem[\protect\citeauthoryear{{Martins}, {Donati}, {Marcolino}, {Bouret},
  {Wade}, {Escolano} \& {Howarth}}{{Martins}
  et~al.}{2010}]{2010MNRAS.407.1423M}
{Martins} F.,  {Donati} J.,  {Marcolino} W.~L.~F.,  {Bouret} J.,  {Wade} G.~A.,
   {Escolano} C.,    {Howarth} I.~D.,  2010, \mnras, 407, 1423

\bibitem[\protect\citeauthoryear{{Martins} \& {Plez}}{{Martins} \&
  {Plez}}{2006}]{2006A&A...457..637M}
{Martins} F.,  {Plez} B.,  2006, \aap, 457, 637

\bibitem[\protect\citeauthoryear{{Martins}, {Schaerer} \& {Hillier}}{{Martins}
  et~al.}{2005}]{2005A&A...436.1049M}
{Martins} F.,  {Schaerer} D.,    {Hillier} D.~J.,  2005, \aap, 436, 1049

\bibitem[\protect\citeauthoryear{{Matthews} \& {Bohlender}}{{Matthews} \&
  {Bohlender}}{1991}]{1991A&A...243..148M}
{Matthews} J.~M.,  {Bohlender} D.~A.,  1991, \aap, 243, 148

\bibitem[\protect\citeauthoryear{{McSwain}}{{McSwain}}{2008}]{2008ApJ...686.1269M}
{McSwain} M.~V.,  2008, \apj, 686, 1269

\bibitem[\protect\citeauthoryear{{McSwain}, {Huang}, {Gies}, {Grundstrom} \&
  {Townsend}}{{McSwain} et~al.}{2008}]{2008ApJ...672..590M}
{McSwain} M.~V.,  {Huang} W.,  {Gies} D.~R.,  {Grundstrom} E.~D.,    {Townsend}
  R.~H.~D.,  2008, \apj, 672, 590

\bibitem[\protect\citeauthoryear{{Mermilliod}}{{Mermilliod}}{2006}]{2006yCat.2168....0M}
{Mermilliod} J.~C.,  2006, VizieR Online Data Catalog, 2168

\bibitem[\protect\citeauthoryear{{Merrill} \& {Burwell}}{{Merrill} \&
  {Burwell}}{1943}]{1943ApJ....98..153M}
{Merrill} P.~W.,  {Burwell} C.~G.,  1943, \apj, 98, 153

\bibitem[\protect\citeauthoryear{{Mewe}, {Raassen}, {Cassinelli}, {van der
  Hucht}, {Miller} \& {G{\"u}del}}{{Mewe} et~al.}{2003}]{2003A&A...398..203M}
{Mewe} R.,  {Raassen} A.~J.~J.,  {Cassinelli} J.~P.,  {van der Hucht} K.~A.,
  {Miller} N.~A.,    {G{\"u}del} M.,  2003, \aap, 398, 203

\bibitem[\protect\citeauthoryear{{Mikul{\'a}{\v s}ek}, {Krti{\v c}ka}, {Henry},
  {Jan{\'{\i}}k} \& {Zverko}}{{Mikul{\'a}{\v s}ek}
  et~al.}{2011}]{2011A&A...534L...5M}
{Mikul{\'a}{\v s}ek} Z.,  {Krti{\v c}ka} J.,  {Henry} G.~W.,  {Jan{\'{\i}}k}
  J.,    {Zverko} J.,  2011, \aap, 534, L5

\bibitem[\protect\citeauthoryear{{Mullan}}{{Mullan}}{1986}]{1986A&A...165..157M}
{Mullan} D.~J.,  1986, \aap, 165, 157

\bibitem[\protect\citeauthoryear{{Najarro}, {Hanson} \& {Puls}}{{Najarro}
  et~al.}{2011}]{2011A&A...535A..32N}
{Najarro} F.,  {Hanson} M.~M.,    {Puls} J.,  2011, \aap, 535, A32

\bibitem[\protect\citeauthoryear{{Naz{\'e}}}{{Naz{\'e}}}{2009}]{2009A&A...506.1055N}
{Naz{\'e}} Y.,  2009, \aap, 506, 1055

\bibitem[\protect\citeauthoryear{{Naz{\'e}}, {Bagnulo}, {Petit}, {Rivinius},
  {Wade}, {Rauw} \& {Gagn{\'e}}}{{Naz{\'e}} et~al.}{2012a}]{2012MNRAS.423.3413N}
{Naz{\'e}} Y.,  {Bagnulo} S.,  {Petit} V.,  {Rivinius} T.,  {Wade} G.,  {Rauw}
  G.,    {Gagn{\'e}} M.,  2012a, \mnras, 423, 3413

\bibitem[\protect\citeauthoryear{{Naz{\'e}}, {Broos}, {Oskinova}, {Townsley},
  {Cohen}, {Corcoran}, {Evans}, {Gagn{\'e}}, {Moffat}, {Pittard}, {Rauw},
  {ud-Doula} \& {Walborn}}{{Naz{\'e}} et~al.}{2011}]{2011ApJS..194....7N}
{Naz{\'e}} Y.,  {Broos} P.~S.,  {Oskinova} L.,  {Townsley} L.~K.,  {Cohen} D.,
  {Corcoran} M.~F.,  {Evans} N.~R.,  {Gagn{\'e}} M.,  {Moffat} A.~F.~J.,
  {Pittard} J.~M.,  {Rauw} G.,  {ud-Doula} A.,    {Walborn} N.~R.,  2011,
  \apjs, 194, 7

\bibitem[\protect\citeauthoryear{{Naz{\'e}}, {Rauw}, {Pollock}, {Walborn} \&
  {Howarth}}{{Naz{\'e}} et~al.}{2007}]{2007MNRAS.375..145N}
{Naz{\'e}} Y.,  {Rauw} G.,  {Pollock} A.~M.~T.,  {Walborn} N.~R.,    {Howarth}
  I.~D.,  2007, \mnras, 375, 145

\bibitem[\protect\citeauthoryear{{Naz{\'e}}, {ud-Doula}, {Spano}, {Rauw}, {De
  Becker} \& {Walborn}}{{Naz{\'e}} et~al.}{2010}]{2010A&A...520A..59N}
{Naz{\'e}} Y.,  {ud-Doula} A.,  {Spano} M.,  {Rauw} G.,  {De Becker} M.,
  {Walborn} N.~R.,  2010, \aap, 520, A59

\bibitem[\protect\citeauthoryear{{Naz{\'e}}, {Zhekov} \& {Walborn}}{{Naz{\'e}}
  et~al.}{2012b}]{2012ApJ...746..142N}
{Naz{\'e}} Y.,  {Zhekov} S.~A.,    {Walborn} N.~R.,  2012b, \apj, 746, 142

\bibitem[\protect\citeauthoryear{{Neiner}, {Alecian}, {Briquet}, {Floquet},
  {Fr{\'e}mat}, {Martayan}, {Thizy} \& {the MiMeS Collaboration}}{{Neiner}
  et~al.}{2012}]{2012A&A...537A.148N}
{Neiner} C.,  {Alecian} E.,  {Briquet} M.,  {Floquet} M.,  {Fr{\'e}mat} Y.,
  {Martayan} C.,  {Thizy} O.,    {the MiMeS Collaboration} 2012, \aap, 537,
  A148

\bibitem[\protect\citeauthoryear{{Neiner}, {Geers}, {Henrichs}, {Floquet},
  {Fr{\'e}mat}, {Hubert}, {Preuss} \& {Wiersema}}{{Neiner}
  et~al.}{2003}]{2003A&A...406.1019N}
{Neiner} C.,  {Geers} V.~C.,  {Henrichs} H.~F.,  {Floquet} M.,  {Fr{\'e}mat}
  Y.,  {Hubert} A.,  {Preuss} O.,    {Wiersema} K.,  2003a, \aap, 406, 1019

\bibitem[\protect\citeauthoryear{{Neiner}, {Grunhut}, {Petit}, {ud-Doula} \&
  {Wade}}{{Neiner} et~al.}{2012a}]{neiner_omegaori}
{Neiner} C.,  {Grunhut} J.~H.,  {Petit} V.,  {ud-Doula} A.,    {Wade} G.~A.,
  2012a, \mnras, submitted

\bibitem[\protect\citeauthoryear{{Neiner}, {Henrichs}, {Floquet}, {Fr{\'e}mat},
  {Preuss}, {Hubert}, {Geers}, {Tijani}, {Nichols} \& {Jankov}}{{Neiner}
  et~al.}{2003}]{2003bA&A...411..565N}
{Neiner} C.,  {Henrichs} H.~F.,  {Floquet} M.,  {Fr{\'e}mat} Y.,  {Preuss} O.,
  {Hubert} A.,  {Geers} V.~C.,  {Tijani} A.~H.,  {Nichols} J.~S.,    {Jankov}
  S.,  2003b, \aap, 411, 565

\bibitem[\protect\citeauthoryear{{Neiner}, {Hubert}, {Fr{\'e}mat}, {Floquet},
  {Jankov}, {Preuss}, {Henrichs} \& {Zorec}}{{Neiner}
  et~al.}{2003c}]{2003A&A...409..275N}
{Neiner} C.,  {Hubert} A.-M.,  {Fr{\'e}mat} Y.,  {Floquet} M.,  {Jankov} S.,
  {Preuss} O.,  {Henrichs} H.~F.,    {Zorec} J.,  2003c, \aap, 409, 275

\bibitem[\protect\citeauthoryear{{Neiner}, {Landstreet}, {Alecian},
  {Kochukhov}, {Bohlender} \& {the MiMeS Collaboration}}{{Neiner}
  et~al.}{2012b}]{Neiner2012_96446}
{Neiner} C.,  {Landstreet} J.~D.,  {Alecian} E.,  {Kochukhov} O.,  {Bohlender}
  D.~A.,    {the MiMeS Collaboration} 2012b, \aap, submitted

\bibitem[\protect\citeauthoryear{{Oksala}, {Wade}, {Townsend}, {Owocki},
  {Kochukhov}, {Neiner}, {Alecian} \& {Grunhut}}{{Oksala}
  et~al.}{2012}]{2012MNRAS.419..959O}
{Oksala} M.~E.,  {Wade} G.~A.,  {Townsend} R.~H.~D.,  {Owocki} S.~P.,
  {Kochukhov} O.,  {Neiner} C.,  {Alecian} E.,    {Grunhut} J.,  2012, \mnras,
  419, 959

\bibitem[\protect\citeauthoryear{{Oskinova}, {Todt}, {Ignace}, {Brown},
  {Cassinelli} \& {Hamann}}{{Oskinova} et~al.}{2011}]{2011MNRAS.416.1456O}
{Oskinova} L.~M.,  {Todt} H.,  {Ignace} R.,  {Brown} J.~C.,  {Cassinelli}
  J.~P.,    {Hamann} W.-R.,  2011, \mnras, 416, 1456

\bibitem[\protect\citeauthoryear{{Pedersen}}{{Pedersen}}{1979}]{1979A&AS...35..313P}
{Pedersen} H.,  1979, \aaps, 35, 313

\bibitem[\protect\citeauthoryear{{Petit}}{{Petit}}{2011}]{2010arXiv1010.2248P}
{Petit} V.,  2011, in {C.~Neiner, G.~Wade, G.~Meynet, \& G.~Peters} ed., IAU Symposium Vol.~272,
 Active OB Stars: Structure, Evolution, Mass
Loss, and Critical Limits, Cambridge Univ. Press, Cambridge, p. 208

\bibitem[\protect\citeauthoryear{{Petit}, {Gagne}, {Cohen}, {Townsend},
  {Leutenegger}, {Savoy}, {Fehon} \& {Cartagena}}{{Petit}
  et~al.}{2012}]{2012arXiv1205.3538P}
{Petit} V.,  {Gagne} M.,  {Cohen} D.~H.,  {Townsend} R.~H.~D.,  {Leutenegger}
  M.~A.,  {Savoy} M.~R.,  {Fehon} G.,    {Cartagena} C.~A.,  2012, in Carciofi
  A.,  Rivinius T.,  eds, ASP Conference Series, Circumstellar Dynamics at High Resolution, in
  press (arXiv 1205.3538)

\bibitem[\protect\citeauthoryear{{Petit}, {Massa}, {Marcolino}, {Wade} \&
  {Ignace}}{{Petit} et~al.}{2011}]{2011MNRAS.412L..45P}
{Petit} V.,  {Massa} D.~L.,  {Marcolino} W.~L.~F.,  {Wade} G.~A.,    {Ignace}
  R.,  2011, \mnras, 412, L45

\bibitem[\protect\citeauthoryear{{Petit} \& {Wade}}{{Petit} \&
  {Wade}}{2012}]{2012MNRAS.420..773P}
{Petit} V.,  {Wade} G.~A.,  2012, \mnras, 420, 773

\bibitem[\protect\citeauthoryear{{Petit}, {Wade}, {Drissen}, {Montmerle} \&
  {Alecian}}{{Petit} et~al.}{2008}]{2008MNRAS.387L..23P}
{Petit} V.,  {Wade} G.~A.,  {Drissen} L.,  {Montmerle} T.,    {Alecian} E.,
  2008, \mnras, 387, L23

\bibitem[\protect\citeauthoryear{{Preston}}{{Preston}}{1967}]{1967ApJ...150..547P}
{Preston} G.~W.,  1967, \apj, 150, 547

\bibitem[\protect\citeauthoryear{{Puls}, {Springmann} \& {Lennon}}{{Puls}
  et~al.}{2000}]{2000A&AS..141...23P}
{Puls} J.,  {Springmann} U.,    {Lennon} M.,  2000, \aaps, 141, 23

\bibitem[\protect\citeauthoryear{{Puls}, {Urbaneja}, {Venero}, {Repolust},
  {Springmann}, {Jokuthy} \& {Mokiem}}{{Puls}
  et~al.}{2005}]{2005A&A...435..669P}
{Puls} J.,  {Urbaneja} M.~A.,  {Venero} R.,  {Repolust} T.,  {Springmann} U.,
  {Jokuthy} A.,    {Mokiem} M.~R.,  2005, \aap, 435, 669

\bibitem[\protect\citeauthoryear{{Puls}, {Vink} \& {Najarro}}{{Puls}
  et~al.}{2008}]{2008A&ARv..16..209P}
{Puls} J.,  {Vink} J.~S.,    {Najarro} F.,  2008, \aapr, 16, 209

\bibitem[\protect\citeauthoryear{{Raassen}, {Cassinelli}, {Miller}, {Mewe} \&
  {Tepedelenlio{\v g}lu}}{{Raassen} et~al.}{2005}]{2005A&A...437..599R}
{Raassen} A.~J.~J.,  {Cassinelli} J.~P.,  {Miller} N.~A.,  {Mewe} R.,
  {Tepedelenlio{\v g}lu} E.,  2005, \aap, 437, 599

\bibitem[\protect\citeauthoryear{{Raassen}, {van der Hucht}, {Miller} \&
  {Cassinelli}}{{Raassen} et~al.}{2008}]{2008A&A...478..513R}
{Raassen} A.~J.~J.,  {van der Hucht} K.~A.,  {Miller} N.~A.,    {Cassinelli}
  J.~P.,  2008, \aap, 478, 513

\bibitem[\protect\citeauthoryear{{Ram{\'{\i}}rez}, {Rebull}, {Stauffer},
  {Strom}, {Hillenbrand}, {Hearty}, {Kopan}, {Pravdo}, {Makidon} \&
  {Jones}}{{Ram{\'{\i}}rez} et~al.}{2004}]{2004AJ....128..787R}
{Ram{\'{\i}}rez} S.~V.,  {Rebull} L.,  {Stauffer} J.,  {Strom} S.,
  {Hillenbrand} L.,  {Hearty} T.,  {Kopan} E.~L.,  {Pravdo} S.,  {Makidon} R.,
    {Jones} B.,  2004, \aj, 128, 787

\bibitem[\protect\citeauthoryear{{Reed}}{{Reed}}{2005}]{2005yCat.5125....0R}
{Reed} C.,  2005, VizieR Online Data Catalog, 5125

\bibitem[\protect\citeauthoryear{{Repolust}, {Puls} \& {Herrero}}{{Repolust}
  et~al.}{2004}]{2004A&A...415..349R}
{Repolust} T.,  {Puls} J.,    {Herrero} A.,  2004, \aap, 415, 349

\bibitem[\protect\citeauthoryear{{Rivinius}, {Hummel} \& {Stahl}}{{Rivinius}
  et~al.}{2011}]{2011IAUS..272..539R}
{Rivinius} T.,  {Hummel} C.~A.,    {Stahl} O.,  2011, in {C.~Neiner, G.~Wade, G.~Meynet, \& G.~Peters} ed., IAU Symposium Vol.~272,
 Active OB Stars: Structure, Evolution, Mass
Loss, and Critical Limits, Cambridge Univ. Press, Cambridge,
p. 539

\bibitem[\protect\citeauthoryear{{Rivinius}, {Stahl}, {Baade} \&
  {Kaufer}}{{Rivinius} et~al.}{2003}]{2003IBVS.5397....1R}
{Rivinius} T.,  {Stahl} O.,  {Baade} D.,    {Kaufer} A.,  2003, Information
  Bulletin on Variable Stars, 5397, 1

\bibitem[\protect\citeauthoryear{{Rivinius}, {Townsend}, {Kochukhov},
  {\v{S}tefl}, {Baade}, {Barrera} \& {Szeifert}}{{Rivinius}
  et~al.}{2012}]{RiviniusHR7355_sub}
{Rivinius} T.,  {Townsend} R.~H.~D.,  {Kochukhov} O.,  {\v{S}tefl} S.,  {Baade}
  D.,  {Barrera} L.,    {Szeifert} T.,  2012, \mnras, submitted

\bibitem[\protect\citeauthoryear{{Romanyuk} \& {Kudryavtsev}}{{Romanyuk} \&
  {Kudryavtsev}}{2008}]{2008AstBu..63..139R}
{Romanyuk} I.~I.,  {Kudryavtsev} D.~O.,  2008, Astrophysical Bulletin, 63, 139

\bibitem[\protect\citeauthoryear{{Runacres} \& {Owocki}}{{Runacres} \&
  {Owocki}}{2002}]{2002A&A...381.1015R}
{Runacres} M.~C.,  {Owocki} S.~P.,  2002, \aap, 381, 1015

\bibitem[\protect\citeauthoryear{{Sanz-Forcada}, {Franciosini} \&
  {Pallavicini}}{{Sanz-Forcada} et~al.}{2004}]{2004A&A...421..715S}
{Sanz-Forcada} J.,  {Franciosini} E.,    {Pallavicini} R.,  2004, \aap, 421,
  715

\bibitem[\protect\citeauthoryear{{Shore}, {Bohlender}, {Bolton}, {North} \&
  {Hill}}{{Shore} et~al.}{2004}]{2004A&A...421..203S}
{Shore} S.~N.,  {Bohlender} D.~A.,  {Bolton} C.~T.,  {North} P.,    {Hill}
  G.~M.,  2004, \aap, 421, 203

\bibitem[\protect\citeauthoryear{{Shore} \& {Brown}}{{Shore} \&
  {Brown}}{1990}]{1990ApJ...365..665S}
{Shore} S.~N.,  {Brown} D.~N.,  1990, \apj, 365, 665

\bibitem[\protect\citeauthoryear{{Shultz}, {Wade}, {Grunhut}, {Bagnulo},
  {Landstreet}, {Neiner}, {Alecian}, {Hanes} \& {MiMeS Collaboration}}{{Shultz}
  et~al.}{2012}]{2012ApJ...750....2S}
{Shultz} M.,  {Wade} G.~A.,  {Grunhut} J.,  {Bagnulo} S.,  {Landstreet} J.~D.,
  {Neiner} C.,  {Alecian} E.,  {Hanes} D.,    {MiMeS Collaboration} 2012, \apj,
  750, 2

\bibitem[\protect\citeauthoryear{{Silvester}, {Neiner}, {Henrichs}, {Wade},
  {Petit}, {Alecian}, {Huat}, {Martayan}, {Power} \& {Thizy}}{{Silvester}
  et~al.}{2009}]{2009MNRAS.398.1505S}
{Silvester} J.,  {Neiner} C.,  {Henrichs} H.~F.,  {Wade} G.~A.,  {Petit} V.,
  {Alecian} E.,  {Huat} A.,  {Martayan} C.,  {Power} J.,    {Thizy} O.,  2009,
  \mnras, 398, 1505

\bibitem[\protect\citeauthoryear{{Sim{\'o}n-D{\'{\i}}az}}{{Sim{\'o}n-D{\'{\i}}az}}{2010}]{2010A&A...510A..22S}
{Sim{\'o}n-D{\'{\i}}az} S.,  2010, \aap, 510, A22

\bibitem[\protect\citeauthoryear{{Sim{\'o}n-D{\'{\i}}az},
  {Garc{\'{\i}}a-Rojas}, {Esteban}, {Stasi{\'n}ska}, {L{\'o}pez-S{\'a}nchez} \&
  {Morisset}}{{Sim{\'o}n-D{\'{\i}}az} et~al.}{2011}]{2011A&A...530A..57S}
{Sim{\'o}n-D{\'{\i}}az} S.,  {Garc{\'{\i}}a-Rojas} J.,  {Esteban} C.,
  {Stasi{\'n}ska} G.,  {L{\'o}pez-S{\'a}nchez} A.~R.,    {Morisset} C.,  2011,
  \aap, 530, A57

\bibitem[\protect\citeauthoryear{{Sim{\'o}n-D{\'{\i}}az}, {Herrero}, {Esteban}
  \& {Najarro}}{{Sim{\'o}n-D{\'{\i}}az} et~al.}{2006}]{2006A&A...448..351S}
{Sim{\'o}n-D{\'{\i}}az} S.,  {Herrero} A.,  {Esteban} C.,    {Najarro} F.,
  2006, \aap, 448, 351

\bibitem[\protect\citeauthoryear{{Stahl}, {Wade}, {Petit}, {Stober} \&
  {Schanne}}{{Stahl} et~al.}{2008}]{2008A&A...487..323S}
{Stahl} O.,  {Wade} G.,  {Petit} V.,  {Stober} B.,    {Schanne} L.,  2008,
  \aap, 487, 323

\bibitem[\protect\citeauthoryear{{Stelzer}, {Flaccomio}, {Montmerle}, {Micela},
  {Sciortino}, {Favata}, {Preibisch} \& {Feigelson}}{{Stelzer}
  et~al.}{2005}]{2005ApJS..160..557S}
{Stelzer} B.,  {Flaccomio} E.,  {Montmerle} T.,  {Micela} G.,  {Sciortino} S.,
  {Favata} F.,  {Preibisch} T.,    {Feigelson} E.~D.,  2005, \apjs, 160, 557

\bibitem[\protect\citeauthoryear{{Stibbs}}{{Stibbs}}{1950}]{1950MNRAS.110..395S}
{Stibbs} D.~W.~N.,  1950, \mnras, 110, 395

\bibitem[\protect\citeauthoryear{{Sundqvist}, {Puls}, {Feldmeier} \&
  {Owocki}}{{Sundqvist} et~al.}{2011}]{2011A&A...528A..64S}
{Sundqvist} J.~O.,  {Puls} J.,  {Feldmeier} A.,    {Owocki} S.~P.,  2011, \aap,
  528, A64

\bibitem[\protect\citeauthoryear{{Sundqvist}, {ud-Doula}, {Owocki}, {Townsend},
  {Howarth} \& {Wade}}{{Sundqvist} et~al.}{2012}]{2012MNRAS.423L..21S}
{Sundqvist} J.~O.,  {ud-Doula} A.,  {Owocki} S.~P.,  {Townsend} R.~H.~D.,
  {Howarth} I.~D.,    {Wade} G.~A.,  2012, \mnras, 423, L21

\bibitem[\protect\citeauthoryear{{Townsend}, {Oksala}, {Cohen}, {Owocki} \&
  {ud-Doula}}{{Townsend} et~al.}{2010}]{2010ApJ...714L.318T}
{Townsend} R.~H.~D.,  {Oksala} M.~E.,  {Cohen} D.~H.,  {Owocki} S.~P.,
  {ud-Doula} A.,  2010, \apjl, 714, L318

\bibitem[\protect\citeauthoryear{{Townsend} \& {Owocki}}{{Townsend} \&
  {Owocki}}{2005}]{2005MNRAS.357..251T}
{Townsend} R.~H.~D.,  {Owocki} S.~P.,  2005, \mnras, 357, 251

\bibitem[\protect\citeauthoryear{{Townsend}, {Owocki} \& {Groote}}{{Townsend}
  et~al.}{2005}]{2005ApJ...630L..81T}
{Townsend} R.~H.~D.,  {Owocki} S.~P.,    {Groote} D.,  2005, \apjl, 630, L81

\bibitem[\protect\citeauthoryear{{Townsend}, {Owocki} \& {ud-Doula}}{{Townsend}
  et~al.}{2007}]{2007MNRAS.382..139T}
{Townsend} R.~H.~D.,  {Owocki} S.~P.,    {ud-Doula} A.,  2007, \mnras, 382, 139

\bibitem[\protect\citeauthoryear{{Townsend}, {Rivinius}, {Rowe}, {Moffat},
  {Bohlender}, {Neiner}, {Telting}, {Guenther}, {Kallinger}, {Kuschnig},
  {Matthews}, {Rucinski}, {Sasselov} \& {Weiss}}{{Townsend}
  et~al.}{2012}]{Townsend2012_MOST}
{Townsend} R.~H.~D. et~al., 2012, \apj, submitted

\bibitem[\protect\citeauthoryear{{Turner}}{{Turner}}{1990}]{1990PASP..102.1331T}
{Turner} D.~G.,  1990, \pasp, 102, 1331

\bibitem[\protect\citeauthoryear{{ud-Doula}}{{ud-Doula}}{2008}]{2008cihw.conf..125U}
{ud-Doula} A.,  2008, in {Hamann} W.-R.,  {Feldmeier} A.,   {Oskinova} L.~M.,
  eds, Clumping in Hot-Star Winds, Electronic publication of the University Potsdam,
p.~125

\bibitem[\protect\citeauthoryear{{ud-Doula} \& {Owocki}}{{ud-Doula} \&
  {Owocki}}{2002}]{2002ApJ...576..413U}
{ud-Doula} A.,  {Owocki} S.~P.,  2002, \apj, 576, 413

\bibitem[\protect\citeauthoryear{{ud-Doula}, {Owocki} \& {Townsend}}{{ud-Doula}
  et~al.}{2008}]{2008MNRAS.385...97U}
{ud-Doula} A.,  {Owocki} S.~P.,    {Townsend} R.~H.~D.,  2008, \mnras, 385, 97

\bibitem[\protect\citeauthoryear{{ud-Doula}, {Owocki} \& {Townsend}}{{ud-Doula}
  et~al.}{2009}]{2009MNRAS.392.1022U}
{ud-Doula} A.,  {Owocki} S.~P.,    {Townsend} R.~H.~D.,  2009, \mnras, 392,
  1022

\bibitem[\protect\citeauthoryear{{Uytterhoeven}, {Harmanec}, {Telting} \&
  {Aerts}}{{Uytterhoeven} et~al.}{2005}]{2005A&A...440..249U}
{Uytterhoeven} K.,  {Harmanec} P.,  {Telting} J.~H.,    {Aerts} C.,  2005,
  \aap, 440, 249

\bibitem[\protect\citeauthoryear{{van Leeuwen}}{{van
  Leeuwen}}{2007}]{2007A&A...474..653V}
{van Leeuwen} F.,  2007, \aap, 474, 653

\bibitem[\protect\citeauthoryear{{Vink}, {de Koter} \& {Lamers}}{{Vink}
  et~al.}{2000}]{2000A&A...362..295V}
{Vink} J.~S.,  {de Koter} A.,    {Lamers} H.~J.~G.~L.~M.,  2000, \aap, 362, 295

\bibitem[\protect\citeauthoryear{{Vink}, {de Koter} \& {Lamers}}{{Vink}
  et~al.}{2001}]{2001A&A...369..574V}
{Vink} J.~S.,  {de Koter} A.,    {Lamers} H.~J.~G.~L.~M.,  2001, \aap, 369, 574

\bibitem[\protect\citeauthoryear{{Wade}, {Alecian}, {Bohlender}, {Bouret} \&
  {Cohen}}{{Wade} et~al.}{2011a}]{2010arXiv1009.3563W}
{Wade} G.~A. et~al.,
  2011a, in {C.~Neiner, G.~Wade, G.~Meynet, \& G.~Peters} ed., IAU Symposium Vol.~272,
 Active OB Stars: Structure, Evolution, Mass
Loss, and Critical Limits, Cambridge Univ. Press, Cambridge,
p.~118

\bibitem[\protect\citeauthoryear{{Wade}, {Apell{\'a}niz}, {Martins}, {Petit} \&
  {Grunhut}}{{Wade} et~al.}{2012a}]{2012MNRAS.425.1278W}
{Wade} G.~A. et~al.,  2012a, \mnras, 425, 1278

\bibitem[\protect\citeauthoryear{{Wade}, {Bagnulo}, {Kochukhov}, {Landstreet},
  {Piskunov} \& {Stift}}{{Wade} et~al.}{2001}]{2001A&A...374..265W}
{Wade} G.~A.,  {Bagnulo} S.,  {Kochukhov} O.,  {Landstreet} J.~D.,  {Piskunov}
  N.,    {Stift} M.~J.,  2001, \aap, 374, 265

\bibitem[\protect\citeauthoryear{{Wade}, {Bohlender}, {Brown}, {Elkin},
  {Landstreet} \& {Romanyuk}}{{Wade} et~al.}{1997}]{1997A&A...320..172W}
{Wade} G.~A.,  {Bohlender} D.~A.,  {Brown} D.~N.,  {Elkin} V.~G.,  {Landstreet}
  J.~D.,    {Romanyuk} I.~I.,  1997, \aap, 320, 172

\bibitem[\protect\citeauthoryear{{Wade}, {Fullerton}, {Donati}, {Landstreet},
  {Petit} \& {Strasser}}{{Wade} et~al.}{2006}]{2006A&A...451..195W}
{Wade} G.~A.,  {Fullerton} A.~W.,  {Donati} J.-F.,  {Landstreet} J.~D.,
  {Petit} P.,    {Strasser} S.,  2006, \aap, 451, 195

\bibitem[\protect\citeauthoryear{{Wade}, {Grunhut}, {Gr{\"a}fener}, {Howarth},
  {Martins}, {Petit}, {Vink}, {Bagnulo}, {Folsom}, {Naz{\'e}}, {Walborn},
  {Townsend} \& {Evans}}{{Wade} et~al.}{2012b}]{2012MNRAS.419.2459W}
{Wade} G.~A. et~al., 2012b, \mnras,
  419, 2459

\bibitem[\protect\citeauthoryear{{Wade}, {Howarth}, {Townsend}, {Grunhut},
  {Shultz}, {Bouret}, {Fullerton}, {Marcolino}, {Martins}, {Naz{\'e}},
  {ud-Doula}, {Walborn}, {Donati} \& {the MiMeS Collaboration}}{{Wade}
  et~al.}{2011b}]{2011MNRAS.416.3160W}
{Wade} G.~A. et~al., 2011b, \mnras, 416, 3160

\bibitem[\protect\citeauthoryear{{Walborn}}{{Walborn}}{1975}]{1975PASP...87..613W}
{Walborn} N.~R.,  1975, \pasp, 87, 613

\bibitem[\protect\citeauthoryear{{Walborn} \& {Nichols}}{{Walborn} \&
  {Nichols}}{1994}]{1994ApJ...425L..29W}
{Walborn} N.~R.,  {Nichols} J.~S.,  1994, \apjl, 425, L29

\bibitem[\protect\citeauthoryear{{Wang}, {Townsley}, {Feigelson}, {Broos},
  {Getman}, {Rom{\'a}n-Z{\'u}{\~n}iga} \& {Lada}}{{Wang}
  et~al.}{2008}]{2008ApJ...675..464W}
{Wang} J.,  {Townsley} L.~K.,  {Feigelson} E.~D.,  {Broos} P.~S.,  {Getman}
  K.~V.,  {Rom{\'a}n-Z{\'u}{\~n}iga} C.~G.,    {Lada} E.,  2008, \apj, 675, 464

\bibitem[\protect\citeauthoryear{{Weber} \& {Davis} Jr.}{{Weber} \&
  {Davis}}{1967}]{1967ApJ...148..217W}
{Weber} E.~J.,  {Davis} Jr. L.,  1967, \apj, 148, 217

\bibitem[\protect\citeauthoryear{{Wolff}}{{Wolff}}{1990}]{1990AJ....100.1994W}
{Wolff} S.~C.,  1990, \aj, 100, 1994

\bibitem[\protect\citeauthoryear{{Wolff}, {Strom} \& {Hillenbrand}}{{Wolff}
  et~al.}{2004}]{2004ApJ...601..979W}
{Wolff} S.~C.,  {Strom} S.~E.,    {Hillenbrand} L.~A.,  2004, \apj, 601, 979

\bibitem[\protect\citeauthoryear{{Yakunin}, {Romanyuk}, {Kudryavtsev} \&
  {Semenko}}{{Yakunin} et~al.}{2011}]{2011AN....332..974Y}
{Yakunin} I.,  {Romanyuk} I.,  {Kudryavtsev} D.,    {Semenko} E.,  2011,
  Astronomische Nachrichten, 332, 974

\bibitem[\protect\citeauthoryear{{Zacharias}, {Monet}, {Levine}, {Urban},
  {Gaume} \& {Wycoff}}{{Zacharias} et~al.}{2005}]{2005yCat.1297....0Z}
{Zacharias} N.,  {Monet} D.~G.,  {Levine} S.~E.,  {Urban} S.~E.,  {Gaume} R.,
   {Wycoff} G.~L.,  2005, VizieR Online Data Catalog, 1297

\bibitem[\protect\citeauthoryear{{Zboril} \& {North}}{{Zboril} \&
  {North}}{2000}]{2000CoSka..30...12Z}
{Zboril} M.,  {North} P.,  2000, Contributions of the Astronomical Observatory
  Skalnate Pleso, 30, 12

\bibitem[\protect\citeauthoryear{{Zboril}, {North}, {Glagolevskij} \&
  {Betrix}}{{Zboril} et~al.}{1997}]{1997A&A...324..949Z}
{Zboril} M.,  {North} P.,  {Glagolevskij} Y.~V.,    {Betrix} F.,  1997, \aap,
  324, 949

\end{thebibliography}

\appendix

\section{Notes on individual stars}
\label{sec:notes}

This section contains additional remarks about certain stars, marked by a dagger in Table \ref{tab:stars}, concerning our choice of parameters, providing alternative calculation in case of disagreements in the literature, or specific information about binarity and other relevant characteristics. 
The sections are numbered according to the ID number of each stars, as given in column (1) of Table \ref{tab:stars}.

\setcounter{subsection}{2}
\subsection{HD\,37022 ($\theta^1$\,Ori\,C)}

$\theta^1$\,Ori\,C is a single-lined spectroscopic and astrometric binary with an 11-yr period and $e\approx0.6$ \citep{2009A&A...497..195K}.

\setcounter{subsection}{3}
\subsection{HD\,191612}

HD\,191612 is a double-lined spectroscopic binary with $P=1548$\,d and $e=0.5$, and a early B-type companion \citep{2007MNRAS.381..433H,2011MNRAS.416.3160W}. 

\setcounter{subsection}{5}
\subsection{HD\,47129 (Plaskett's star)}

Plaskett's star is a high-mass ($M_\mathrm{tot}\sin i=93\msol$) O+O double-lined spectroscopic binary with a short period ($P=14$\,d) circular orbit \citep{2008A&A...489..713L}. The magnetic field is associated with the rapidly rotating ($\vsini=300$\,km\,\sm) secondary star \citep{Grunhut2012_plaskett}. 

\setcounter{subsection}{7}
\subsection{ALS\,15218 (Tr\,16-22)}

We use the effective temperature and luminosity derived from the cluster photometry analysis of \citet{2011ApJS..194....5G}. We assume $\log(g)=4.0$. 

\setcounter{subsection}{9}
\subsection{HD\,37742 ($\zeta$\,Ori\,Aa)}

HD\,37742 is an astrometric double-lined spectroscopic binary \citep{2000ApJ...540L..91H} with a O9\,Ib primary and an early B-type star companion. The preliminary dynamic mass derived by \cite{2011IAUS..272..539R} ($2.48\pm5.6$\,$\msol$) is smaller than the mass derived by \citet{2008MNRAS.389...75B} ($39\pm8$\,$\msol$).

A comparison of the disentangled component spectra with the published Zeeman magnetic signature confirms that the signature cannot originate from component Ab, since its lines are too narrow (Th. Rivinius, priv. com).

\setcounter{subsection}{11}
\subsection{HD\,37061 (NU\,Ori)}

NU\,Ori is a double-lined spectroscopic binary with $P=19$\,d and $e=0.14$ \citep{1991ApJ...367..155A}, with a magnetic early B-type primary and a late B-type companion \citep{2008MNRAS.387L..23P}.

\setcounter{subsection}{17}
\subsection{HD\,205021 ($\beta$\,Cep)}

$\beta$\,Cep is a double-lined spectroscopic binary. The magnetic primary (component A) is the prototype of a class of pulsating hot stars. The secondary (component Aa) is an H$\alpha$-emitting classical Be star \citep{2008MNRAS.387..759C}.

\setcounter{subsection}{18}
\subsection{ALS\,15211 (Tr16-13)}

We use the effective temperature and luminosity derived from the cluster photometry analysis of \citet{2011ApJS..194....5G}. We assume $\log(g)=4.0$. 	

\setcounter{subsection}{19}
\subsection{HD\,122451 ($\beta$\,Cen)}

$\beta$\,Cen is a double-lined spectroscopic binary with $P=356$\,d and $e=0.8$ \citep{2006A&A...455..259A}, and components of nearly identical mass. The magnetic field detection is associated with the narrow-line primary star \citep{2011A&A...536L...6A}. 

\setcounter{subsection}{23}
\subsection{HD\,96446 (V\,430\,Car)}

\citet{1991A&A...243..148M} observed photometric variations with a period of 0.85\,d, as well as other shorter periods interpreted as $\beta$\,Cep-type pulsations. However, other photometric periods are possible and compatible with the low-amplitude variations of the longitudinal field measurements \citep{Neiner2012_96446}. We use the long period of 5.73\,d. The shortest period of 0.85\,d would yield $W=0.46$ and $\rk=1.7\rs$.

\setcounter{subsection}{29}
\subsection{HD\,37017 (V\,1046\,Ori)}

V\,1046\,Ori is a double-lined spectroscopic binary with a 18.6\,d period ($e=0.4$). The field detection is associated with the B2 He-strong primary \citep{1998A&A...337..183B}. The companion is a late B-type star. 

\setcounter{subsection}{31}
\subsection{HD\,149277}

An inspection of archival HARPS and MiMeS ESPaDonS spectra revealed that HD\,149277 is a double-lined spectroscopic binary with the magnetic field detection associated with the lower $\vsini$ component (Petit et al. in prep). 

\setcounter{subsection}{33}
\subsection{HD\,37776 (V\,901\,Ori)}

We choose the temperature and luminosity from \citet{2007A&A...470..685L}, in order to be consistent with the radius used by \citet{2011ApJ...726...24K} in their magnetic analysis. \citet{2011ApJ...726...24K} have shown that the field structure is much more complex than a dipole. Their surface magnetic field reconstruction displays a surface field varying from 5\,kG to 30\,kG across the stellar surface (see their Fig.~4). As their associated mean magnetic field modulus varies between 13 and 16\,kG, we use a dipolar strength of 15\,kG to estimate the wind confinement. It is important to keep in mind that the resulting magnetospheric structure will be complex, and cannot be described in detail by a global $\ra$, as testified by the complex H$\alpha$ variations (Shultz et al. in prep). 

\setcounter{subsection}{34}
\subsection{HD\,136504 ($\epsilon$\,Lup)}

$\epsilon$\,Lup is an eccentric double-lined spectroscopic binary \citep[$P=4.6$\,d, $e=0.28$;][]{2005A&A...440..249U}, with similar components. The magnetic field measurements found in the literature \citep{2009AN....330..317H,2012ApJ...750....2S} do not specify which component is magnetic, but follow-up MiMeS observations show that the field is associated with the primary star and that the published longitudinal field measurements are underestimated by a factor of two. 

\citet{2005A&A...440..249U} also found a possible period of 1.2\,d for the primary. This would lead to $W=0.38$ and $\rk=1.9\rs$.

\setcounter{subsection}{39}	
\subsection{HD\,186205}

The SIMBAD database gives a spectral type of B5, although \cite{1975PASP...87..613W} classified it as B2Vp He-strong. The effective temperature determinations are varied: 17\,kK \citep{2000CoSka..30...12Z} and 23.5\,kK \citep{1977A&A....60..259L}. An analysis of a MiMeS observation suggests $\teff=20$\,kK and $\log g=4.0$.

\setcounter{subsection}{41}
\subsection{HD\,200775 (V\,3780\,Cep)}

HD\,200775 is a double-lined spectroscopic binary with a period $\sim4$\,yr ($e=0.3$) with components of similar temperatures. The magnetic field is associated with the sharp-lined primary star \citep{2008MNRAS.385..391A}.
	
\setcounter{subsection}{47}
\subsection{HD\,58260}

According to the various parameter determinations reported by \citet{2007A&A...468..263C} and \citet{1989ApJ...346..459B}, we opt for $\teff=20$\,kK and $\log g=3.55$. 

\citet{1979A&AS...35..313P} report a possible period of 1.657\,d, based on spectrophotometry of the He\textsc{i} $\lambda$4026\,\AA\ line, with a variation of the order of 0.01\,mag. \citet{2005A&A...430.1143B} phased the dozen available longitudinal field measurements with this period, however the amplitude of the variation is relatively small. We therefore use the $\vsini$ as the lower limit on the equatorial velocity. The 1.6-d period, if confirmed, would yield $W=0.55$ and $\rk=1.5\rs$, and an inclination angle of the rotation axis near zero.

\setcounter{subsection}{48}
\subsection{HD\,36485 ($\delta$\,Ori\,C)}	
	
$\delta$\,Ori\,C is a double-lined spectroscopic binary with a 30-d period ($e=0.32$). \citet{2010MNRAS.401.2739L} determined $T_\mathrm{eff}=20$\,kK and $M=7$\,M$_\odot$ for the magnetic primary, and $T_\mathrm{eff}=10$\,kK and $M=2.8$\,M$_\odot$ for the secondary star (with a $\Delta V$ difference of 1.3\,mag). 

\citet{2010MNRAS.401.2739L} determined a dipolar field strength between 7.3 and 12\,kG. We use a mean value of 10\,kG for our calculations. 

\setcounter{subsection}{49}
\subsection{HD\,208057 (16\,Peg)}

The star was reported once to display H$\alpha$ emission \citep{1943ApJ....98..153M}, but there exists no confirmation of this emission \citep{2009IAUS..259..393H}. 

\setcounter{subsection}{51}
\subsection{HD\,25558 (40\,Tau)}

A magnetic field detection was reported by \citet{2009AN....330..317H}, but was refuted by \citet{2012A&A...538A.129B} based on a re-analysis of the same dataset. However, a weaker field was detected with MiMeS observations.

\setcounter{subsection}{52}
\subsection{HD\,35298}

Although \citet{2005A&A...430.1143B} reported a longitudinal field with extremum at 3\,kG, \citet{2011AN....332..974Y} found some larger values up to 5\,kG but also reported large variation in the field determined with lines from different elements. 
The photometric period used by \citet{2005A&A...430.1143B} was confirmed by additional longitudinal field measurements at the Dominion Astrophysical Observatory, with an extremum around 5\,kG (D. Bohlender, priv com).
We opt for a conservative lower limit on the dipolar strength of 9\,kG. 

\setcounter{subsection}{54}	
\subsection{HD\,142990 (V\,913\,Sco)}

There are many different effective temperature determinations in the literature, ranging from 16.5 to 18.5\,kK \citep[e.g.][]{2007A&A...468..263C}. We use 17.5\,kK as a mean value and assume $\log(g)=4.0$.

\setcounter{subsection}{55}	
\subsection{HD\,37058 (V\,359\,Ori)}

\citet{2007AstBu..62..319G} determined $T_\mathrm{eff}=17$\,kK and $\log(g)=3.80$ from spectral fitting. 
\citet{2007A&A...470..685L} found a higher temperature of 20\,kK. 
We however prefer the lower temperature given that the star is He-weak. 

A period of 14\,d was reported by \citet{1979A&AS...35..313P} based on spectrophotometry of He lines. On the other hand, \citet{2005A&A...430.1143B} phased the sparse longitudinal field measurements from the literature with a period of 1.022\,d. An additional measurement was taken by \citet{2006A&A...450..777B}, but the period is not precise enough to test the phasing. We therefore use the conservative 14\,d period. The shorter period would lead to $W=0.57$ and $\rk=1.5\rs$. 

\setcounter{subsection}{56}	
\subsection{HD\,35502}

HD\,35502 is a hierarchical spectroscopic triple system with a broad-lined magnetic B-type primary (HD\,35502\,A) and a companion (HD\,35502\,Bab) composed of two sharp-lined A-type stars (Bohlender et al. in prep).
\citet{1981ApJ...249L..39B} lists a possible period of 1.7\,d, however this period is not compatible with the new longitudinal field measurements ($P=0.85$\,d).

\setcounter{subsection}{59}	
\subsection{HD\,61556 (HR\,2949)}

HD\,61555/6 is a visual pair, the light of which is combined in the Hipparcos identifier HIP\,37229 \citep{2003IBVS.5397....1R}. Both components were observed separately in the context of the MiMeS Project, and a magnetic field was detected only for HD\,61556. 

\setcounter{subsection}{60}	
\subsection{HD\,175362 (Wolff's star)}

There is a large scatter of effective temperature determinations in the literature, from 14 to 17\,kK. We adopt the temperature ($\teff=15$\,kK) derived by \citet{1997A&A...320..257L} and $\log g=4.0$.

\setcounter{subsection}{61}	
\subsection{HD\,105382 (HR\,4618)}

Although often classified at a Be star, \citet{2001A&A...366..121B} have shown that this is not the case and the Be classifications probably resulted from accidentally observing the very nearby, well known Be star $\delta$\,Cen.

\end{document}